\documentclass[traditabstract]{aa}

\usepackage{natbib}
\usepackage{graphicx}

\newcommand{\FWHM}{{\it{FWHM}}}

\usepackage{natbib}

\defcitealias{Fischera20013a}{Paper~I}

\begin{document}

\title{On the probability distribution function of the mass surface density of molecular clouds.~II}
\titlerunning{On the PDF of the  mass surface density of molecular clouds.~II}

\author{J\"org Fischera}
\institute{Canadian Institute for Theoretical Astrophysics, University of Toronto, 60 St. George Street, Toronto, ON M5S3H8, 
        Canada\label{inst1}
        }

%\email{fischera@cita.utoronto.ca}

\abstract{The probability distribution function (PDF) 
        of the mass surface density of molecular clouds provides essential information
        about the structure of molecular cloud gas and condensed structures out of which stars may form.
        In general, the PDF shows two basic components: a broad distribution around the maximum
        with resemblance to a log-normal function,
        and a tail at high mass surface densities attributed to turbulence and
        self-gravity. In a previous paper, the PDF of condensed structures
        has been analyzed and an analytical formula presented based on a truncated radial density profile,
        $\rho(r) = \rho_{\rm c}/(1+(r/r_0)^2)^{n/2}$ with central density $\rho_c$ and inner radius $r_0$,
        widely used in astrophysics as a generalization of physical density profiles. 
        In this paper, the results are applied to analyze the PDF of self-gravitating, isothermal,
        pressurized, spherical (Bonnor-Ebert spheres) and cylindrical condensed structures with
        emphasis on the dependence of the PDF
        on the external pressure $p_{\rm ext}$ and on the overpressure $q^{-1} =p_{\rm c}/p_{\rm ext}$,
        where $p_{\rm c}$ is the central pressure. 
        Apart from individual clouds, we also consider ensembles of spheres or cylinders,
        where effects caused by a variation of pressure ratio, a distribution of condensed cores
        within a turbulent gas, and (in case of cylinders) a distribution of
        inclination angles on the mean PDF are analyzed.
        The probability distribution of pressure ratios $q^{-1}$ is assumed to be given by
        $P(q^{-1}) \propto q^{-k_1}/(1+(q_0/q)^\gamma)^{(k_1+k_2)/\gamma}$ 
        , where $k_1$, $\gamma$, $k_2$, and $q_0$ are fixed parameters.
        The PDF of individual spheres with overpressures below $\sim 100$
        is well represented by the PDF of a sphere with an analytical density profile with $n=3$. At higher 
        pressure ratios, the PDF at mass surface densities $\Sigma\ll\Sigma(0)$, where $\Sigma(0)$ is the
        central mass surface density, asymptotically
approaches the PDF of a sphere with $n=2$. Consequently, the
        power-law asymptote at mass surface densities above the peak steepens from 
        $P_{\rm sph}(\Sigma)\propto \Sigma^{-2}$
        to $P_{\rm sph}(\Sigma)\propto \Sigma^{-3}$. The corresponding asymptote of the PDF of cylinders for the large $q^{-1}$ is approximately given by 
        $P_{\rm cyl}(\Sigma)\propto \Sigma^{-4/3}(1-(\Sigma/\Sigma(0))^{2/3})^{-1/2}$.
        The distribution of overpressures $q^{-1}$ produces a power-law asymptote
        at high mass surface densities given by $\left<P_{\rm sph}(\Sigma)\right>\propto \Sigma^{-2 k_2-1}$ (spheres) 
        or $\left<P_{\rm cyl}(\Sigma)\right>\propto \Sigma^{-2 k_2}$ (cylinders). 
        }

\keywords{ISM: molecular clouds numerical methods: statistics, analytical}

\maketitle

\section{Introduction}

The observed probability distribution function (PDF) of the mass surface density 
of molecular clouds suggests
a combination of two separate  components that produce a broad distribution around
the peak, which is generally modeled using a log-normal function 
and a tail at high mass surface densities that often has a power-law form 
(e.g., \citet{Kainulainen2009, Kainulainen2013,Schneider2012,Schneider2013}).
The broad distribution around the maximum is typically attributed to turbulence of the 
molecular gas, the tail to self-gravitating structures. 
The relative amount of gas in gravitationally
bound structures seems to be indicative of the star formation rate in the cloud.
The PDF of star-forming molecular clouds like Taurus or Ophiuchus, for example,
is characterized by a strong tail that seems to be very low or absent in
clouds with no apparent star formation, such as  the 'Coalsack' \citep{Kainulainen2009}.

Images of the column densities of molecular clouds \citep{Schneider2012,Kainulainen2013} show that the high column density 
values are not randomly distributed within the cloud, but are indeed related to small regions,
as expected for cold condensed structures surrounded by warmer or more
turbulent gas. In a previous paper \citep[Paper~I]{Fischera2014a} the PDFs of spherical 
and cylindrical condensed structures have been analyzed assuming a truncated smooth 
radial density profile used in astrophysics as a generalization of physical density profiles. It was found
that the PDF of such a profile can be described by a simple implicit analytical function. 
In Paper~I, we showed how the geometric shape, the radial density profile,
and the ratio$\rho_{\rm c}/\rho(r_{\rm cl})$
of the central density and the density at the peripheral regions affect the functional form of the PDF.
For example, we found that the PDF of a sphere is truncated, 
while the PDF of a cylinder has a pole at the highest mass surface density. 

In this paper, the results of Paper~I are applied to analyze the PDF of isothermal 
self-gravitating pressurized spheres and cylinders as an approximation of individual condensations
in giant molecular clouds. In the case of the global statistical properties of molecular clouds,
the condensed structures probably show a variety of 
different physical conditions. The consequences of a number of distributions on the
mean PDF are therefore also addressed.

The paper is divided into two main sections.
In the first part (Sect.~\ref{sect_pdfmain}), the properties of the PDFs of individual spheres
and cylinders are discussed. We show how the highest position and the asymptotes at
low and high mass surface densities are affected by the pressure ratio $p_{\rm c}/p_{\rm ext}$
and the external pressure $p_{\rm ext}$ surrounding the structures.
The second part (Sect.~\ref{sect_pdfmean}) is about
the properties of a mean PDF for a variety of distributions as they may apply to molecular clouds.
In Sect.~\ref{sect_pdfdistribution},
we analyze the mean PDF of an ensemble of spheres and cylinders for a distribution of
gravitational states. In Sect.~\ref{sect_pdfcylmeanangle}, the effect
caused by a distribution of cylinders with a variety of inclination angles is addressed,
and in Sect.~\ref{sect_pdfsphmeanbgrd}, we consider a mean PDF
of cores that are uniformly distributed within a turbulent medium.
The results are discussed in Sect.~\ref{sect_discussion} and compared with observed PDFs,
as derived by \citet{Kainulainen2009}.
The paper is summarized in Sect.~\ref{sect_summary}.

\section{\label{sect_pdfmain}Probability distribution function of the mass surface density of condensed structures}

\subsection{\label{sect_densityprofile}Density profile of the isothermal condensed structures}

In Paper~I, we have analyzed the one-point statistic for condensed spheres and cylinders
assuming an analytical density profile given by
\begin{equation}
        \label{eq_densityprofile}
        \rho(r) = \frac{\rho_{\rm c}}{(1+(r/r_0)^2)^{n/2}},\end{equation}
where $\rho_{\rm c}$ is the central density and $r_0$ the inner radius. 
In the following we show how the density profile is related to the density profile of self-gravitating
isothermal spheres, known as Bonnor-Ebert spheres \citep{Ebert1955,Bonnor1956}, and cylinders.

In isothermal clouds, the gas pressure and density are proportional, $p = K\rho$, where the proportionality 
constant is given by $K = kT / ({\mu}m_{\rm H})$, where $k$, $T$, $\mu$, and $m_{\rm H}$ are the Boltzmann constant, 
the effective temperature, the mean molecular weight, and the   mass of a hydrogen atom. 
For an idealized gas we have $K=c^2_{\rm s}$, where $c_{\rm s}$ is the sound speed.

To analyze isothermal self-gravitating clouds, it is convenient to introduce a unit-free radius $\theta = rA$.
The constant is given by $A^2=(4\pi G\rho_{\rm c})/K$, where $G$ is the gravitational constant. 
The radius $r_{0}$ is related to the effective temperature $T$ and the central density 
through $r_0 = \sqrt{\xi_n} /A,$ where $\xi_n$ is a fixed appropriate constant for given $n$
so that $r/r_0 = \theta/\sqrt{\xi_n}$. The parameters for the considered density profiles are 
listed in Table~\ref{table_densprofconst}. For Bonnor-Ebert spheres
the analytical density profiles are only representations of the correct density profile 
at low and and in the limit of high $\theta$. The corresponding constants $\xi_n$ are chosen,
as described in Sect.~\ref{sect_isosphere}, to
match the correct density profiles where the approximations are valid.

The pressure ratio $q^{-1} = p_{\rm c}/p_{\rm ext}$
of the central pressure $p_{\rm c}$ and the external pressure $p_{\rm ext}$ are referred to as `overpressure'.
As a reference pressure we assume a mean interstellar medium
(ISM) pressure $p_{\rm ext}/k=2\times 10^4~{\rm K\,cm^{-3}}$
, as previously suggested by \citet{Curry2000} based on the study of \citet{Boulares1990}, where
the pressure is attributed to thermal, turbulent, and magnetic pressure components. The same value
has recently been inferred from the physical parameters and profiles 
of filaments \citep{Fischera2012a,Fischera2012b}.

\begin{figure}[htbp]
        \includegraphics[width=\hsize]{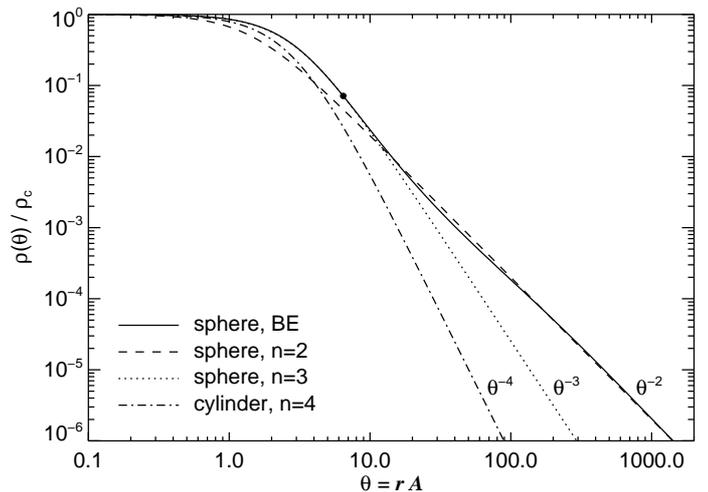}
        \caption{\label{fig_densityprofile}Normalized radial density profiles of some cases considered in this study
        given as a function of the
        unit-free radius $\theta=rA$ where $A^2 = (4\pi G \rho_c)/K$. 
        The filled symbol marks the condition for a critically stable
        sphere where $\theta_{\rm crit}\approx 6.451$ and $\rho_{\rm c}/\rho(\theta_{\rm crit})\approx 14.04$.
        The curves of the analytical density profile given by Eq.~\ref{eq_densityprofile} 
        are labeled with their asymptotic behavior for $\theta\gg \sqrt{\xi_n}$.} 
\end{figure}

For completeness, we discuss in App.~\ref{app_pdfdens} the PDF of the local density for spheres and cylinders.

\begin{table}[htbp]
        \caption{\label{table_densprofconst}Parameters of analytical density profiles}
        \begin{tabular}{cc|cc}
                & isothermal cylinder & \multicolumn{2}{c}{Bonnor-Ebert sphere}\\
                & (exact)       &$\theta<10$ & $\theta\gg 10$ \\
        $n$     & 4      & $3$  & $2$  \\
        $\xi_n$ & 8     & $8.63$ & $2$\\
        \end{tabular}
\end{table}

\subsubsection{\label{sect_isosphere}Bonnor-Ebert sphere}

The density profile of an isothermal self-gravitating sphere, or Bonnor-Ebert sphere, 
is given by 
\begin{equation}
        \rho(\theta) = \rho_{\rm c} e^{-\omega(\theta)},
\end{equation}
where $\omega=\phi/K$ is the unit-free potential 
of the gravitational potential $\phi$
determined by the Lane-Emden equation,
\begin{equation}
        \label{eq_laneemden}
        \frac{1}{\theta^2}\frac{{\rm d}}{{\rm d}\theta}\left[\theta^2\frac{{\rm d}\omega}{{\rm d}\theta}\right] = e^{-\omega}.
\end{equation} 

Spheres have a well-known critically stable configuration with an overpressure above which the sphere
becomes gravitationally unstable under compression \citep{Ebert1955,Bonnor1956,McCrea1957, 
Curry2000,Fischera2008,Fischera2011,Fischera2012a} (see App.~\ref{sect_critmass}). The critically 
stable configuration is characterized through
an overpressure of $p_{\rm c}/p_{\rm ext}\sim 14.04$ at $\theta_{\rm crit}=6.451$. However, theoretically,
equilibrium solutions
with higher overpressures can also be constructed. As in previous papers \citep{Fischera2011,Fischera2012a},
we refer to spheres with overpressures below the critical value as subcritical and spheres with higher overpressure
as supercritical. { The term `supercritical' is often used in the literature to characterize clouds with  masses
above the critically stable   mass. The  supercritical spheres in this paper
have masses that are \emph{lower} than the mass of a critically stable sphere if we let the temperature and external pressure remain constant, as discussed by \citet{Fischera2012a} (see also App.~\ref{app_isocomparison}). 

As shown in Fig.~\ref{fig_densityprofile}, the profile of the Bonner-Ebert sphere
below $\theta=10$ or $\rho(r)>\sim 0.01\,\rho_{\rm c}$ is well represented by
an analytical density profile as given by Eq.~\ref{eq_densityprofile} with $n=3$. 
For the constant $\xi_3$ we have chosen a value that
is consistent with the overpressure and the size of a critically stable sphere:
\begin{equation}
        {\xi_3} = \left(\theta_{\rm crit} q_{\rm crit}^{1/3}/\sqrt{1-q_{\rm crit}^{2/3}}\right)^2\approx 8.63.
\end{equation}
This value is used throughout the paper.

At large sizes $\theta\gg 10$ the density profile of the Bonnor-Ebert sphere
approaches asymptotically approaches a power-law profile $\rho(\theta)/\rho_{\rm c}\propto C/\theta^2$ 
where $C$ is a constant. Inserting the power-law profile in Eq.~\ref{eq_laneemden}, we find $C=2$.
The same asymptotic behavior is found for an analytical density profile as given in Eq.~\ref{eq_densityprofile} 
with $n=2$ and $\xi_2={2}$ in the limit of large sizes $z\gg\sqrt{\xi_2}$.
We consider this profile as the asymptotic profile of Bonnor-Ebert spheres for high overpressures. 
Although stable configurations with overpressure above the critical value probably do not exist, the 
profiles of sub- and supercritical solutions 
seem to reflect within the observed uncertainties the profiles not only of stable clouds, but also seem to mimic the density profile of collapsing clouds \citep{Kandori2005,Keto2010,Keto2014}.

Physical parameters of Bonnor-Ebert spheres such as mass and radius are discussed 
in more detail in App.~\ref{app_isocomparison} in comparison with spheres with a density 
profile with $n=3$ or $n=2$.
In App.~\ref{app_pdfdensbonnor}, we show the variation of the local density PDF with overpressure.

\subsubsection{Self-gravitating cylinders}
For $n=4$
the density profile is identical to the profile of an isothermal self-gravitating cylinder with 
$\xi_4 = 8$ \citep{Stodolkiewicz1963,Ostriker1964}. The physical parameters of pressurized 
cylinders have been studied in great detail by \citet{Fischera2012b}. 
Here, we wish to describe the most important properties
for understanding the PDF of the mass surface density of cylinders. 

As an important characteristic, a highest possible
  mass line density $(M/l)_{\rm max}=2K/G$ exists for cylinders,
which corresponds to a cylinder with infinite overpressure and 
infinitely small size. As shown in \citet{Fischera2012b}, for
example, physical properties such as size, full width at half
maximum (\FWHM),
and stability considerations can be expressed through the   mass ratio 
$f_{\rm cyl}=(M/l)/(M/l)_{\rm max}$ of the   mass line density
and the highest possible   mass line density. It is related to the pressure ratio $q$ by
\begin{equation}
        q = (1-f_{\rm cyl})^2.
\end{equation}
Cylinders with high  mass ratio $f_{\rm cyl}$ have a steep density profile $\rho(r)\propto r^{-4}$ at radii $\theta\gg \sqrt{8}$.
However, Fig.~\ref{fig_densityprofile} shows that this asymptotic profile is only established 
at a relatively large radius $\theta\gg \sim 10,$ which approximately corresponds to cylinders with
an overpressure $q^{-1}>\sim 100$ or equally to cylinders with mass ratio $f_{\rm cyl}>\sim 0.9$. Measurements of 
the density profile of observed filaments in general indicate a flatter density profile more consistent with $r^{-2}$ , as discussed by
\citet{FiegePudritz2000a},
suggesting significantly lower mass ratios. Indications for a filament with a high  mass ratio and a possible steep 
density profile are found for a filament in the cloud IC~5146 \citep{Fischera2012b}.\footnote{
On the other hand, there are claims that star-forming filaments in particular have mass ratios 
partly considerably higher than the theoretical highest value \citep{Arzoumanian2011}. However, this might be related to 
the low temperature considered in observational studies \citep{Fischera2012a}.}

\subsection{\label{sect_msurfprofile}Profile of the mass surface density}

The profile of the  mass surface density of a pressurized, isothermal, self-gravitating sphere or cylinder seen
at inclination angle $i$ is  in general given by
\begin{equation}
        \label{eq_msurfiso}
        \Sigma(\theta_{\bot}) = \frac{2}{\cos^\lambda i}\sqrt{\frac{1}{q}} \sqrt{\frac{p_{\rm ext}}{4\pi G}} 
                \int_{\theta_{\bot}}^{\theta_{\rm cl}}{d}\theta\, \frac{\theta_{\bot}}{\sqrt{\theta^2-\theta_{\bot}^2}}
                \frac{\rho\left(\theta\right)}{\rho_{\rm c}},
\end{equation}
where $\theta_\bot$ and $\theta_{\rm cl}$ are the projected unit-free radius and cloud radius. $\lambda$ is 
a power index with $\lambda=0$ for
spheres and $\lambda=1$ for cylinders. In this convention the inclination angle is $i=0^\circ$ for a cylinder seen from the side.
Profiles for spheres and cylinders
for a range of overpressures can be found in the work of \citet{Fischera2012a}. 
It is noticeable that the profile does not depend on the gas temperature. Consequently, isothermal clouds 
in a certain pressure region with the same overpressure will have the same profile $\Sigma(\theta)$
independently of their mass or mass line density, which vary for given overpressure with
$M_{\rm sph}\propto T^2/\sqrt{p_{\rm ext}}$ or $[M/l]_{\rm cyl}\propto T/\sqrt{p_{\rm ext}}$ 
(App.~\ref{app_cloudmass}).  Their profiles only vary in terms of size as for fixed overpressure 
$r_{\rm cl}\propto T/\sqrt{p_{\rm ext}}$ (App.~\ref{app_cloudradius}).

The profiles of the mass surface density for a sphere and a cylinder with a truncated radial density profile as given
by Eq.~\ref{eq_densityprofile} were presented in Paper~I. It was found to be convenient to introduce a unit-free
mass surface density $X_n$ defined by
\begin{equation}
        \label{eq_defxn}
        \Sigma_n = \frac{2}{\cos^\lambda i} r_0\rho_{\rm c} q^{\frac{n-1}{n}} X_n.
\end{equation}
The unit-free mass surface density is given by
\begin{equation}
        \label{eq_defxn2}
        X_n(y_n) = (1-y_n)^{\frac{1-n}{2}}\int_0^{u_{\rm max}} {\rm d}u\, \frac{1}{(1+u^2)^{n/2}},
\end{equation}
where $y_n=(1-q^{2/n})(1-x^2)$ and where $x=r_\bot /r_{\rm cl}$ is the ratio of the projected radius $r_\bot$
and the cloud radius $r_{\rm cl}$ , which we refer to as the
normalized impact parameter.
The upper limit of the integral is $u_{\rm max}=\sqrt{y_n/(1-y_n)}$. 
For density profiles with $n=2$, 3, and 4 the profiles are simple analytical
functions and can be found in Paper~I. 

For isothermal self-gravitating pressurized clouds, we can replace the inner radius using $r_{0}=\sqrt{\xi_n}/A$
and the mass surface density becomes
\begin{equation}
        \label{eq_xnsnrelation}
        \Sigma_n = \frac{2}{\cos^\lambda i} \sqrt{\frac{\xi_n p_{\rm ext}}{4\pi G}} q^{\frac{n-2}{2 n}}\,X_n.
\end{equation}
We note that for $n=2$ the relation is independent of the pressure ratio $q$. 

The mass surface density profiles through Bonnor-Ebert spheres is shown in Fig.~\ref{fig_msurfprofilesph}.
At small projected radius $\theta_{\bot}\ll 10$ the profile is dominated by the innermost part of the radial density profile and we
can use the profile of the analytical density profile with $n=3$ as approximation. At large projected radius $\theta_\bot$ the profile
of a Bonnor-Ebert sphere fluctuates asymptotically toward a power law $\Sigma\propto 1/{\theta_\bot}$.
With the exception of small $\theta_\bot,$ the mass surface densities of a Bonnor-Ebert sphere is approximately 
described by a sphere with an analytical density profile with $n=2$.

\begin{figure}[htbp]
        \includegraphics[width=\hsize]{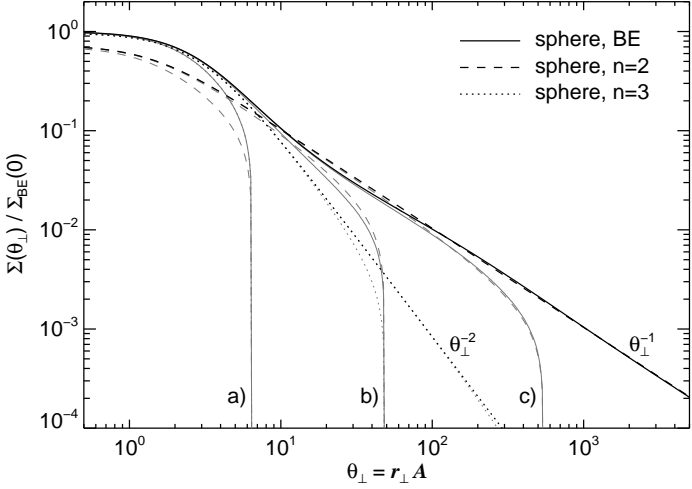}
        \caption{\label{fig_msurfprofilesph}Profiles of the mass surface density through Bonnor-Ebert
        spheres and spheres with an analytical density profile as given in Eq.~\ref{eq_densityprofile}.
        The gray curves are profiles for different cloud radii $\theta_{\rm cl}$. 
        For Bonnor-Ebert spheres the cloud radius
        corresponds to $q=q_{\rm crit}$ (a), $q=10^{-2} q_{\rm crit}$ (b), and $q=10^{-4}q_{\rm crit}$ (c)
,        where $q_{\rm crit}=1/14.04$. For each assumed cloud radius the central pressure $p_{\rm c}$
        in the considered spheres is the same. The profiles are normalized by the corresponding 
        central mass surface density through a Bonnor-Ebert sphere.}
\end{figure}

To compare the PDF of Bonnor-Ebert spheres and the PDF of the approximations 
it is convenient to measure the mass surface density in units of
\begin{equation}
        \label{eq_msurfconstant}
        \hat\Sigma= 2\sqrt{\frac{p_{\rm ext}}{4\pi G}}= 0.00363\sqrt{\frac{p_{\rm ext}/k}{2\times 10^{4}~{\rm K\,cm^{-3}}}}~{\rm g\,cm^{-2}}.
\end{equation}
For the truncated analytical density profile with index $n$ the mass surface density becomes
\begin{equation}
        \label{eq_msurfnormalized}
        \Sigma_{n}(y_n)  = \frac{1}{\cos^\lambda i}\hat \Sigma\,q^{\frac{n-2}{2n}}\sqrt{\xi_n} X_n(y_n).
\end{equation}

It is common to measure the projected density in visual extinction $A_V$ assuming the
optical properties derived for the diffuse interstellar medium. To provide an estimate of the mass surface
density in units of extinction $A_V$ we assumed a gas-to-dust ratio of  
$N_{\rm H}/E(B-V)=5.8\times 10^{21}~{\rm cm^{-2}mag^{-1}}$ \citep{Bohlin1978} and an 
absolute-to-relative extinction $R_V$ of 3.1 \citep{Fitzpatrick1999} so that
\begin{equation}
        A_{\rm V}/\Sigma = \frac{R_V}{\bar\mu m_{\rm H}}\left(\frac{N_{\rm H}}{E(B-V)}\right)^{-1}\approx 228~{\rm mag\, g^{-1}\,cm^{2}},
\end{equation}
where $\bar \mu=1.4$ is the assumed mean atomic gas mass in units of a hydrogen atom with mass $m_{\rm H}$.

The critically stable Bonnor-Ebert sphere ($q_{\rm crit}=1/14.04$) has a normalized central   mass surface density of $\Sigma_{\rm crit}(0)\sim10.0705\hat\Sigma,$ which
corresponds to a central visual extinction of $A_V\sim 8.3~{\rm mag}\sqrt{p_{\rm ext}/k/2\times 10^{4}~{\rm K\,cm^{-3}}}$.

As a special important case we give the behavior of the central mass surface density in the limit of 
high overpressures. 
In this regime the central mass surface density through Bonnor-Ebert spheres is approximately given by 
(Eq.~C.4, \citealt{Fischera2012a})
\begin{equation}
        \Sigma_{\rm BE}(0) \approx \hat \Sigma \frac{1}{\sqrt{q}} 3.028.
\end{equation}
For spheres and cylinders with analytical density profiles as given in Eq.~\ref{eq_densityprofile},
the upper limit of the integral in Eq.~\ref{eq_defxn2} behaves as $u_{\rm max}\rightarrow \infty$ 
and the central mass surface density becomes approximately
\begin{equation}
        \label{eq_msurfapproxhigh}
        \Sigma_n(0) 
         = \frac{1}{\cos^\lambda i}\hat\Sigma \sqrt{\frac{\xi_n}{q}} \frac{1}{2}{\rm B}\left(\frac{n-1}{2},\frac{1}{2}\right),
\end{equation}
where 
\begin{equation}
        {\rm B}(a,b) = \frac{\Gamma(a)\Gamma(b)}{\Gamma(a+b)}
\end{equation}
is the beta-function and where $\Gamma(x)$ is the $\Gamma$-function.
For a sphere with $n=3$, we obtain with the constant $\xi_3=8.63$ in the
limit of high overpressures
\begin{equation}
        \Sigma_3(0)\approx \hat\Sigma\frac{1}{\sqrt{q}}\,2.94,
\end{equation}
which is only $3\%$ lower than the correct asymptotic value of the Bonnor-Ebert sphere.

\subsection{\label{sect_pdfisosphere}Probability distribution function of Bonnor-Ebert spheres}

The probability of measuring a mass surface density $\Sigma$ of an individual sphere
in the range from $\Sigma ... \Sigma+{\rm d}\Sigma$ is given by
\begin{equation}
        \label{eq_pdf}
        P(\Sigma)\,{\rm d}{\Sigma} = P(r_\bot)\left(-\frac{{\rm d}\Sigma}{{\rm d}r_\bot}\right)^{-1} {\rm d}\Sigma,
\end{equation}
where $P(r_\bot)$ is the probability of measuring a projected radius $r_\bot$. For a sphere,
we have $P(r_\bot) = 2\pi r_\bot / (\pi r_{\rm cl}^2)$.

The derivative of the mass surface density of a cloud with a radial density profile 
is in general given by
\begin{eqnarray}
        \label{eq_dmsurfgeneral}
        \frac{{\rm d} \Sigma }{{\rm d}r_\bot}(r_\bot) &=& 2
                \Bigg\{\int\limits_{r_\bot}^{r_{\rm cl}}{\rm d}r\, \frac{{\rm d} \rho(r) }{{\rm d}r} \frac{r_{\bot}}{\sqrt{r^2-r_{\bot}^2}}\nonumber\\
                &&\quad\quad-\,\frac{r_\bot}{\sqrt{r_{\rm cl}^2-r_\bot^2}} \rho(r_{\rm cl}) \Bigg\},
\end{eqnarray}
where $r_{\rm cl}$ is the cloud radius and $r_\bot$ the projected radius. The integral only applies to the first term in brackets.
After substituting the density with $\rho(\theta)=\rho_{\rm c}e^{-\omega(\theta)}$  , we obtain
\begin{eqnarray}
        \label{eq_dmsurfiso}
        \frac{{\rm d}\Sigma}{{\rm d}r_\bot}(\theta_\bot) &=& 
        -2\rho_{\rm c}
        \Bigg\{
        \int_{\theta_\bot}^{\theta_{\rm cl}} {\rm d}\theta\,e^{-\omega(\theta)}\frac{{\rm d} \omega}{{\rm d}\theta}\,
                \frac{\theta_\bot}{\sqrt{\theta^2-\theta_\bot^2}}\nonumber\\
        &&\quad\quad+\frac{\theta_\bot}{\sqrt{\theta_{\rm cl}^2-\theta_\bot^2}}\,e^{-\omega(\theta_{\rm cl})}\Bigg\},
\end{eqnarray}
where we have replaced $r$, $r_{\bot}$, and $r_{\rm cl}$ through the corresponding unit-free radii $\theta=rA$,
$\theta_{\bot}=r_\bot A$, and $\theta_{\rm cl}=r_{\rm cl} A$ with $A = \sqrt{(4\pi G\rho_{\rm c})/K}$.
Inserting the expression in Eq.~\ref{eq_pdf} provides for the PDF of the Bonnor-Ebert sphere
\begin{eqnarray}
        \label{eq_pdfisosph}
        P_{\rm BE}(\Sigma(\theta_\bot)) & = & \sqrt{\frac{4\pi G}{p_{\rm ext}}}\frac{\sqrt{q}}{\theta_{\rm cl}^2}
                        \,\Bigg\{\int\limits_{\theta_\bot}^{\theta_{\rm cl}}{\rm d}\theta\,\frac{e^{-\omega(\theta)}}{\sqrt{\theta^2-\theta_{\bot}^2}}
                        \frac{{\rm d}\omega}{{\rm d}\theta}\nonumber\\&&\quad\quad\quad\quad
                        +\frac{e^{-\omega(\theta_{\rm cl})}}{\sqrt{\theta_{\rm cl}^2-\theta_\bot^2}}\Bigg\}^{-1},
\end{eqnarray}
where $e^{-\omega(z_{\rm cl})}=q,$ and where we have again replaced the radius and the projected radius 
by the corresponding unit-free expressions
and $K \rho_{\rm c} $ with $p_{\rm ext}/q$. 

The PDF of Bonnor-Ebert spheres is compared with spheres with an analytical density profile 
as given in Eq.~\ref{eq_densityprofile} with $n=3$ and $n=2$. The PDF of spheres with such
a profile with arbitrary $n$ is given by the implicit function (Paper~I)
\begin{equation}
        \label{eq_pdfsphere}
        P_{\rm sph}(\Sigma_n(y_n)) = \left(2 r_0\rho_{\rm c} q^{\frac{n-1}{n}}\right)^{-1} P_{\rm sph}(X_n(y_n)),
\end{equation}
where 
\begin{equation}
        \label{eq_pdfsphere2}
        P_{\rm sph}(X_n(y_n)) = \frac{2}{1-q^{2/n}}\,\frac{\sqrt{y_n}(1-y_n)}{[1+(n-1) \sqrt{y_n} X_n(y_n)]}.\end{equation}

It is straightforward to show that for the approximations of the radial density profile of Bonnor-Ebert spheres
the two different expressions \ref{eq_pdfisosph} and \ref{eq_pdfsphere} with \ref{eq_pdfsphere2} for the PDF are identical
(App.~\ref{app_pdfisoasymptote}).

\begin{figure}[htbp]
        \includegraphics[width=\hsize]{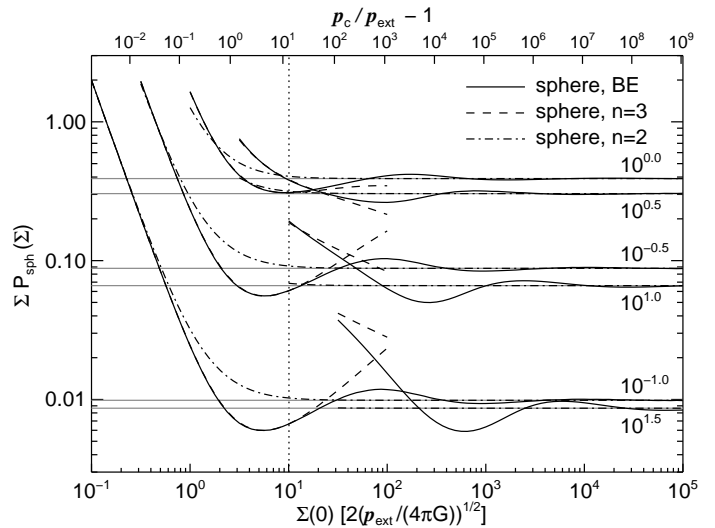}
        \caption{\label{fig_pdfasymptotes}Probabilities of a number of mass surface densities through Bonnor-Ebert spheres 
        shown as a function of the central mass surface density. 
        The curves are labeled by the corresponding mass surface density given in units of $\hat \Sigma=2\sqrt{p_{\rm ext}/(4\pi G)}$.
        The probabilities are compared with those of analytical density profiles as given in Eq.~\ref{eq_densityprofile} with
        $n=3$ and $n=2$. The probabilities for $n=3$ are only shown for $\Sigma/\hat \Sigma<100$ where the
        analytical density profile becomes a valid approximation of the density profile of a Bonnor-Ebert sphere.
        The overpressures given in the upper
        axis are only valid for Bonnor-Ebert spheres. The gray lines show the asymptotic probability for Bonnor-Ebert spheres
        in the limit of infinite overpressure. The vertical dotted line marks the location of the critically stable Bonnor-Ebert sphere.}
\end{figure}

The behavior of the PDF of Bonnor-Ebert spheres with increasing overpressure
is shown in Fig.~\ref{fig_pdfasymptotes}. At low overpressures the probabilities are well described
by a sphere with an analytical density profile with $n=3$. The probability for increasing overpressure fluctuates
asymptotically toward the value of a sphere with an analytical density profile with n=2 in the limit
of infinite overpressure.

It is common in observational studies of molecular clouds to measure the logarithmic PDF $P(\ln \Sigma)$ 
of the mass surface density by estimating the number
of mass surface densities $\Sigma$ within a logarithmic mass surface density bin $\Delta \ln \Sigma$.
This is motivated by the large scales in mass surface densities involved, but also by the fact that
turbulent cloud structures should show a log-normal density distribution (see Sect.~\ref{sect_pdfsphmeanbgrd}).
As  $P(\Sigma)\,{\rm d}\Sigma = \Sigma P(\Sigma)\,{\rm d}\Sigma/\Sigma = \Sigma P(\Sigma)\,{\rm d}\ln \Sigma$
we have $P(\ln \Sigma) = \Sigma P(\Sigma)$. In Paper~I, $P(\Sigma)$ is therefore referred to as linear 
PDF and $\Sigma P(\Sigma)$ as logarithmic PDF of the mass surface density $\Sigma$. It follows from Eqs.~\ref{eq_msurfiso} 
and~\ref{eq_pdfisosph} that the amplitude of the logarithmic PDF of Bonnor-Ebert spheres
is independent of the external pressure. The same is true for clouds with analytical density profiles 
as $X_n P(X_n) = \Sigma_n P(\Sigma_n)$. The statistical properties of the clouds are thus
visualized using the logarithmic PDF.

\begin{figure}
        \includegraphics[width=\hsize]{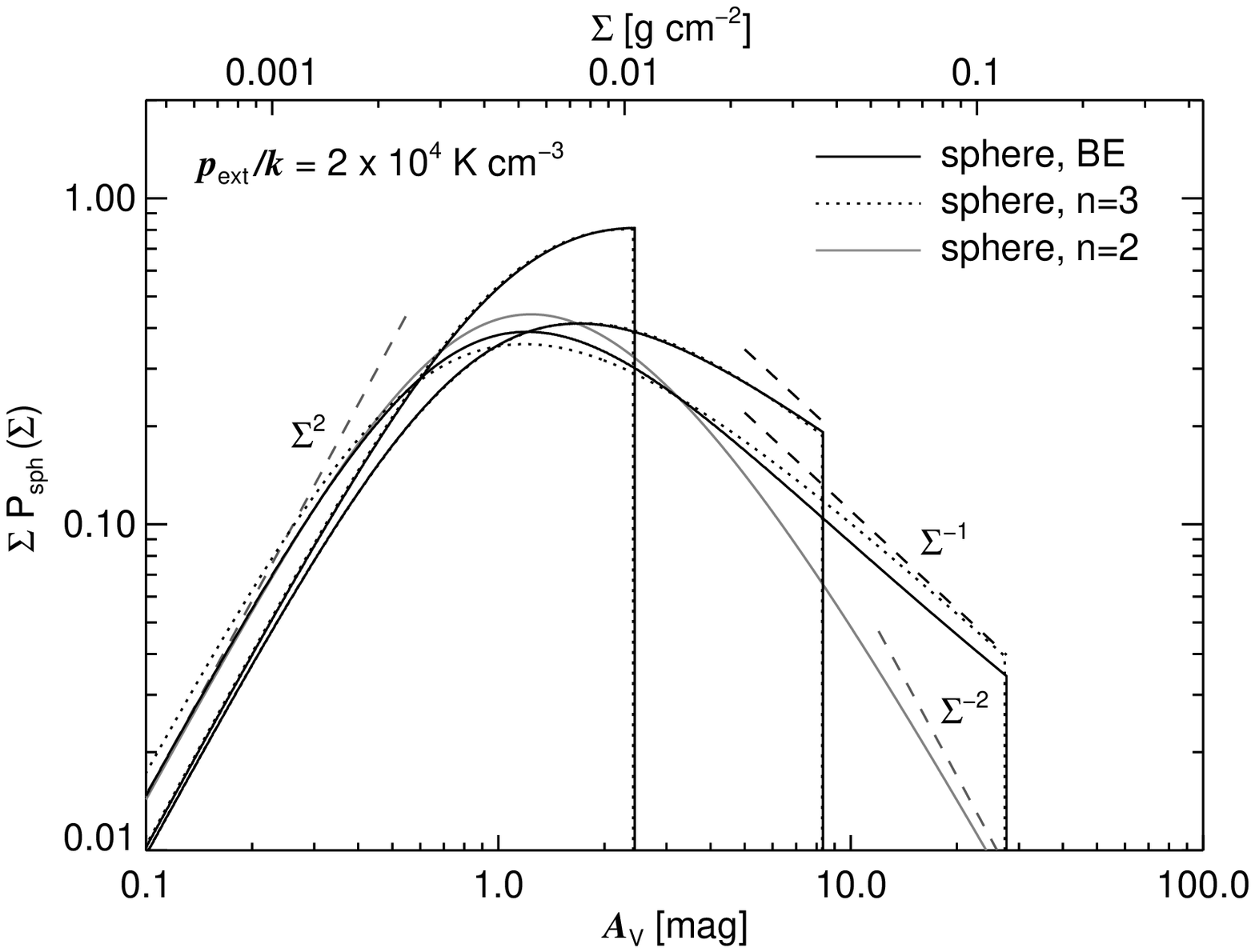}
        \caption{\label{fig_pdfisosphere}PDF of the mass surface density of Bonnor-Ebert spheres embedded in
        a medium with $p_{\rm ext}/k = 2\times 10^{4}~{\rm K\,cm^{-3}}$ for different gravitational states (solid black curves). 
        The curves correspond to overpressures  $q^{-1} = 1+(q^{-1}_{\rm crit}-1) \times 10^i$ with $i=-1,0,1$ where
        $q^{-1}_{\rm crit} = 14.04$ is the overpressure of a critically stable Bonnor-Ebert sphere.
        The PDFs are directly compared with those
        of spheres with a simple density profile (Eq.~\ref{eq_densityprofile}) with $n=3$ (dotted curves).
        The black dashed lines are asymptotes of the PDF $\Sigma_3P_{\rm sph}(\Sigma_3)$
        in the limit of high mass surface densities (Eq.~\ref{eq_pdfsphapproxlargen3}). The PDF of a Bonnor-Ebert sphere
        in the limit of infinite overpressure is shown as a gray curve and is derived assuming a sphere with an analytical
        density profile with $n=2$. The gray dashed lines are the corresponding power-law asymptotes 
        in the limit of low (Eq.~\ref{eq_pdfsphapproxsmall}) and high (Eq.~\ref{eq_pdfsphapproxlargen2}) mass surface densities. 
        }
\end{figure}

\begin{figure}
        \includegraphics[width=\hsize]{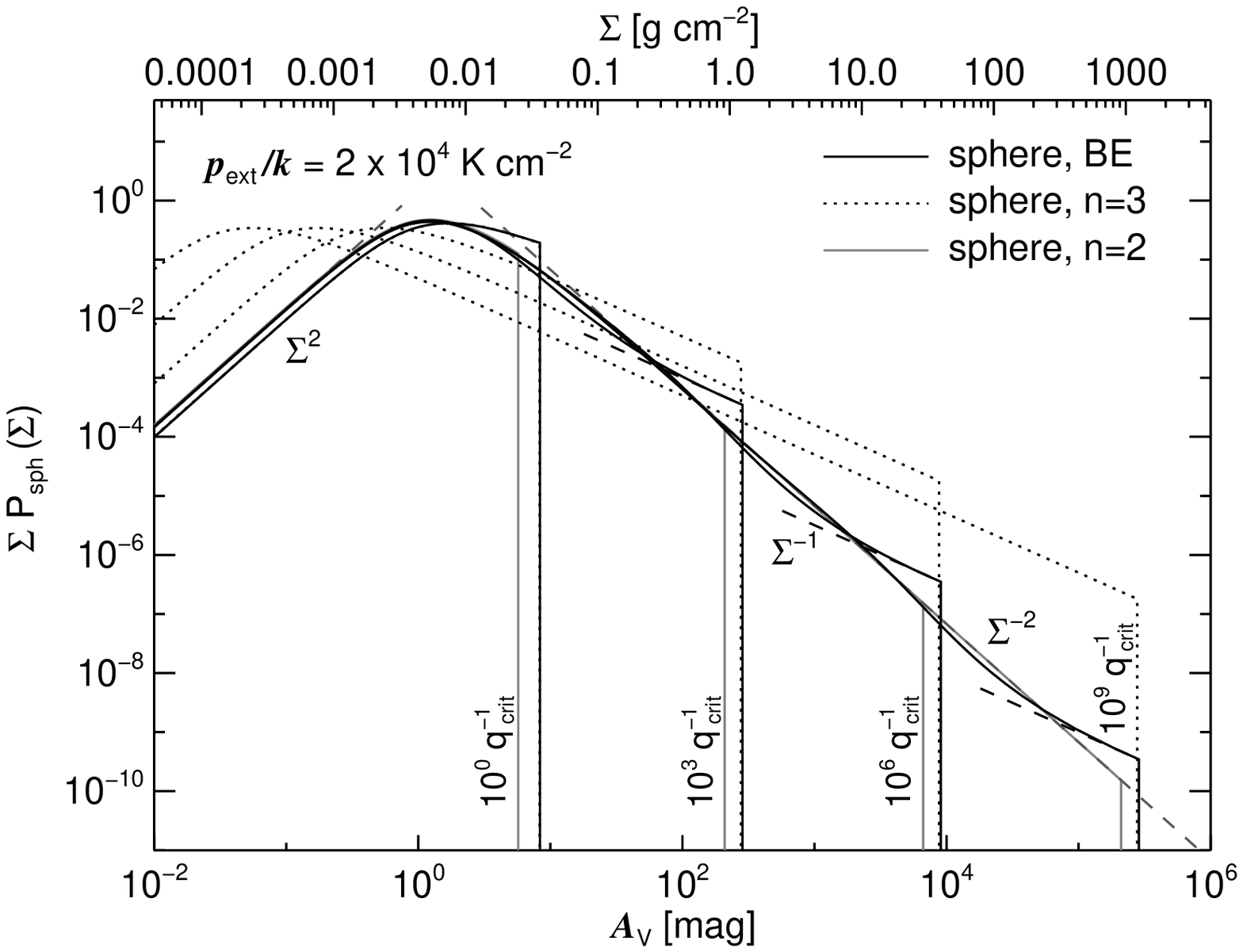}
        \caption{\label{fig_pdfisosphere2}Similar to Fig.~\ref{fig_pdfisosphere}, but for a wider range
        of pressure ratios. We show the transition from a PDF of a critically stable Bonnor-Ebert sphere 
        to the PDFs of highly supercritical spheres. The overpressure is varied 
        in units of the critical value $q_{\rm crit}^{-1}=14.04$. Also shown are the corresponding
        PDFs for a sphere with the analytical density profile with $n=2$ and $n=3$. The power-law asymptotes 
        close to the truncation point (black dashed lines) are scaled to the PDF of the Bonnor Ebert spheres
        at central mass surface density. 
        }
\end{figure}

The PDFs of Bonnor-Ebert spheres for a series of overpressures 
are shown in Figs.~\ref{fig_pdfisosphere} and \ref{fig_pdfisosphere2}. 
We might interpret the series of curves
as those of an initially stable cloud that becomes gravitationally unstable through cooling or
mass accretion and finally collapses.

The PDF shows characteristics discussed in Paper~I for spheres with a truncated analytical density profiles as given
in Eq.~\ref{eq_densityprofile}, for example, the cutoff at the central mass surface
density, the broad maximum above a certain overpressure, and 
the power-law asymptote in the limit of low mass surface densities where $\Sigma P(\Sigma)\propto \Sigma^2$. 
For high overpressures, the PDF shows oscillations at high mass surface densities related to
the oscillations of the radial density profile (Fig.~\ref{fig_densityprofile}) 
or the profile of the mass surface densities (Fig.~\ref{fig_msurfprofilesph}).
With decreasing mass surface density, the same PDFs fluctuate  
asymptotically toward the PDF of a sphere with an $n=2$-profile. 
A quantitative analysis of the highest position and the  
properties at low and high mass surface densities of the PDF of Bonnor-Ebert spheres is given
in Sects. \ref{sect_pdfmaximum}, \ref{sect_pdfsphapproxsmall}, and \ref{sect_pdfsphapproxlarge}.

\subsubsection{\label{sect_pdfmaximum}Maxima position}

\begin{figure*}[htbp]
%        \includegraphics[width=0.486\hsize]{pdfiso_xmax.eps}
%        \hfill
%        \includegraphics[width=0.49\hsize]{pdfiso_msurfmax.eps}
        \includegraphics[width=0.486\hsize]{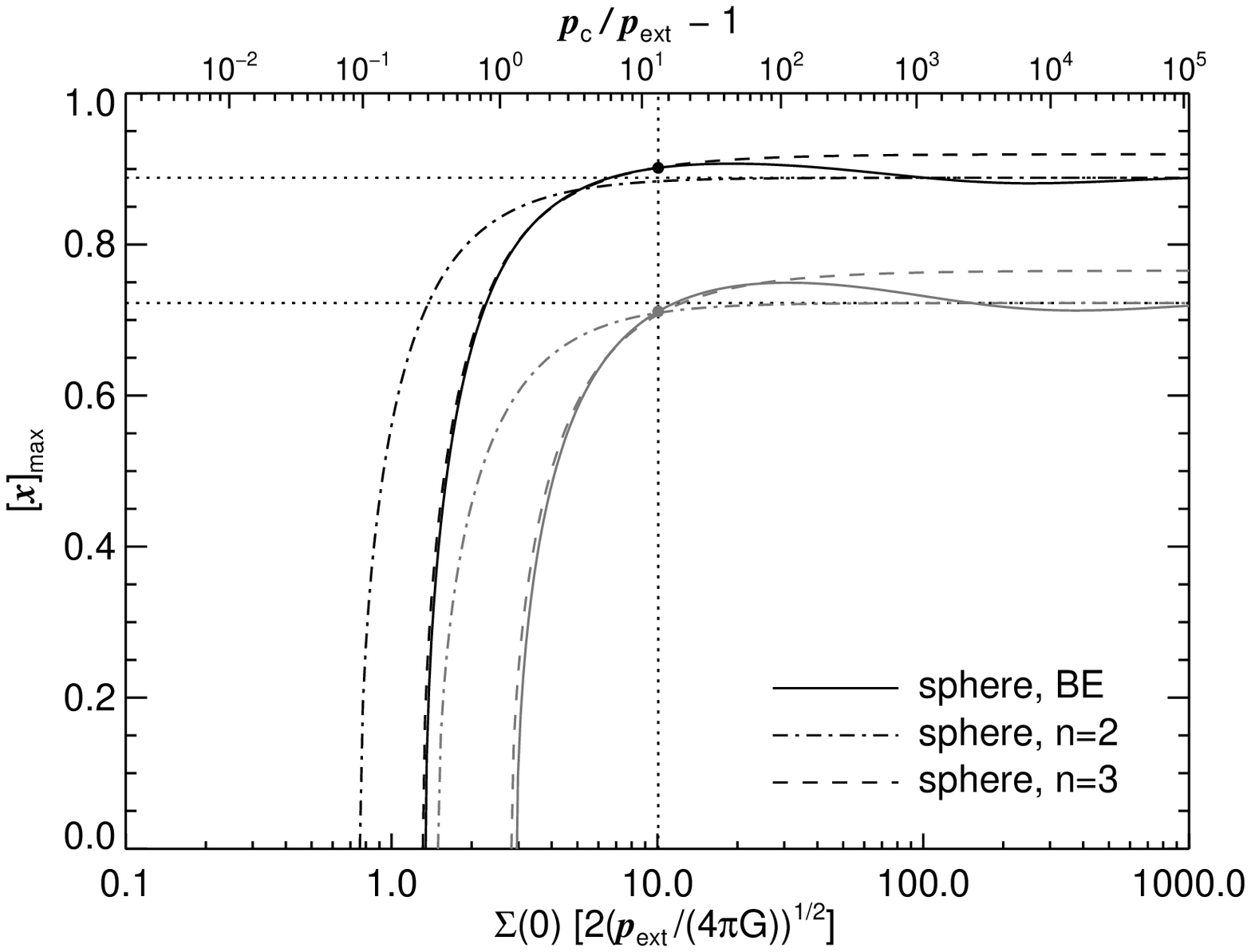}
        \hfill
        \includegraphics[width=0.49\hsize]{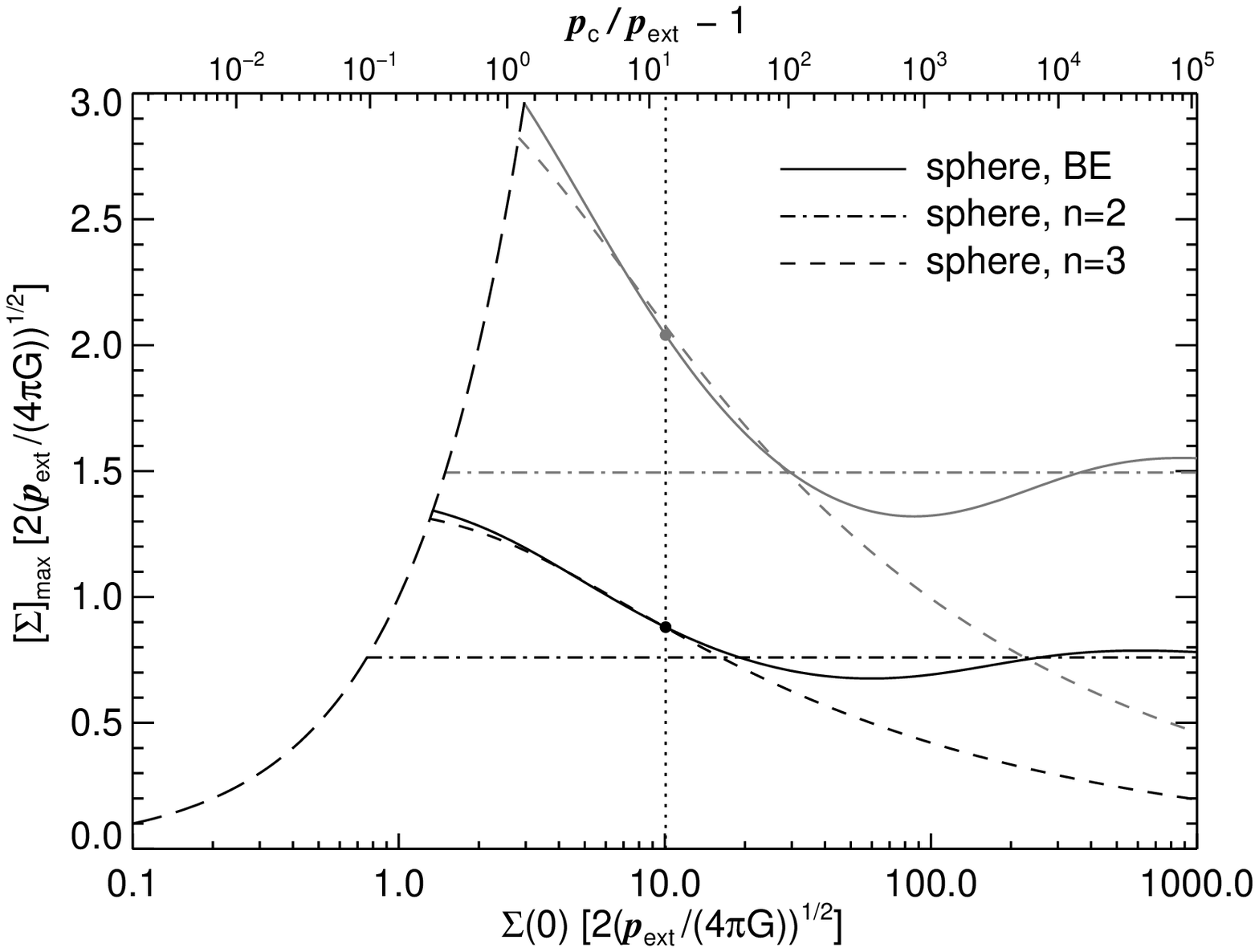}
        \caption{\label{fig_pdfisomax}
        Normalized impact parameter (left-hand panel) and   mass surface density (right-hand panel) 
        at the maximum of the PDF of the   mass surface densities of Bonnor-Ebert spheres, 
        given as a function of the central   mass surface density.
        The gray and black curves correspond to the logarithmic and the 
        linear pdf. The  values are compared with the corresponding 
        impact parameters and mass surface densities
        of pressurized spheres with a density profile as given in Eq.~\ref{eq_densityprofile}
        with $n=2$ and $n=3$. At low central mass surface densities 
        the maximum of the PDF is at impact parameter $[x]_{\rm max}=0$ and 
        the curves in the right-hand panel follow the long dashed line where $[\Sigma]_{\rm max}=\Sigma(0)$.
        The location of the critically stable Bonnor-Ebert sphere
        is marked by a vertical dotted line and the corresponding 
        normalized impact parameters and mass surface densities at the PDF maximum as filled
        circles. The horizontal dotted lines in the left panel are the asymptotic values of the 
        impact parameters of the PDF maximum of Bonnor-Ebert spheres in the limit of high overpressure. 
        The overpressures shown in the upper axis are only valid for the Bonnor-Ebert sphere.
        }
\end{figure*}

In this section we quantitatively analyze the highest position of the PDF of Bonnor-Ebert 
spheres. The normalized impact parameters $[x]_{\rm max}$ and the 
mass surface density $[\Sigma]_{\rm max}$ at the maxima of the linear 
and logarithmic PDF of Bonnor-Ebert spheres are shown in Fig.~\ref{fig_pdfisomax}. 
The values were 
derived by solving the conditional equations for maxima positions as given in App.~\ref{app_pdfmaximum}.

The behavior of the PDF maxima of Bonnor-Ebert spheres 
are compared with the maxima positions of the corresponding PDFs of spheres
with analytical density profiles as given in Eq.~\ref{eq_densityprofile} with n=3 and n=2.
As found in Paper~I, above a certain overpressure the local maxima of the PDF of spheres with
analytical density profiles
is related to a fixed parameter $[y_n]_{\rm max}$ , and consequently, considering Eq.~\ref{eq_defxn2},
to a fixed normalized mass surface density $[X_n]_{\rm max}$. The 
corresponding values for $n=2$ and $n=3$ are given in Table~\ref{table_pdfmaxima_iso}. 
The mass surface density $[\Sigma_n]_{\rm max}$ at the PDF maximum 
is obtained using the relation given in Eq.~\ref{eq_msurfnormalized}.

At low overpressures, the maximum coincides in general with the central mass surface density, so that the 
corresponding impact parameter is $[x]_{\rm max}=0$. A local maximum with $[x]_{\rm max}>0$ occurs 
at a sufficiently high overpressure $[q^{-1}]_{0,\rm max}$. 
For analytical density profiles the lowest overpressure for a local maximum is given by
\begin{equation}
        [q^{-1}]_{0,\rm max} = \left(1-[y_n]_{\rm max}\right)^{-\frac{n}{2}}.
\end{equation}
From Eq.~\ref{eq_densityprofile} we find the corresponding unit-free cloud radius 
\begin{equation}
        [\theta_{\rm cl}]_{0,\rm max} = \sqrt{\xi_n}\sqrt{\frac{[y_n]_{\rm max}}{1-[y_n]_{\rm max}}}.
\end{equation}
A local maximum with $x>0$ is otherwise related to a central mass surface density 
higher than
\begin{equation}
        [\Sigma]_{0,\rm max}/\hat\Sigma = (1-[y_n]_{\rm max})^{\frac{n-2}{4}}\sqrt{\xi_n}[X_n]_{\rm max}.
\end{equation}

Above the lowest overpressure the normalized impact parameter at the 
PDF maximum of spheres with an analytical density profile behaves as
\begin{equation}
        \label{eq_impactmax}
        [x]_{\rm max} =\sqrt{ \frac{1-[y_n]_{\rm max}-q^{2/n}}{1-q^{2/n}}}.
\end{equation}
The impact parameter strongly moves outward with increasing overpressure
and asymptotically approaches a constant value at high overpressure given by
\begin{equation}
        [x]_{\rm max} = \sqrt{1-[y_n]_{\rm max}}.
\end{equation}

We see from Fig.~\ref{fig_pdfisomax} that at low overpressure $q^{-1}<100$, the 
values of the normalized impact parameter and the mass surface density
at PDF maximum of the Bonnor-Ebert sphere follow the values of a sphere with $n=3$.
In this regime, a higher overpressure is
related to a lower   mass surface density at the PDF maximum. 
For the logarithmic PDF we find for the maximum 
\begin{eqnarray}
        [\Sigma_3]_{\rm max} \approx  0.0075 \sqrt{\frac{p_{\rm ext}/k}{2\times 10^4~{\rm K\,cm^{-3}}}}
                                                \left(\frac{q^{-1}}{14.04}\right)^{-\frac{1}{6}}\,{\rm g\, cm^{-2}}.
\end{eqnarray}
Assuming diffuse dust properties, this relates to a visual extinction
\begin{equation}
        [A_V]_{\rm max} = 1.72\sqrt{\frac{p_{\rm ext}/k}{2\times 10^{4}\,{\rm K\,cm^{-3}}}}
                \left(\frac{q^{-1}}{14.04}\right)^{-\frac{1}{6}}~{\rm mag}.
\end{equation}

At higher overpressure, the impact parameter and the  mass surface density 
at the maximum of the PDF of the Bonnor-Ebert sphere
fluctuate asymptotically to the corresponding constant value of a sphere 
with an analytical density profile with n=2. 
If we consider again the logarithmic PDF, the asymptotic 
position of the PDF maximum is given by
\begin{eqnarray}
        [\Sigma_2]_{\rm max} =  0.0054 \sqrt{\frac{p_{\rm ext}/k}{2\times 10^4~{\rm K\,cm^{-3}}}}\,{\rm g\,cm^{-2}}
\end{eqnarray}
or 
\begin{eqnarray}
        [A_V]_{\rm max} =  1.24 \sqrt{\frac{p_{\rm ext}/k}{2\times 10^4~{\rm K\,cm^{-3}}}}\,{\rm mag}.
\end{eqnarray}

Characteristic values of the maxima of the PDF are summarized in Tab.~\ref{table_pdfmaxima_iso}.
We also add the corresponding values of critically stable conditions both for the Bonnor-Ebert sphere ($q_{\rm crit}^{-1}=14.04$)
and the sphere with n=3 where $q^{-1}_{\rm crit}=13.46$ (see Sect.~\ref{sect_critmass}). Isothermal
self-gravitating pressurized cylinders are known to be stable under compression for all overpressures 
as ${\rm d}p(r_{\rm cl})/{\rm d}r_{\rm cl}<0$ (see \citet{Fischera2012a}
and references therein). The same would apply for a sphere with a hypothetical analytical profile with $n=2$.

\begin{table*}[htbp]    

        \caption{\label{table_pdfmaxima_iso}Location of the PDF
maxima}
        \begin{tabular}{c|cccccc|cc}
        \hline
                & \multicolumn{6}{c|}{sphere} & \multicolumn{2}{c}{cylinder} \\
                        & BE & n=3 & n=2  & BE & n=3 & n=2 & n=4 & n=4 \\
                        & \multicolumn{3}{c}{linear PDF} & \multicolumn{3}{c|}{logarithmic PDF} & lin. & log.      \\
                \hline
        $[y_n]_{\rm max}$\tablefootmark{a} & --- & 0.155& 0.211& ---&  0.414  & 0.478 & 0.141& 0.497 \\
        $[X_n]_{\rm max}$\tablefootmark{a} & --- & 0.465 & 0.537& --- & 1.099 & 1.057 & 0.461 & 1.797 \\
                \hline
                \multicolumn{9}{c}{onset for maxima with $[x]_{\rm max}>0$}\\
                $[\theta_{\rm cl}]_{0,\rm max}$ & 1.287 & 1.257 & 0.731 & 2.568 & 2.470 & 1.353  & 1.147 & 2.811 \\
                $[\Sigma]_{0,\rm max}$\tablefootmark{b} &  1.343 & 1.311 & 0.760 &  2.961 & 2.824 & 1.494  & 1.208 & 3.605 \\
                $[q^{-1}]_{0,\rm max}$ & 1.291 & 1.287 &  1.267 & 2.319  & 2.230 & 1.916 & 1.356 & 3.952  \\
                \hline
                \multicolumn{9}{c}{critically stable cloud\tablefootmark{c}}\\
                $[x]_{\rm max}$ & 0.9014 & 0.9012 & --- & 0.7114 & 0.7049  & --- & --- & ---\\           
                $[\Sigma]_{\rm max}$\tablefootmark{b} & 0.8800 & 0.8863 & --- & 2.0402 & 2.0928 & --- & --- & ---\\
                \hline
                \multicolumn{9}{c}{maxima in the limit $q\rightarrow 0$}\\
                $[x]_{\rm max}$ & 0.888 & 0.919 & 0.888 & 0.722 & 0.765 & 0.722 & 0.927 & 0.709 \\              
                $[\Sigma]_{\rm max}$\tablefootmark{b} & 0.760 & 0.000 & 0.760 &1.494 & 0.000 & 1.494 & 0.000 & 0.000 \\
        \end{tabular}
        \tablefoottext{a}{Maxima values with $x>0$ for clouds with an analytical density profile as given in Eq.~\ref{eq_densityprofile}.
        The values for cylinders refer to the \emph{asymptotic} PDF without a pole (Eq.~\ref{eq_pdfcylasymptote}).}
        \tablefoottext{c}{Critical stability in this paper is defined as critically stable under compression where ${\rm d}p(r_{\rm cl})/{\rm d}r_{\rm cl}=0$.}
        \tablefoottext{b}{Mass surface density in units of $(2/\cos^{\lambda} i)\sqrt{p_{\rm ext}/(4\pi G)}$ with $\lambda=1$ for cylinders
                and $\lambda=0$ for spheres.}

\end{table*}

\subsubsection{\label{sect_pdfsphapproxsmall}Asymptotic behavior for $\Sigma\ll [\Sigma]_{\rm max}$}

The behavior of the PDF at low mass surface densities is related to the truncation of the density profile
and the limit $\Sigma\ll[\Sigma]_{\rm max}$ is equivalent to $\theta_\bot\rightarrow \theta_{\rm cl}$.
In this regime the integrand in Eq.~\ref{eq_msurfiso} for the mass surface
density of Bonnor-Ebert spheres becomes approximately constant 
with $\rho/\rho_{\rm c}\approx e^{-\omega(\theta_{\rm cl})}=q$ and it follows that
\begin{equation}
        \Sigma \approx 2\sqrt{q}\sqrt{\frac{p_{\rm ext}}{4 \pi G}} \sqrt{\theta_{\rm cl}^2-\theta_\bot^2}.
\end{equation}
Applying Eq.~\ref{eq_pdf} and using $r_\bot =\theta_\bot/A,$ we obtain for the PDF
\begin{equation}
        P_{\rm BE}(\Sigma) \approx \frac{2\pi G}{p_{\rm ext}} \frac{1}{q\,\theta_{\rm cl}^2} \Sigma.
\end{equation}
The replacement of the unit-free radius with $\theta_{\rm cl}^2=r^2_{\rm cl} A^2 = \xi_n q^{-\frac{2}{n}}(1-q^{\frac{2}{n}})$
provides
\begin{equation}
        \label{eq_pdfsphapproxsmall}
        P_{\rm BE}(\Sigma) \approx \frac{2\pi G}{\xi_n  p_{\rm ext}} \frac{q^{\frac{2-n}{n}}}{1-q^{2/n}}\Sigma .
\end{equation}
In the limit of infinite overpressure we obtain with $n=2$ and $\xi_2=2$ the asymptote
\begin{equation}
        P_{\rm BE}(\Sigma) = \frac{\pi G}{p_{\rm ext}} \Sigma.
\end{equation}

\subsubsection{\label{sect_pdfsphapproxlarge}Asymptotic behavior for $\Sigma\gg [\Sigma]_{\rm max}$}

In Paper~I we showed that in the regime of high mass surface densities 
the PDF of spheres with a truncated radial 
density profile as given Eq.~\ref{eq_densityprofile} asymptotically
approaches a power law 
\begin{equation}
        \label{eq_pdfsphasymptotelarge}
        P_{\rm sph}(X_n) \approx \frac{1}{n-1}\frac{2}{1-q^{2/n}} X_n^{-\frac{n+1}{n-1}}\zeta_n^{\frac{2}{n-1}},
\end{equation}
where 
\begin{equation}
        \zeta_n = \frac{1}{2}\, {\rm B}\left(\frac{n-1}{2},\frac{1}{2}\right), 
\end{equation}
where ${\rm B}(a,b)$ is the beta-function. Using the relation Eq.~\ref{eq_xnsnrelation} for the unit-free 
mass surface density $X_n$ and the relation 
Eq.~\ref{eq_pdfsphere} provides for the approximations of Bonnor-Ebert spheres the asymptotes
\begin{eqnarray}
        \label{eq_pdfsphasymptotelarge2}
        P_{\rm sph}(\Sigma_n) &\approx& \frac{1}{n-1}\frac{2^{\frac{n+1}{n-1}}}{1-q^{2/n}}
                \left(\frac{\xi_n p_{\rm ext}}{4\pi G}\right)^{\frac{1}{n-1}}\nonumber\\
                &&\times q^{\frac{n-2}{n(n-1)}}\zeta_n^{\frac{2}{n-1}}
                \Sigma_n^{-\frac{n+1}{n-1}}.
\end{eqnarray}
For the logarithmic PDF, the asymptote becomes $\Sigma P(\Sigma)\propto \Sigma^{-\beta}$ with
$\beta = 2/(n-1)$. The slope agrees with what is expected for spheres with simple power-law profiles $\rho\propto r^{-n}$ 
of the local
density \citep{Kritsuk2011, Federrath2013}.

We showed (Fig.~\ref{fig_msurfprofilesph}) that the mass surface density at low impact parameters $\theta_\bot$
is dominated by the innermost part of the radial density profile of Bonnor-Ebert spheres approximately described 
by the $n=3$-profile,
which explains the slope  of the asymptote ($\beta=1$) close to the highest mass surface density.
For Bonnor-Ebert spheres with low overpressure we find using Eq.~\ref{eq_pdfsphasymptotelarge2}
\begin{equation}
        \label{eq_pdfsphapproxlargen3}
        P_{\rm BE}(\Sigma) \approx \frac{2}{1-q^{2/3}}\sqrt{\frac{\xi_3 p_{\rm ext}}{4\pi G}}q^{\frac{1}{6}}\Sigma^{-2}.
\end{equation}
Figure~\ref{fig_pdfisosphere}
shows that for critically stable spheres the asymptotic behavior is not established. The slope of the logarithmic
PDF is more consistent
with $\beta = 0.7$ instead of $1$. As Fig.~\ref{fig_pdfisosphere2}
shows,
Eq.~\ref{eq_pdfsphapproxlargen3} is not applicable for Bonnor-Ebert spheres
with high overpressure.

To approximate a Bonnor-Ebert sphere in the limit of high overpressure ($n=2$, $\xi_2=2$, $q\approx 0$)
we obtain for the asymptote using Eq.~\ref{eq_pdfsphasymptotelarge2}
\begin{equation}
        \label{eq_pdfsphapproxlargen2}
        P_{\rm BE}(\Sigma)\approx  \frac{\pi p_{\rm ext}}{G} \Sigma^{-3}.
\end{equation}

The fluctuation of the PDF at high mass surface densities of Bonnor-Ebert spheres with high overpressure 
 (Fig.~\ref{fig_pdfisosphere2})
are also seen in collapsing-cloud models \citep{Kritsuk2011}. The slope in these simulations is somewhat
steeper than for a simple Bonnor-Ebert sphere ($\beta \approx 2.5$), suggesting a flatter density profile ($n\approx 1.8$,
for negligible background).

\subsection{Probability distribution function of self-gravitating cylinders}

\begin{figure}
        \includegraphics[width=\hsize]{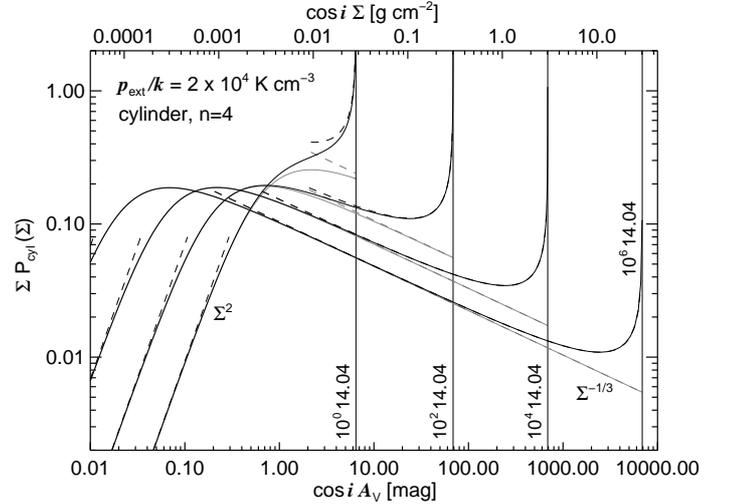}
        \caption{\label{fig_pdfisocylinder}PDF of the mass surface density of isothermal self-gravitating pressurized
                cylinders for various overpressures. The vertical line marks the pole position of the PDFs and is labeled
                with the corresponding pressure ratio $q^{-1}$.
                The asymptotes at high and low mass surface density are shown 
                as black dashed lines. The gray curves are the corresponding asymptotic PDFs 
                (Eq.~\ref{eq_pdfcylasymptote}) where the pole is removed. 
                The power-law asymptotes at high mass surface densities are
                shown as gray dashed lines.
                }
\end{figure}

The PDF of the mass surface density of cylinders with an analytical density profile as
given in Eq.~\ref{eq_densityprofile}  
is given by the implicit function (Paper~I)
\begin{equation}
        \label{eq_pdfcylinder}
        P_{\rm cyl}(\Sigma_n(y_n)) = \left(\frac{2}{\cos i} r_0\rho_c q^{\frac{n-1}{n}}\right)^{-1} P_{\rm cyl}(X_n(y_n)),
\end{equation}
where 
\begin{equation}
        \label{eq_pdfcylinder2}
        P_{\rm cyl}(X_n(y_n)) = \frac{1}{2x} P_{\rm sph}(X_n(y_n)).
\end{equation}
$P_{\rm sph}(X_n)$ is the PDF of spheres of the unit-free mass surface density 
given by Eq.~\ref{eq_pdfsphere2} and $x$ is the normalized impact parameter 
given by $x=r_\bot /r_{\rm cl}$.

The PDFs of the mass surface density of isothermal self-gravitating cylinders where $n=4$ with various 
pressure ratios are shown in Fig.~\ref{fig_pdfisocylinder}. The pressure ratios and the external pressure are the
same as assumed for spheres in Fig.~\ref{fig_pdfisosphere2}. 
Compared with the PDFs of Bonnor-Ebert spheres,
the PDFs of cylinders have a pole at the highest mass surface density. The PDF of 
cylinder with high overpressure have a local maximum at low mass surface densities that continues to shift to 
lower mass surface densities for higher overpressures. 
For the same overpressure the central mass surface density is slightly lower than for spheres (see also \citet{Fischera2011}).
For comparison, the asymptotic PDFs of cylindrical clouds are shown given by (Paper~I)
\begin{equation}
        \label{eq_pdfcylasymptote}
        P_{\rm cyl}^{(a)}(\Sigma) = \frac{x}{\sqrt{1-y_n}} P_{\rm cyl}(\Sigma).
\end{equation}
Figure~\ref{fig_pdfisocylinder} shows that the asymptotic PDF provides the correct functional
behavior of the cylindrical PDF apart from the pole, which is removed. For example, the asymptotic
PDF provides the correct power-law asymptotes at low and high mass surface densities as well as the
correct position of the local maximum of the PDF maximum for high overpressures.

\subsubsection{Highest position}

Because the pole of the PDF of cylinders are close to the highest mass
surface density, the analysis of the highest positions of the PDF 
for isothermal self-gravitating pressurized cylinders have been based upon the asymptotic PDF.
Qualitatively, the situation for the asymptotic PDF of cylinders is similar to the PDF of spheres. At low
overpressure the local maximum coincides with the central mass surface density. Above
a characteristic overpressure the asymptotic PDF develops a local maximum with $[x]_{\rm max}>0,$ which
corresponds to a fixed $[y_n]_{\rm max}$ and $[X_n]_{\rm max}$. The values listed in Table~\ref{table_pdfmaxima_iso} for
cylinders with an analytical density profile with $n=4$ are taken from Paper~I. The characteristic parameters
can be derived as shown for spheres with an analytical density profile.

The impact parameters and
the mass surface densities at the local maximum of the asymptotic PDFs with $[x]_{\rm max}>0$
are shown in Fig.~\ref{fig_pdfcylmaxima}. The normalized impact parameter of the local maximum
follows Eq.~\ref{eq_impactmax} for $n=4$. It follows from Eq.~\ref{eq_msurfnormalized} that the local maxima $[\Sigma_4]_{\rm max}$ decrease proportional to $q^{1/4}$ with 
increasing overpressure. The dependence is slightly stronger than for Bonnor-Ebert spheres
with a low overpressure ($q^{-1}<100$).
The local maxima of the logarithmic asymptotic PDF 
are located at
\begin{equation}
        [\Sigma_4]_{\rm max} = \frac{0.0095}{\cos i}\sqrt{\frac{p_{\rm ext}/k}{2\times 10^4~{\rm K\,cm^{-3}} }}
                \left(\frac{q}{1/14.04}\right)^{\frac{1}{4}}{\rm mag}
\end{equation}
or
\begin{equation}
        [A_V]_{\rm max}  =\frac{2.17}{\cos i} \sqrt{\frac{p_{\rm ext}/k}{2\times 10^4~{\rm K\,cm^{-3}}}}
                \left(\frac{q}{1/14.04}\right)^{\frac{1}{4}}~{\rm mag}.
\end{equation}
For low overpressures, the maxima of the asymptotic logarithmic PDFs
agree within a factor of two with the maxima of  Bonnor-Ebert spheres. 
Figure~\ref{fig_pdfisocylinder} shows that
at low overpressures the correct PDF of cylinders is still strongly dominated by the pole. In this case,
the highest position of the asymptotic PDF approximately corresponds to the location of a shoulder
in the correct PDF.

\subsubsection{\label{sect_pdfcylapproxsmall}Asymptotic behavior for $\Sigma\ll [\Sigma]_{\rm max}$ }
In Paper~I, we found for low mass surface densities the asymptote 
\begin{equation}
        P_{\rm cyl}(X_n)\approx \frac{1}{1-q^{2/n}}X_n.
\end{equation}
For isothermal cylinders we obtain
\begin{equation}
        \label{eq_msurfcylapproxlow}
        P_{\rm cyl}(\Sigma) \approx \frac{\cos^2 i}{1-\sqrt{q}}\frac{\pi G}{\xi_4 p_{\rm ext}}q^{-1/2} \Sigma.
\end{equation}

\begin{figure*}[htpb]
%        \includegraphics[width=0.49\hsize]{pdfcyl_xmax.eps}
%        \hfill
%        \includegraphics[width=0.481\hsize]{pdfcyl_msurfmax.eps}
        \includegraphics[width=0.49\hsize]{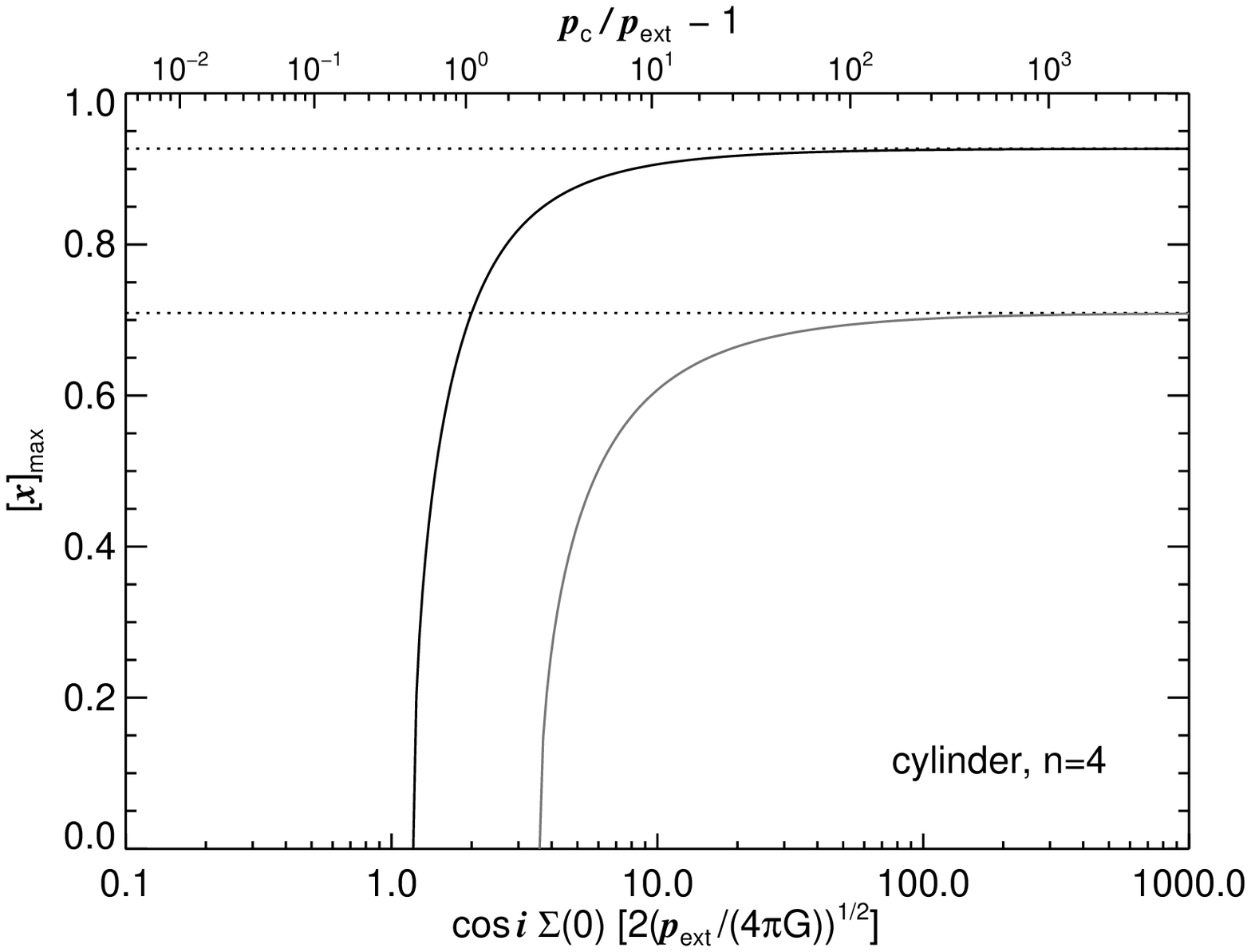}
        \hfill
        \includegraphics[width=0.481\hsize]{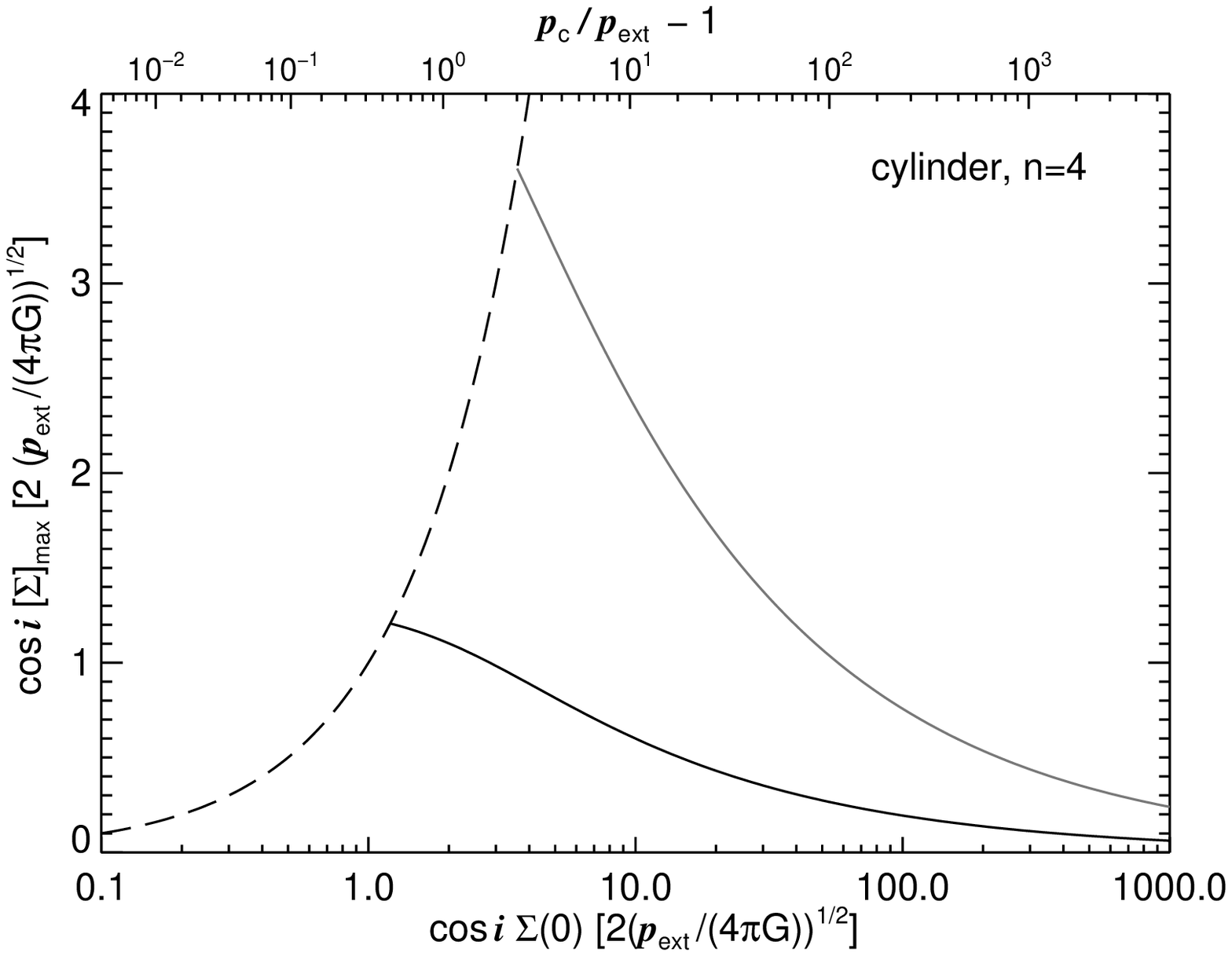}
        \caption{\label{fig_pdfcylmaxima} Same as Fig.~\ref{fig_pdfisomax}, but for the
        asymptotic PDF of isothermal self-gravitating cylinders where the PDF has no pole 
        at the highest mass surface density.
        }
\end{figure*}

\subsubsection{Asymptotic behavior for $\Sigma\gg [\Sigma]_{\rm max}$}

For high overpressure the PDF of a cylinder with the analytical density profile with $n>1$
behaves at high mass surface densites $\Sigma_n\gg [\Sigma_n]_{\rm max}$ as (Eq. B.7, Paper~I)
\begin{eqnarray}
        P_{\rm cyl}(X_n)\approx \frac{1}{n-1}\frac{1}{\sqrt{1-q^{2/n}}}X_n^{-\frac{n}{n-1}}\zeta_n^{\frac{1}{n-1}}\nonumber\\
                        \times \left[1-\left(\frac{X_n}{X_n(0)}\right)^{\frac{2}{n-1}}\right]^{-\frac{1}{2}}.
\end{eqnarray}
For self-gravitating isothermal cylinders we obtain using the defintion Eq.~\ref{eq_defxn} for $X_n$ 
and the relation Eq.~\ref{eq_pdfcylinder} for the different PDFs for $X_n$ and $\Sigma_n$ the
asymptote
\begin{eqnarray}
        \label{eq_msurfcylapproxhigh}
        P_{\rm cyl}(\Sigma_4)& \approx & \frac{1}{3}\frac{1}{\sqrt{1-\sqrt{q}}}\Sigma_4(0)^{\frac{1}{3}}
                                \nonumber\\
                                &&\times\Sigma_4^{-\frac{4}{3}} q^{\frac{1}{4}}
                                \left[1-\left(\frac{\Sigma_4}{\Sigma_4(0)}\right)^{\frac{2}{3}}\right]^{-\frac{1}{2}}
,\end{eqnarray}
where 
\begin{equation}
        \label{eq_msurfapproxhigh_n4}
        \Sigma_4(0) = \frac{\hat \Sigma}{\cos i} \sqrt{\frac{\xi_4}{q}}\frac{\pi}{4}
\end{equation}
is the mass surface density through the center of a cylinder in the limit of high overpressure (Eq.~\ref{eq_msurfapproxhigh}).
For sufficiently high overpressure the PDF shows a power-law slope 
for $[\Sigma]_{\rm max}\ll\Sigma\ll \Sigma(0)$ given by
\begin{equation}
        \label{eq_msurfcylapproxhigh2}
        P_{\rm cyl}(\Sigma) \approx \frac{1}{3}\left(\frac{\pi}{2\cos i}
                        \sqrt{\frac{\xi_4 p_{\rm ext}}{4\pi G}}\right)^{1/3}\frac{q^{1/12}}{\sqrt{1-q^{1/2}}}\,\Sigma^{-4/3}.
\end{equation}
Figure~\ref{fig_pdfisocylinder} shows that this power-law behavior is only established for cylinders
with overpressures greater than $\sim 10^3$.

\section{\label{sect_pdfmean}Averaged probability distribution function}

The origin of the observed PDF of the   mass surface density of giant molecular clouds is very complex.
In general, observations indicate that the global PDF of star-forming molecular clouds is
a combination of different components of randomly moving gas, generally attributed to turbulence, 
and the condensed structures. But structures related to wind-blown bubbles of massive stars 
or star clusters might also contribute to the global PDF.

The parameters determining the condensed spherical or elongated structures 
in a molecular cloud are not constant, but show certain variations. The main parameters for the pressurized 
structures of a certain geometry and density profile (power index $n$) are the pressure ratio $q$,
the external pressure $p_{\rm ext}$, and the temperature $T_{\rm cl}$. The condensed clouds
are located on a certain background $\Sigma_{\rm b}$ that
might also vary for different clouds. The bounding pressure of the condensed structures
is probably related to a pressure profile
in the larger molecular cloud. Moreover, the condensed structures
might form in a hierarchical fashion so that the bounding pressure of the innermost structure is also the central
pressure of the surrounding cloud. 

Observations \citep{Schneider2012,Schneider2013} suggest that as a first approach the condensations can be
treated as individual structures within the larger molecular cloud. In this case, we can obtain the 
mean PDF for the cloud through an average of the PDFs of individual condensations. 
We furthermore assume that the different
parameters determining the individual structures are independent (uncorrelated) variables.

We refer to the PDF of a single condensation under the condition of parameters $\vec a$ as
$P(\Sigma , \vec a),$ where $\vec a = (q,p_{\rm ext},\Sigma_{\rm b})$. Under these
assumptions, the averaged probability distribution over
the distribution functions $P(\vec a)$ is then given by
\begin{equation}
        \label{eq_pdfmean}
        \left<P_{\rm cl}(\Sigma)\right>_{\vec a} =\frac{\int{\rm d}\vec a\,P(\vec a)\,r_{\rm cl}^\kappa(\vec a) \,P_{\rm cl}(\Sigma,\vec a)}
                {\int{\rm d}\vec a\,P(\vec a) \,r_{\rm cl}^\kappa(\vec a)},
\end{equation}
where $\kappa=2$ for spheres and $\kappa=1$ for cylinders.

\subsection{Temperature distribution}

The   mass of a core for a given external pressure and overpressure varies strongly with 
the kinetic temperature ($M_{\rm sph}\propto T^2$) (see App.~\ref{app_cloudmass}).
But as we showed in the previous section, the temperature has no effect on the PDF of the column density
if both the external pressure and the overpressure are fixed. All critically stable cores in a certain pressure
region, for example, have the same PDF of their mass surface density independent of their temperature. However, because the total size scales
with temperature (see App.~\ref{app_cloudradius}), the temperature distribution affects the covering factor and therefore the ratio between the PDFs of the cores and the PDF of the surrounding gas.

\subsection{\label{sect_pdfdistribution}Distribution of the gravitational states}

In general, the condensed cores in  giant molecular clouds are probably 
in various different gravitational states that are characterized by different pressure ratios $q$.
The situation of the condensed cores might be  similar to the situation of Bok globules,
which, according to the study of \citet{Kandori2005},
show a variety of different pressure ratio $q$. 

To demonstrate the effect of a distribution of the gravitational state on the average PDF of the column density
we consider a large sample of cores that are pressurized by the same external pressure, which is taken to be
$p_{\rm ext}/k = 2\times 10^4~{\rm K\,cm^{-3}}$.
In addition, we consider the cores to have the same effective temperature and the same $K$. 
As shown in App.~\ref{app_cloudmass}, the masses of supercritical cores, for example, are
smaller than the mass of the critically stable core by up to a factor 2. By fixing $p_{\rm ext}$ and $K,$ the   mass of the cores does not vary monotonically with overpressure.

It seems reasonable to assume that the distribution of gravitational states of 
subcritical and stable spheres ($q^{-1}<q_{\rm crit}^{-1}\sim14.04$) 
follows functional forms different from the distribution of clouds in
a supercritical state. As indicated in the study of Bok globules by \citet{Kandori2005},
most cores are probably close to the critical value. This might be related to the fact
that clouds with a deeper gravitational potential become more stable with a deeper
gravitational potential against external disturbances. It might also be a simple observational
bias because the central extinction through Bonnor-Ebert spheres strongly
increases when the mass approaches 
the critical value \citep{Fischera2008}.
The distribution of supercritical states is most likely related to gravitational collapse, as 
was also considered in the study by \citet{Kandori2005}. Because
a detailed analysis goes beyond the scope of this paper, we simply assumed
that the probability of supercritical states decreases with increasing central density
or overpressure $q^{-1}$.

To be able to consider two different distributions for spheres with low and high overpressure
we assume for the probability function of the overpressures the function
\begin{equation}
        \label{eq_probstates}
        P(q^{-1}) = C \,q^{-k_1}\left(1+(q^{-1}/q_0^{-1})^{\gamma}\right)^{-\frac{k_1+k_2}{\gamma}},
\end{equation}
where $k_2>1$.
The constant $C$ can be neglected in this study
because of the additional renormalization of Eq.~\ref{eq_pdfmean}. 
The parameter $q_0^{-1}$ is a characteristic overpressure 
determining the transition between the two regimes.
The distribution at $q\gg q_0$ and $q\ll q_0$
are power laws given by $P(q^{-1}) \sim C\,q^{-k_1}$ and $P(q^{-1})\sim C\,(q/q_0)^{k_2} q_0^{-k_1}$. 
The parameter $\gamma$ determines the smoothness of the transition between the two power laws. In the
limit $\gamma\rightarrow \infty$ the distribution, for example, becomes a broken power law. In this study we assume
$\gamma=2$. For the distribution we assume a power index $k_1=2,$ which produces a 
decreasing probability at lower gravitational states (smaller
$q^{-1}$) for $q^{-1}\ll q^{-1}_0$ .

The distribution does not vanish for $q^{-1}\rightarrow 1,$ which seems unrealistic. However, the product of the surface area with the probability distribution of the overpressure does indeed vanish as
$r_{\rm sph}(q) \sim r_{n}(q) \rightarrow 0$ when $q\rightarrow 1$ according to Eq.~\ref{eq_cloudradius}.

\begin{figure*}
        \includegraphics[width=0.74\hsize]{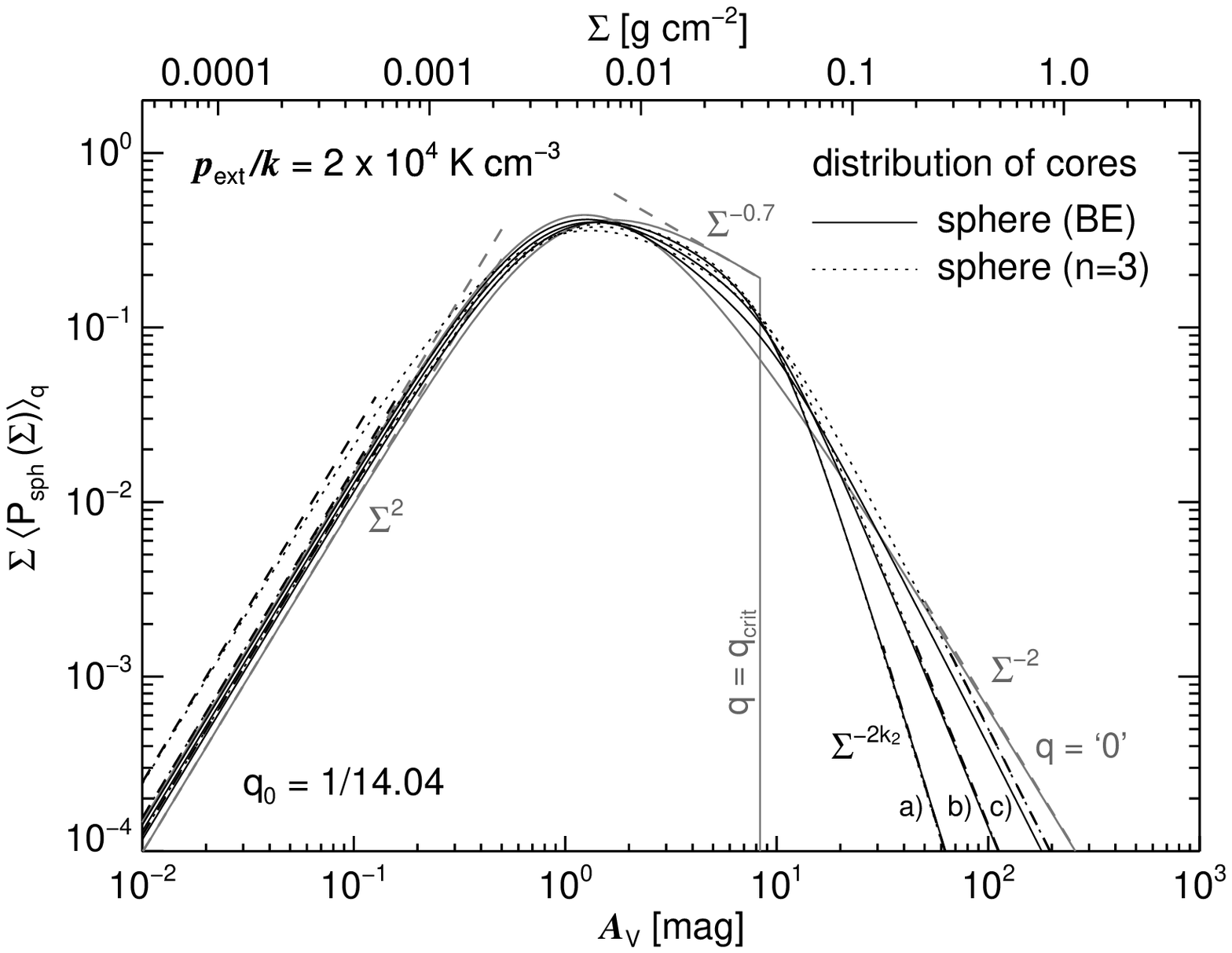}\\
        \includegraphics[width=0.74\hsize]{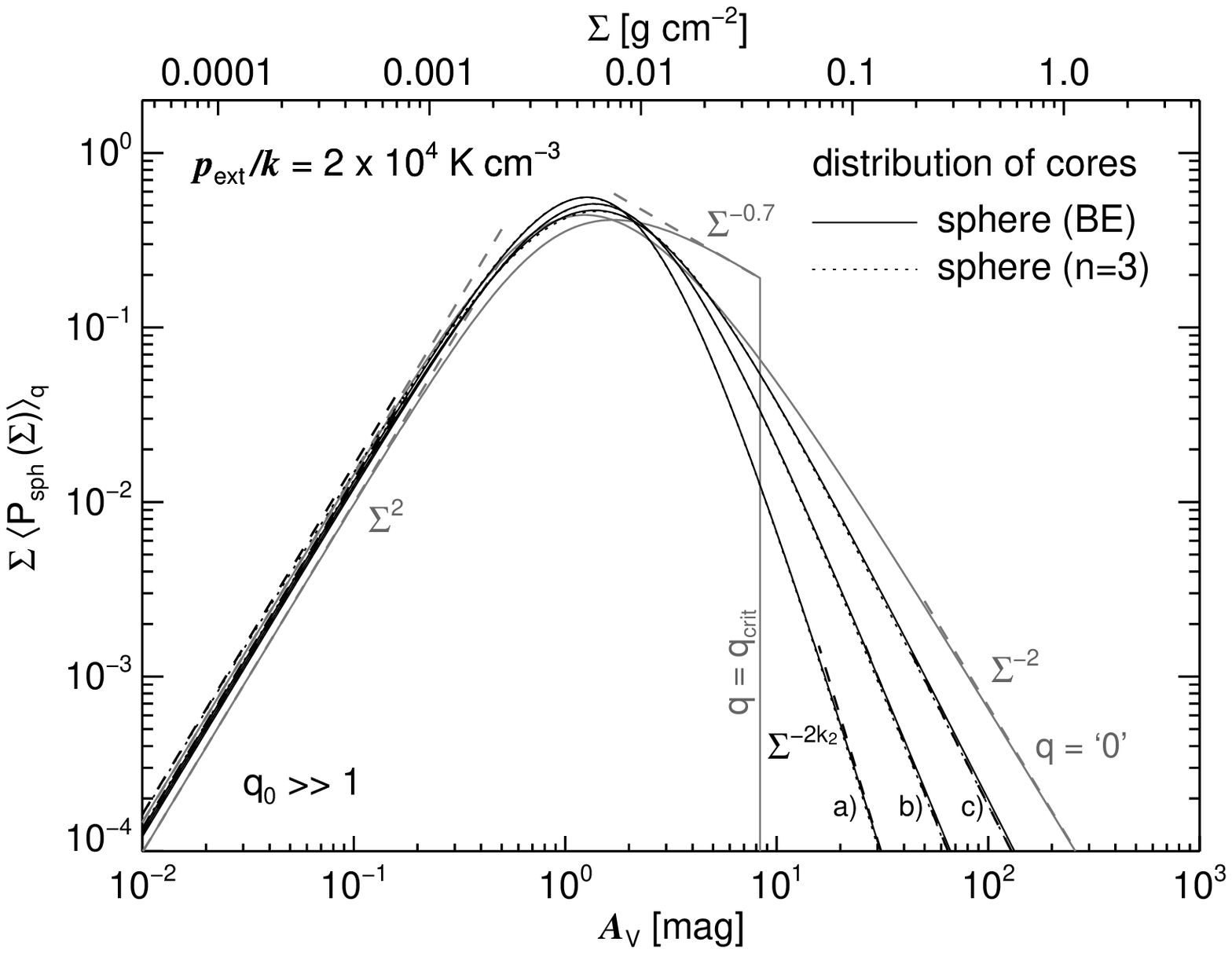}
        \caption{\label{fig_pdfisosphmean}Mean PDF of the  mass surface density 
        for an ensemble of cores assumed to be Bonnor-Ebert spheres 
        with a distribution of gravitational states or pressure ratios $q^{-1}$
        as given in Eq.~\ref{eq_probstates}.
        For comparison the mean PDF is also derived for spheres with an analytical density profile  
        as given in Eq.~\ref{eq_densityprofile} with $n=3$. 
        The characteristic pressure ratio of the probability distribution of the gravitational states is either 
        $q_0=q_{\rm crit}\sim 1/14.04$ of a 
        critically stable sphere (top panel) or $q_0\gg 1$ 
        (bottom panel). The various curves labeled $a)$, $b)$, and $c)$ correspond to different
        power indices $k_2$ assumed to be $2.0$, $1.5$, and $1.2$. 
        The power $k_1=2$ and the smoothness parameter $\gamma=2$ are the same in all calculations.
        The black dashed lines are power-law approximations of the PDFs in the limit of low and high mass surface
        densities for an analytical density profile with $n=3$ (App~\ref{app_meanpdfsphasymptotes}). 
        The external pressure is assumed to be $p_{\rm ext}/k=2\times 10^4~{\rm K\,cm^{-3}}$.
        Also shown are the PDFs of a critically stable sphere and a sphere with infinite overpressure 
        (gray lines and gray annotation). The corresponding power-law asymptotes of the PDFs of single cores
        in the limit of low (Eq.~\ref{eq_pdfsphapproxsmall}) and high (Eq.~\ref{eq_pdfsphapproxlargen3} 
        and Eq.~\ref{eq_pdfsphapproxlargen2}) mass surface densities are shown as gray dashed lines.
        }
\end{figure*}

To show the effect of the characteristic overpressure $q_0^{-1}$ we considered two different cases. In a first
set of calculations the characteristic overpressure is taken to be the overpressure of a critically stable sphere ($q_0^{-1}=14.04$). 
In a second set of calculations the power law of the high $q^{-1}$-states extends down to lowest gravitational states by 
choosing $q_0^{-1}\ll1$.
For a direct comparison, the same assumptions were made for both spheres and cylinders.
The power index $k_2$ was varied for given geometry and chosen characteristic
overpressure $q_0^{-1}$ to visualize its dependence on the mean PDF. For spheres the assumed 
power indices $k_2$ were $1.2$, $1.5$, and $2.0$ and for cylinders $1.5$, $2.0$, and $3.0$.

\subsubsection{Mean PDF of spheres}

The mean PDFs for distributions of Bonnor-Ebert spheres 
are shown in Fig.~\ref{fig_pdfisosphmean}. They are compared
with the mean PDFs of spheres with analytical density profiles as given in Eq.~\ref{eq_densityprofile} 
with $n=3$.

The mean PDF is characterized by a broad peak at mass surface densities between the highest position
of a critically stable sphere and a sphere with infinite overpressure. For the reference pressure the maxima
lie at $A_V\sim (1-2)~{\rm mag}$. At low and high mass surface densities the mean PDF asymptotically approaches 
the  power laws discussed quantitatively in App.~\ref{app_meanpdfasymptotes} for analytical density profiles.

The power law at low mass surface densities is identical to the power-law asymptote at low mass surface densities
of pressurized spheres (Sect.~\ref{sect_pdfsphapproxsmall}) and cylinders (Sect.~\ref{sect_pdfcylapproxsmall})
with $\Sigma \left<P_{\rm sph}(\Sigma)\right>_q\propto \Sigma^2$. 
The absolute probabilities follow the probabilities of single highly supercritical Bonnor-Ebert spheres within
a factor of two. 

The power law at high mass surface densities is independent of the power n of the radial density profile 
and is given by $\Sigma \left<P_{\rm sph}(\Sigma)\right>_q\propto \Sigma^{-2 k_2}$. The slope steepens
as one would expect for larger $k_2$ and lower probabilities for spheres with high overpressures. The values of $k_2$
are limited to $k_2>1$. Figure~\ref{fig_pdfisosphmean} shows
that for $k_2\rightarrow 1$ the mean
PDF approaches the PDF of Bonnor-Ebert spheres with infinite overpressure.

The mean PDF for a characteristic pressure ratio $q_0=1/14.04$ deviates considerably from the PDF
of a critically stable Bonnor-Ebert sphere. At the central mass surface density of a critically stable sphere
the mean PDF shows a knee with a flatter curvature at lower and steeper curvature at
higher mass surface densities. The feature is more pronounced for higher $k_2$. The mean
PDF for $q_0\gg 1$ has no additional feature apart from the prominent peak.
  
As we see in the figure, for high $k_2$ and the assumed characteristic pressures $q_0$ 
the mean PDF of a distribution of Bonnor-Ebert spheres
is well represented by the corresponding mean PDF of a sphere with the analytical density profile with $n=3$.
The two curves only start to deviate for low $k_2$ where spheres with higher overpressure than $q^{-1}\sim 100$ 
attribute to the mean PDF where the analytical density profile with $n=3$ is not a valid approximation of 
a Bonnor-Ebert sphere. We have seen that the PDF of an individual sphere with an analytical density profile 
with $q^{-1}>100$ is higher than the PDF of a Bonnor-Ebert sphere 
(Fig.~\ref{fig_pdfisosphere2}) at both low and high mass surface densities. In this case, the mean PDF of the analytical density profile is therefore 
too high with respect to the correct mean PDF, as seen for $k_2=1.2$ and $q_0=q_{\rm crit}$.

\subsubsection{Mean PDF of cylinders}

The mean PDF of an ensemble of self-gravitating isothermal cylinders pressurized by the same 
external pressure, but with a distribution of pressure ratios $q,$ is shown
in Fig.~\ref{fig_pdfcylmean}. The mean PDF is compared with the PDF of a single
cylinder with an overpressure $q^{-1}=14.04$.

One of the main characteristics is that the mean PDF of a distribution of cylinders
has no pole. If we consider the mean PDF for $q_0^{-1} =14.04,$ the PDF has a broad feature located
approximately at the pole position of the PDF of a single cylinder with a pressure ratio $q=q_0$.
The feature broadens and its maximum shifts to higher mass surface densities 
for a flatter distribution of $q^{-1}$ or equivalently lower $k_2$. 
The mean PDF for $q_0\gg 1,$ as is also the case for Bonnor-Ebert spheres, has 
no additional feature apart from the prominent peak at $A_V\sim 1~{\rm mag}$.

At low and high mass surface densities the mean PDF has power-law
asymptotes, similar to the mean PDF
of spheres. The asymptotes shown in Fig.~\ref{fig_pdfcylmean} are
derived in App.~\ref{app_meanpdfcylasymptotes}.
At low mass surface densities the asymptote is again
given by $\Sigma\left<P_{\rm cyl}(\Sigma)\right>_q\propto \Sigma^2$. At high mass surface
densities the slope is flatter than that of a distribution of spheres with 
$\Sigma\left<P_{\rm cyl}(\Sigma)\right>_q\propto \Sigma^{-2 k_2+1}$.

\begin{figure*}
%        \includegraphics[width=0.49\hsize]{pdfcylmeanplawa.eps}
%        \hfill
%        \includegraphics[width=0.49\hsize]{pdfcylmeanplawb.eps}
        \includegraphics[width=0.49\hsize]{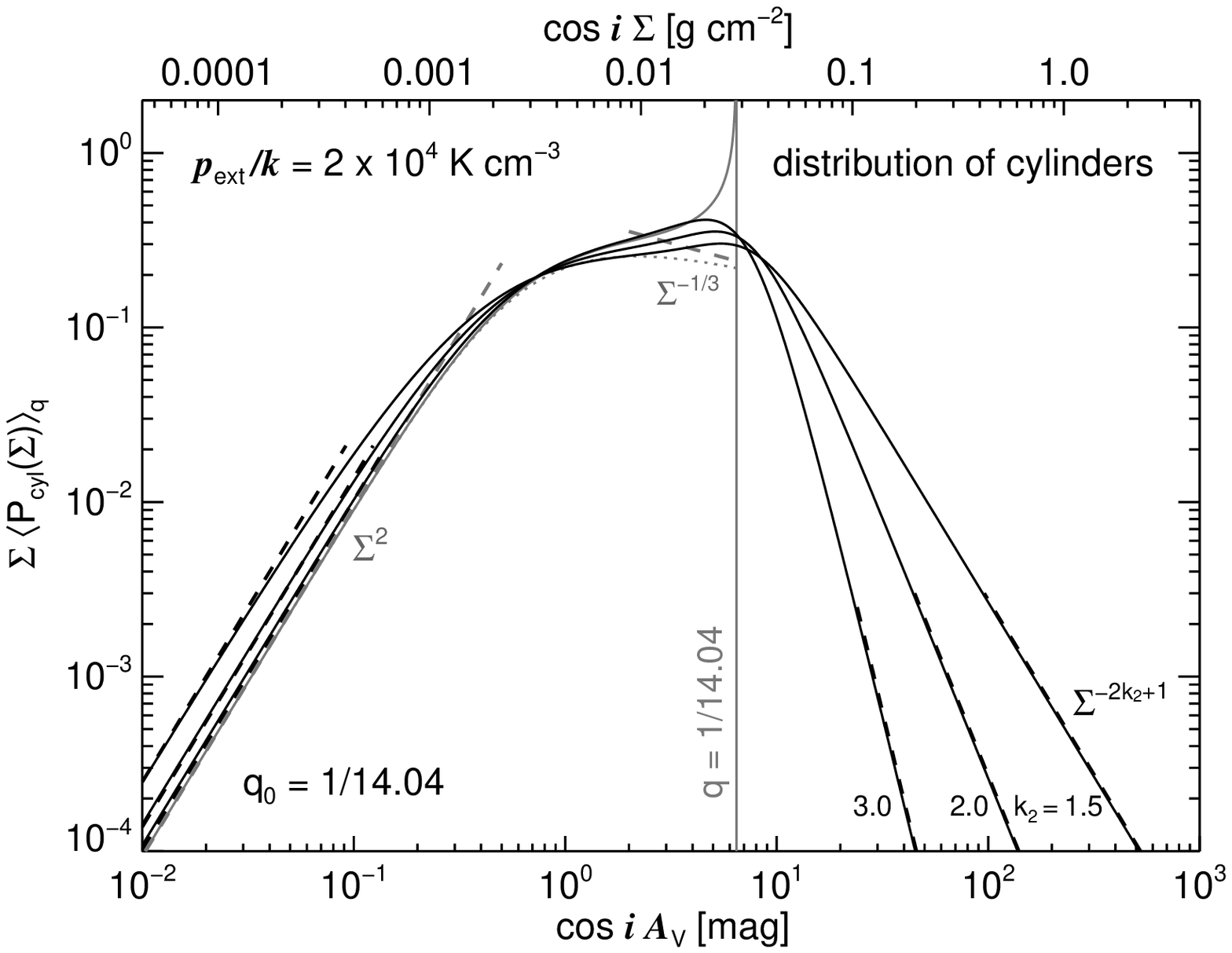}
        \hfill
        \includegraphics[width=0.49\hsize]{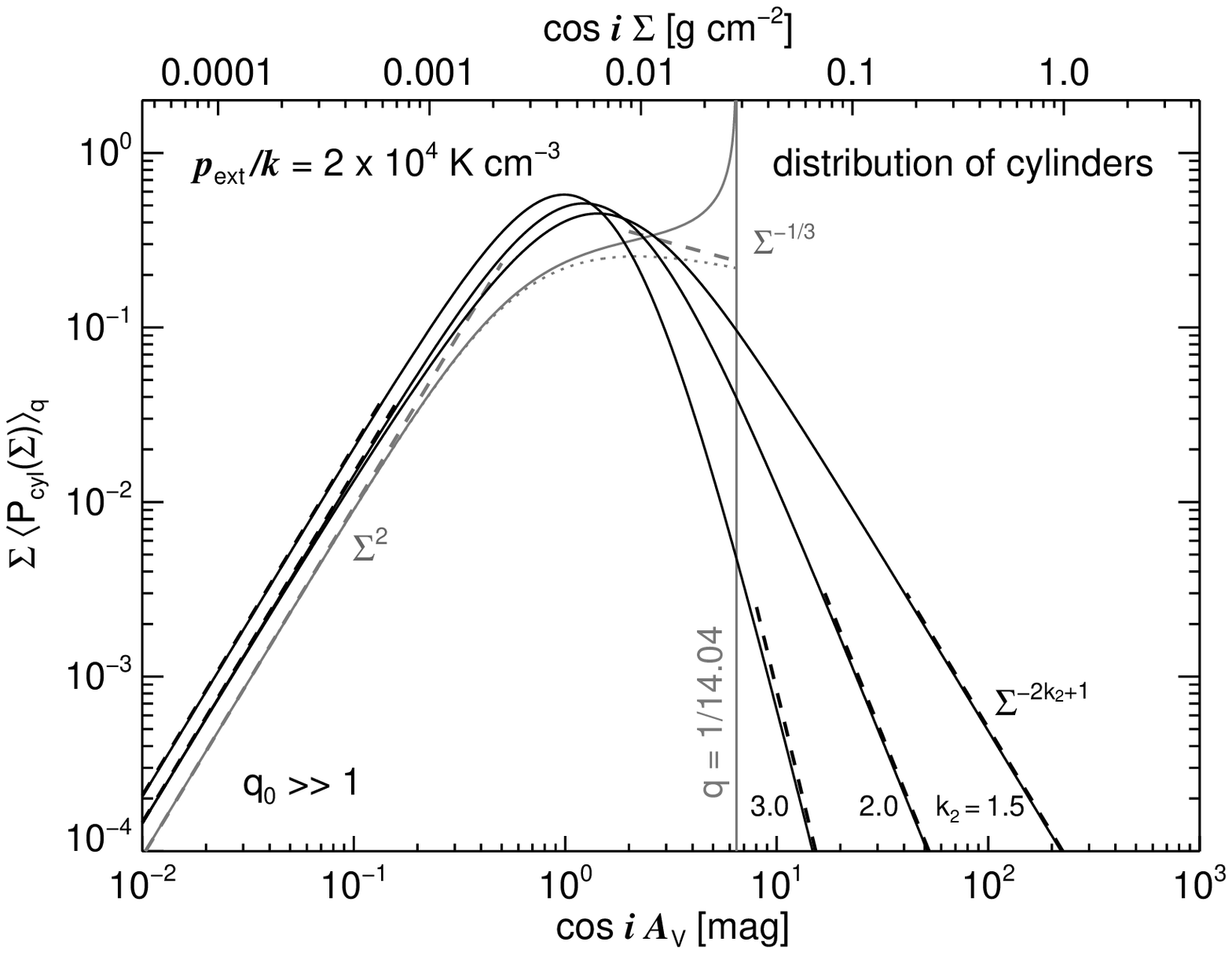}
        \caption{\label{fig_pdfcylmean}Mean PDF of the column density for an ensemble of isothermal self-gravitating cylinders
        with a variety of gravitational states or overpressures $q^{-1}$ , but fixed inclination angle $i$. The probability distribution
        of the overpressures is given by Eq.~\ref{eq_probstates} with fixed $q_0$, $k_1=2$, and $\gamma=2$ 
        where $q_0$
        is either $1/14.04$ (left-hand figure) or $ q_0\gg 1$ (right-hand figure). The power index 
        $k_2$ is assumed to be $1.5$, $2$, and $3$. The 
        cylinders are pressurized by a medium with $p_{\rm ext}/k=2\times 10^4~{\rm K\,cm^{-3}}$.
        The black dashed lines are asymptotes at low and high mass surface densities 
        (App.~\ref{app_meanpdfcylasymptotes}). For comparison, the
        PDF of a single cylinder with overpressure $q^{-1}=14.04$ is shown (gray line). The gray dotted line
        is the asymptotic PDF as defined in Eq.~\ref{eq_pdfcylasymptote}, 
        and the gray dashed lines are the corresponding 
        asymptotes at low (Eq.~\ref{eq_msurfcylapproxlow}) and high (Eq.~\ref{eq_msurfcylapproxhigh2}) 
        mass surface densities.}
\end{figure*}

\subsection{\label{sect_pdfcylmeanangle}Angle-averaged PDF for cylinders}

For filaments, we have the additional complication that the orientation 
most probably varies not only between different filaments, but may also vary along individual
filamentary structures. To analyze its effect 
on the mean PDF we consider an ensemble of cylinders with a certain distribution of inclination angles $i$
that are otherwise identical (same $q$).

For randomly distributed filaments, the probability of the cosine of the
projection angle $i$ is a constant. However, observations reveal that the orientation of
dense massive structures, which probably dominate the PDF, is not completely
random. The molecular cloud IC~5146 \citep{Arzoumanian2011} or the Taurus complex \citep{Kainulainen2009}
are dominated by massive filamentary structures with a certain mean orientation.
We ignore the complication that the orientation can
change at $90^\circ$ angles, as is the case for the Taurus filament.
To show the effect of any distribution of orientation angles, 
we assume that the cosine $\mu=\cos i$ of the inclination angle $i$ has a Gaussian
distribution $P(\mu)$ around a certain mean $\mu_0$ with a standard deviation $\sigma_\mu$. The 
distribution is renormalized so that
\begin{equation}
        \int_0^{1}{\rm d}\mu\,P(\mu)=1.
\end{equation}
The limit $\sigma_{\mu}\gg 1$ produces a flat distribution of
randomly orientated cylinders with $P(\mu)=1$.

If $P_{\rm cyl}(\Sigma)$ is the PDF for cylinders seen edge-on, the PDF of a cylinder seen at an inclination
angle $i$ is given by
\begin{equation}
        P_{\rm cyl}'(\Sigma') = P_{\rm cyl}(\mu\Sigma')\mu.
\end{equation}
The average over all inclination angles is then given by
\begin{equation}
        \left<P_{\rm cyl}(\Sigma')\right>_{\mu} = \int_{\mu_{\rm min}}^{\mu_{\rm max}}{\rm d}\mu\,P(\mu)\,P_{\rm cyl}(\mu\Sigma')\mu,
\end{equation}
where 
\begin{equation}
        \mu_{\rm max}=\left\{
        \begin{array}{ccc}
                \Sigma(0)/\Sigma' & {\rm for}&\Sigma(0)/\Sigma'<1,\\
                1                               & {\rm for}&\Sigma(0)/\Sigma'\ge 1,
        \end{array}                     \right.
\end{equation}
and where $\mu_{\rm min}$ is related to the greatest length of the filament. For 
infinitely long filaments we have $\mu_{\rm min}=0$. The mean PDF
for $\Sigma\le \Sigma(0)$ is an average over all inclination angles. For $\Sigma>\Sigma(0)$
the mean PDF is related to cylinders seen at increasingly high inclination angles.

For randomly distributed cylinders we derive\begin{equation}
        \left<P_{\rm cyl}(\Sigma')\right>_{\mu} = \frac{1}{\Sigma'^2}\int_{0}^{\Sigma'\, \mu_{\rm max}}{\rm d}\Sigma\,\Sigma\,P_{\rm cyl}(\Sigma).
\end{equation}
For $\Sigma'>\Sigma(0)$ the angle-averaged PDF of randomly distributed cylinders 
becomes a simple power law $\left<P_{\rm cyl}(\Sigma')\right>_{\mu} =C \Sigma'^{-2}$ where 
\begin{equation}
        C = \int_0^{\Sigma(0)}{\rm d}\Sigma \,\Sigma\,P_{\rm cyl}(\Sigma).
\end{equation}
This power-law slope is the same as for spheres with a density profile with $n=3$ in the limit of high overpressure 
(Figs.~\ref{fig_pdfisosphere} and~\ref{fig_pdfisosphere2}). 

\begin{figure*}[htbp]
%        \includegraphics[width=0.49\hsize]{pdfcylmeanangle00.eps}
%        \hfill
%        \includegraphics[width=0.49\hsize]{pdfcylmeanangle50.eps}
        \includegraphics[width=0.49\hsize]{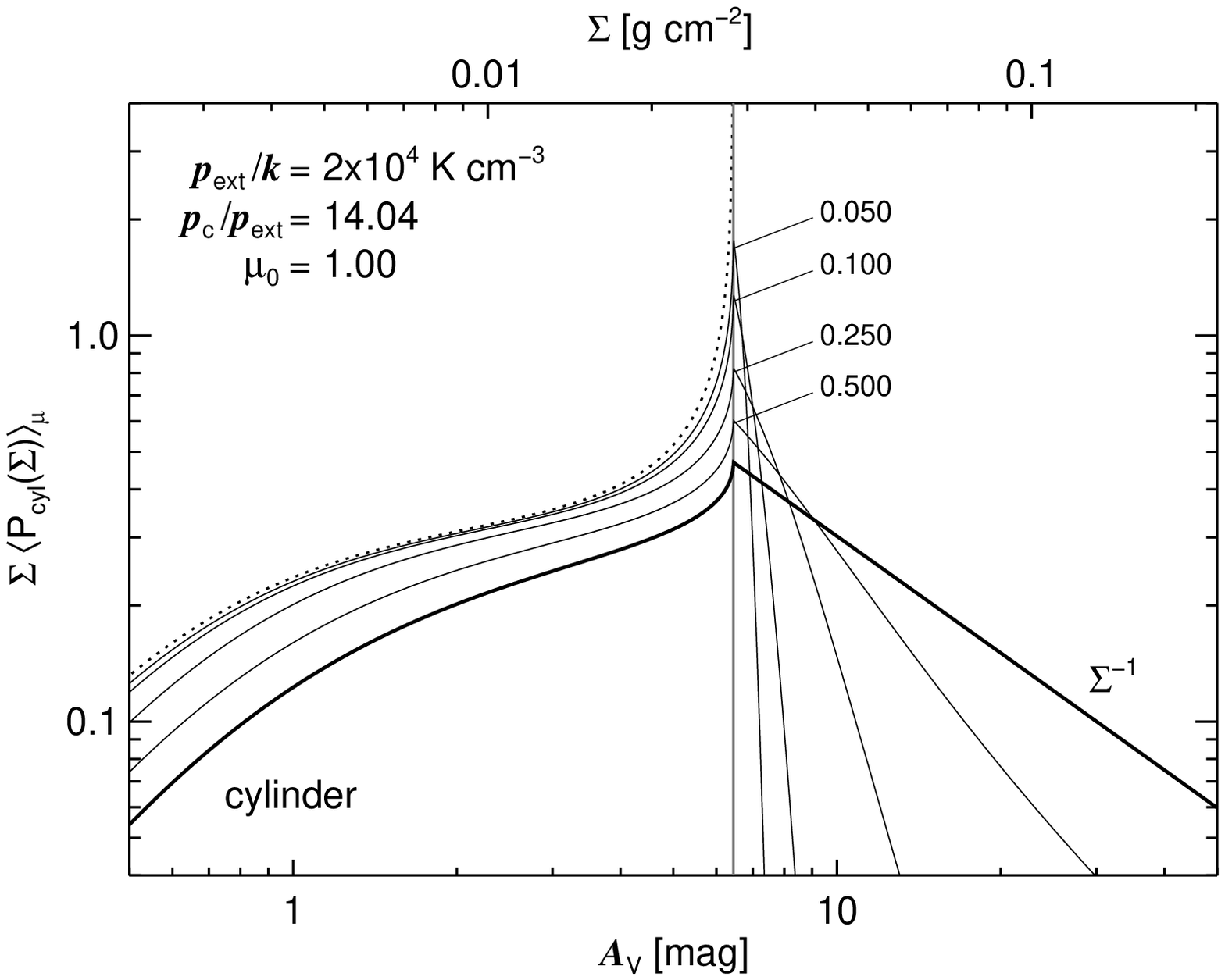}
        \hfill
        \includegraphics[width=0.49\hsize]{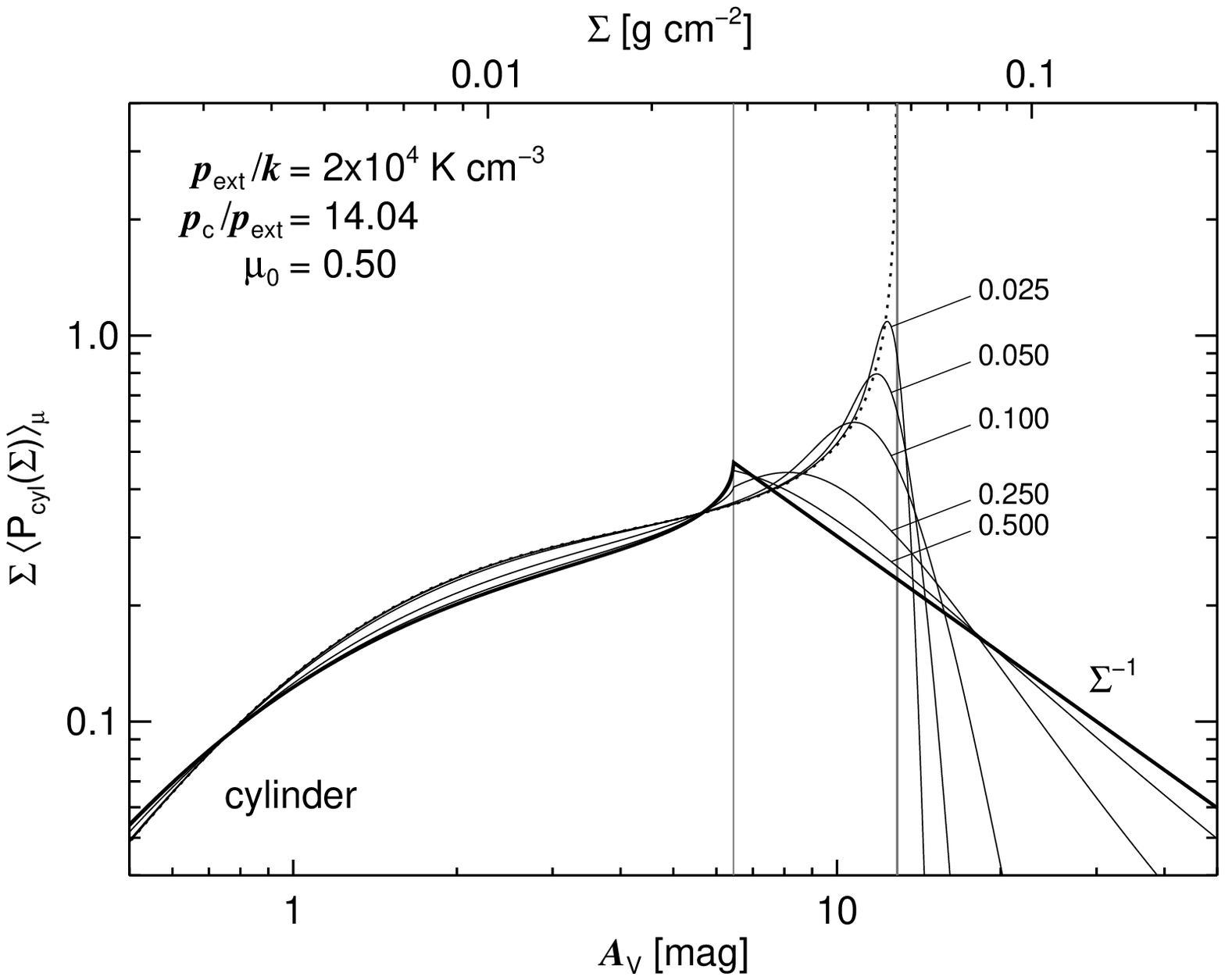}
        \caption{\label{fig_pdfcylmeanangle}Angle-averaged PDF of the column density for cylinders
        with a mean inclination angle of $0^\circ$ (left-hand figure) and $60^\circ$ (right-hand figure). The cosine of the
        inclination angle is assumed to be Gaussian distributed. The curves are labeled with the assumed standard
        deviation $\sigma_{\mu}$. The cylinders are pressurized by a medium with $p_{\rm ext}/k=2\times 10^4~{\rm K\,cm^{-3}}$
        and have an overpressure $p_{\rm c}/p_{\rm ext}=14.04$. The intrinsic PDF for the limit $\sigma_{\mu}\rightarrow 0$ 
        is shown as dotted curves. The gray vertical lines mark the highest extinction values of a cylinder seen under
        the assumed inclination angle and the position of the peak of the averaged PDF of randomly oriented cylinders. 
        The thick curves are the PDFs for randomly distributed cylinders.}
\end{figure*}

The effect of the angular distribution on the PDF for cylinders is shown in Fig.~\ref{fig_pdfcylmeanangle}. A distribution 
of inclination angles reduces the peak at the highest column density and produces a tail 
beyond the peak that flattens for wider distributions up to $P_{\rm cyl}(\Sigma)\propto \Sigma^{-2}$
 for randomly oriented cylinders. The effect on the peak is considerably stronger for cylinders with a mean 
 inclination angle $i>0$ ($\mu_0<1$). Figure~\ref{fig_pdfcylmeanangle}
shows that for $\mu_0<1$
the angular distribution also broadens the feature and produces a shift to lower   mass surface densities. The PDF of
randomly distributed cylinders has a small feature at the central   mass surface density for $i=0$.
Because the filamentary 
structure in the ISM is unlikely to have a well-defined overpressure ($p_{\rm c}/p_{\rm ext}$), this feature might be much less
prominent than shown in Fig.~\ref{fig_pdfcylmeanangle}, or might
even be absent.

\subsection{\label{sect_pdfsphmeanbgrd}Background-averaged PDF}

In the previous sections we have considered the PDFs of the   mass surface density of condensed structures
for a negligible background. However, this is certainly no longer valid if the condensed structures
are located in molecular clouds
or if they are seen through the interstellar medium. As an example, we
consider randomly distributed cores in an otherwise turbulent medium. As we show below, the combined
PDF of the turbulent structure and the condensed structures resemble the main
features of the observed PDF of typical star-forming clouds.

\subsubsection{\label{sect_mixture}Turbulent medium}

As shown in hydrodynamical simulation of driven turbulence, turbulence produces a log-normal density distribution of
the local density \citep{Vazquez1994, Padoan1997b,Passot1998}. 

The log-normal function of a statistical variable $s$ with a mean value $\left<s\right>$ is given by
\begin{equation}
        P(s) = \frac{1}{\sqrt{2\pi} s \, \sigma_{\ln \hat s}} 
                e^{-\frac{1}{2\sigma_{\ln \hat s}^2}\left(\ln \hat s +\frac{1}{2}\sigma_{\ln \hat s}^2\right)^2},
\end{equation}
where $\hat s = s/\left<s\right>$ is the
normalized variable and $\sigma_{\ln \hat s}$ the standard deviation of the log-normal function,
which is related to the standard deviation of the variable $s$ through
\begin{equation}
        \sigma_{s}^2 = \left<s\right>^2 \left(e^{\sigma^2_{\ln\hat s}}-1\right).
\end{equation}

According to simulations of driven turbulence \citep{Padoan1997b,Nordlund1999}, there 
exists a simple linear relation between the standard 
deviation of the local density and the Mach number $M,$ where
\begin{equation}
        \label{eq_densitycontrast}
        \sigma_{\hat \rho} = b M.
\end{equation}
The proportionality factor $b$ and the applicability for the interstellar medium 
is still a debated question, as discussed in Paper~III \citep{Fischera2014c} or by \citet{Kainulainen2013}.
Observational studies of IC~5146 and the Taurus molecular cloud complex suggest a value of $b\sim 0.5$ \citep{Padoan1997b,Brunt2010a}. 
A lower value ($b\sim0.20$) has been proposed by \citet{Kainulainen2013} based on high dynamic-range
extinction mapping of infrared dark clouds. 
Simulations indicate a value between 0.26 and 1
\citep{Padoan1997a,Passot1998,Kritsuk2007,Beetz2008,Federrath2008b,Federrath2010,Price2011}. 

The PDF of the column density through simulated turbulence was also found to be close to 
a log-normal function \citep{Ostriker2001,Vazquez2001,Brunt2010c}. The results are
supported by studies of the 
column density through non-star-forming molecular clouds that agree well with a simple log-normal function, as in the case
of Lupus~V and the Coalsack 
\citep{Kainulainen2009}. 

The statistical properties of the local density are directly related to the properties of the column density. Their
general functional dependence has been described by \citet{Fischera2004a} and \citet{Brunt2010b,Brunt2010c}. 
\citet{Fischera2004a} have studied the relationship between the column density 
and the local density of an isothermal turbulent screen by examining the one- and two-point statistics.
They assumed a simple log-normal function
of the density distribution of the local density and a simple power law $S(k) = |\hat \rho(k)|^2\propto k^{m}$ 
of the power spectrum, where $\hat \rho(k)$ is the Fourier coefficient of the 
local density $\rho$ and $k$ is the wavenumber.

It has been found that if the screen thickness 
$\Delta$ is greater than the highest turbulent scale $L_{\rm max}$ , the variance of the column density of 
thick screens varies as \citep{Fischera2004a}
\begin{equation}
        \label{eq_msurfcontrast}
        \sigma_{\Sigma/\left<\Sigma\right>}^2 = \sigma_{\rho/\left<\rho\right>}^2\frac{1}{2}\frac{L_{\rm max}}{\Delta} \frac{[m+3]}{[m+2]}
                \frac{[1-(k_{\rm max}/k_{\rm min})^{2+m}]}{[1-(k_{\rm max}/k_{\rm min})^{3+m}]},
\end{equation}
where $m\ne -2$ and $m\ne -3$.
For isolated molecular clouds the thickness should be $\Delta \ge L_{\rm max}$.
Relation \ref{eq_msurfcontrast} 
allows an estimate of the variance of the local density by measuring both the variance of the column density
and the power $m$ of the power spectrum.
For a Kolmogorov-like power spectrum of the local density\footnote{This assumption is correct 
for a contaminant driven by a Kolmogorov
velocity field as might apply to dust particles \citep{Lazarian2000, Fischera2004a}.}
, where $m=-11/3,$ and for a wide range of turbulent length scales ($k_{\rm max}\gg k_{\rm min}$), 
we have, for example,\footnote{\citet{Fischera2004a} took the power index of the
Kolmogorov velocity field to be $m=-10/3,$ which leads to 
$\sigma_{\Sigma/\left<\Sigma\right>}=\sigma_{\rho/\left<\rho\right>}\sqrt{L_{\rm max}/(8 \Delta)}$.} 
\begin{equation}
        \label{eq_msurfcontrastkolmogorov}
        \sigma_{\Sigma/\left<\Sigma\right>}=\sigma_{\rho/\left<\rho\right>}\sqrt{L_{\rm max}/(5\Delta)}.
\end{equation}

\subsubsection{Combined PDF of cores and turbulent gas}

The cores are considered to be small relative to the molecular cloud so that
the variations of the column density of the background of individual cores can be neglected. 
In addition, we consider the simplified case where the cores do not overlap and are
homogeneously distributed within the turbulent molecular cloud.
The background-averaged PDF of the mass surface density of the cores is then given by the convolution
\begin{equation}
        \left<P_{\rm cl}(\Sigma)\right>_{\Sigma_{\rm b}} = \int_0^{\Sigma}{\rm d}\Sigma_{\rm b}\,P_{\rm cl}(\Sigma-\Sigma_{\rm b})\,P_{\rm turb}(\Sigma_{\rm b}),
\end{equation}
where $P_{\rm cl}(\Sigma)$ is the PDF of a single core and $P_{\rm turb}(\Sigma)$ the PDF of the turbulent medium
given by the log-normal function discussed in the previous section.
For a narrow distribution of the mass surface density caused by the turbulent medium, the PDF of the background
becomes a delta function, and the mean is simply given by
\begin{equation}
        \left<P_{\rm cl}(\Sigma)\right>_{\Sigma_{\rm b}} \sim P_{\rm cl}(\Sigma-\Sigma_{\rm b}).
\end{equation}

The cores are assumed to cover a certain fraction $f_A$ of the area of the cloud.
The total PDF of the molecular cloud is then given by the combined PDF of the turbulent background and
the background-averaged PDF of the cores, so that
\begin{equation}
        P_{\rm MC}(\Sigma) = (1-f_A) P_{\rm turb}(\Sigma) + f_A \left<P_{\rm cl}(\Sigma)\right>_{\Sigma_{\rm b}}.
\end{equation}

Examples of the combined PDF of cores and the turbulent medium are shown in Fig.~\ref{fig_pdfbgrd}
for two different gravitational states of the cores and three different assumptions of the
turbulent medium. The cores are either critically stable or highly supercritical
Bonnor-Ebert spheres to mimic collapsing cores.
The density profiles are approximated using the corresponding analytical
density profiles for low ($n=3$) and high ($n=2$) overpressure. 
The pressure within the turbulent medium is again assumed to be 
$p_{\rm ext}/k=2\times 10^4~{\rm K\,cm^{-3}}$. The mean background
of the turbulent gas is related to the central mass surface density of a critically stable Bonnor-Ebert sphere.
The ratios $\left<\Sigma_{\rm b}\right>/\Sigma(0)$ for the mean backgrounds $\left<\Sigma_{\rm b}\right>$ 
and central mass surface density $\Sigma(0)$ through a critically stable sphere are $10^{-1.0}$, $10^{-0.5}$, and $10^{0.0}$ to
cover the range from a low to a very high background level.
The lowest background is close to the peak position for a number of PDFs of molecular clouds studied by 
\citet{Kainulainen2009}. 

To assign a standard deviation of the mass surface density to the different mean backgrounds
we considered an idealized turbulent slab with a clearly defined mean density $\left<\rho\right>$ and a fixed greatest
turbulent length scale  $L_{\rm max}$ so that
$\sigma^2_{\Sigma/\left<\Sigma\right>}\propto 1/\left<\Sigma\right>$. 
With Eq.~\ref{eq_msurfcontrast}, a higher mean background is then related to a lower fluctuation 
of the normalized mass surface densities $\Sigma/\left<\Sigma\right>$. 
As a typical standard deviation we assumed for the lowest mean background 
$\sigma_{\Sigma/\left<\Sigma\right>}=0.40$ (Table~1, \citet{Kainulainen2009}), which provides
a standard deviation of $0.22$ and $0.13 $ for the next higher
backgrounds.
This implies, considering Eq.~\ref{eq_msurfcontrastkolmogorov}, a standard
deviation of the local density of $\sigma_{\rho/\left<\rho\right>}=0.89\sqrt{\Delta/L_{\rm max}}$. Assuming
for the correlation constant $b=0.5$ between $\sigma_{\rho/\left<\rho\right>}$ and $M,$ this suggests
a Mach number of $M = 1.8\sqrt{\Delta/L_{\rm max}}$. According to \citet{Kainulainen2013}, molecular
clouds typically have a Mach number of about $M\sim 10,$ which would imply $\Delta\gg L_{\rm max}$
, or, considering a thickness $\Delta\approx L_{\rm max}$,
a smaller correlation coefficient or possibly a flatter power spectrum.

The combined PDF resembles several  features of the global PDF of star-forming molecular clouds.
The PDF is characterized by a broad peak and a tail at high mass surface densities.
The mean PDF of critically stable cores embedded in the turbulent medium 
is furthermore characterized by a strong decrease of probabilities at high mass surface densities.

 The background clearly modifies the shape of the PDF of individual cores.
The background-averaged PDF appears squeezed compared with the intrinsic PDF,
where the probabilities are shifted to higher mass surface densities that are visible, for example,
in the shift of the highest position or in the shift of the knee in the case of critically stable cores.
Naturally, the effect on the PDF of the cores increases with 
background level and is strongest at low mass surface densities.

As discussed in Sect.~\ref{sect_pdfsphapproxlarge} and shown in Fig.~\ref{fig_pdfbgrd}, 
the high end tail of the PDF of individual cores without background  only becomes a power law
in the limit of high mass surface densities. 
However, a low background can 
reduce the difference between the PDF of the cores and 
the asymptote and may even produce a tail at high mass surface densities below a possible
knee that approximately is a power law.
On the other hand, for higher backgrounds with $\left<\Sigma_{\rm b}\right>/\Sigma(0)\gg 0.1$,
the tail at high mass surface densities develops a curved shape where the probabilities at the low end are
considerably higher than the power-law asymptote.

As a result of the additive nature of two different components, 
the functional form of the mean PDF is not necessarily
a simple composition of a turbulent part at low mass surface densities and a tail at high mass surface densities.
The appearance will depend on the covering factor and the background level for the condensed cores.
Indeed, the theoretical curves of the mean PDFs derived for the high backgrounds show a smooth transition
of the log-normal part and a curved tail.
For a low background, however, the tail at high mass surface densities
shows a broad feature at the low end that is caused by the shifted peak of the background-averaged PDF.

\begin{figure*}[htbp]
%        \includegraphics[width=0.49\hsize]{pdfbgrd_sphn3.eps}
%        \hfill
%        \includegraphics[width=0.49\hsize]{pdfbgrd_sphn2.eps}
        \includegraphics[width=0.49\hsize]{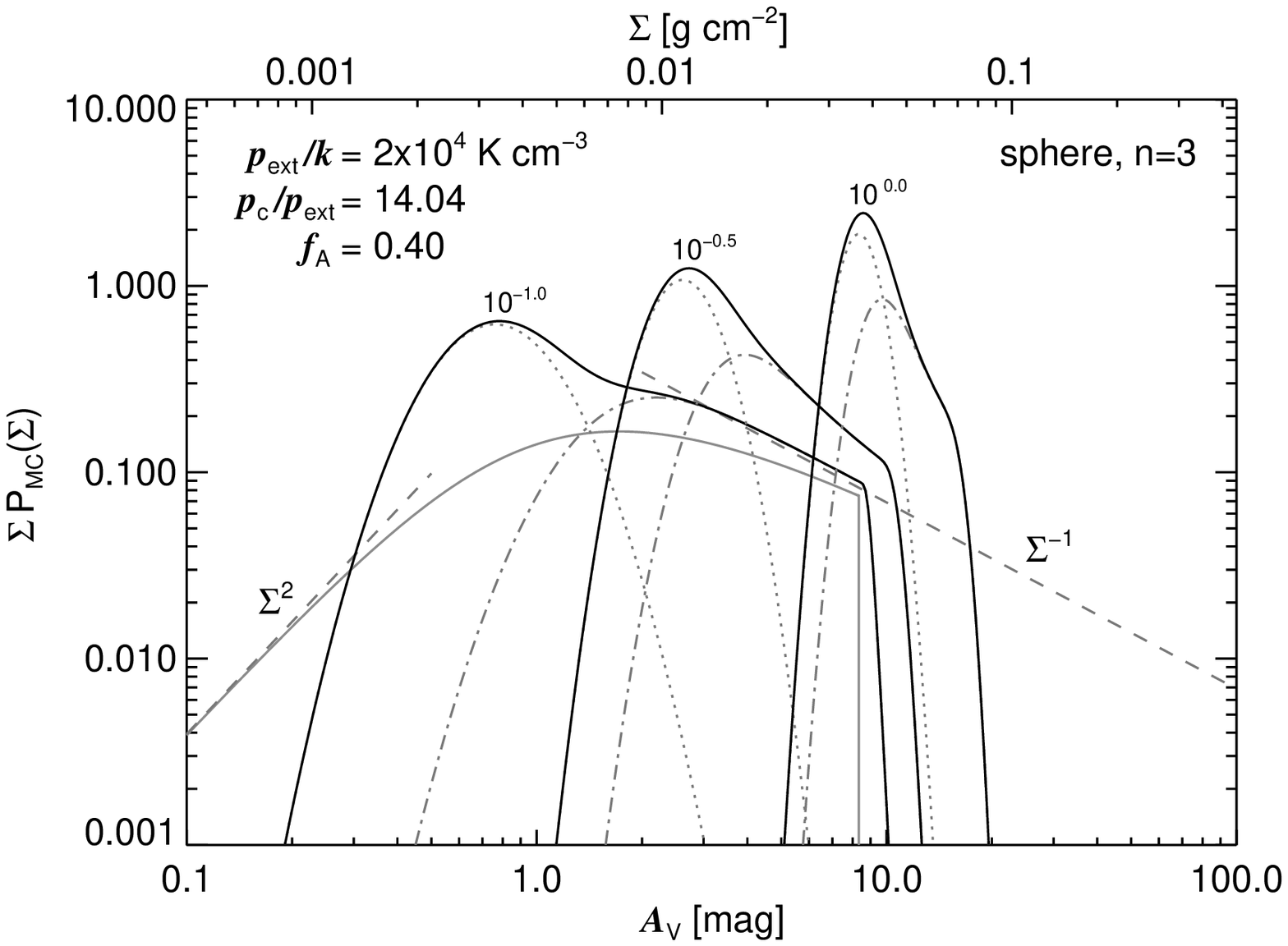}
        \hfill
        \includegraphics[width=0.49\hsize]{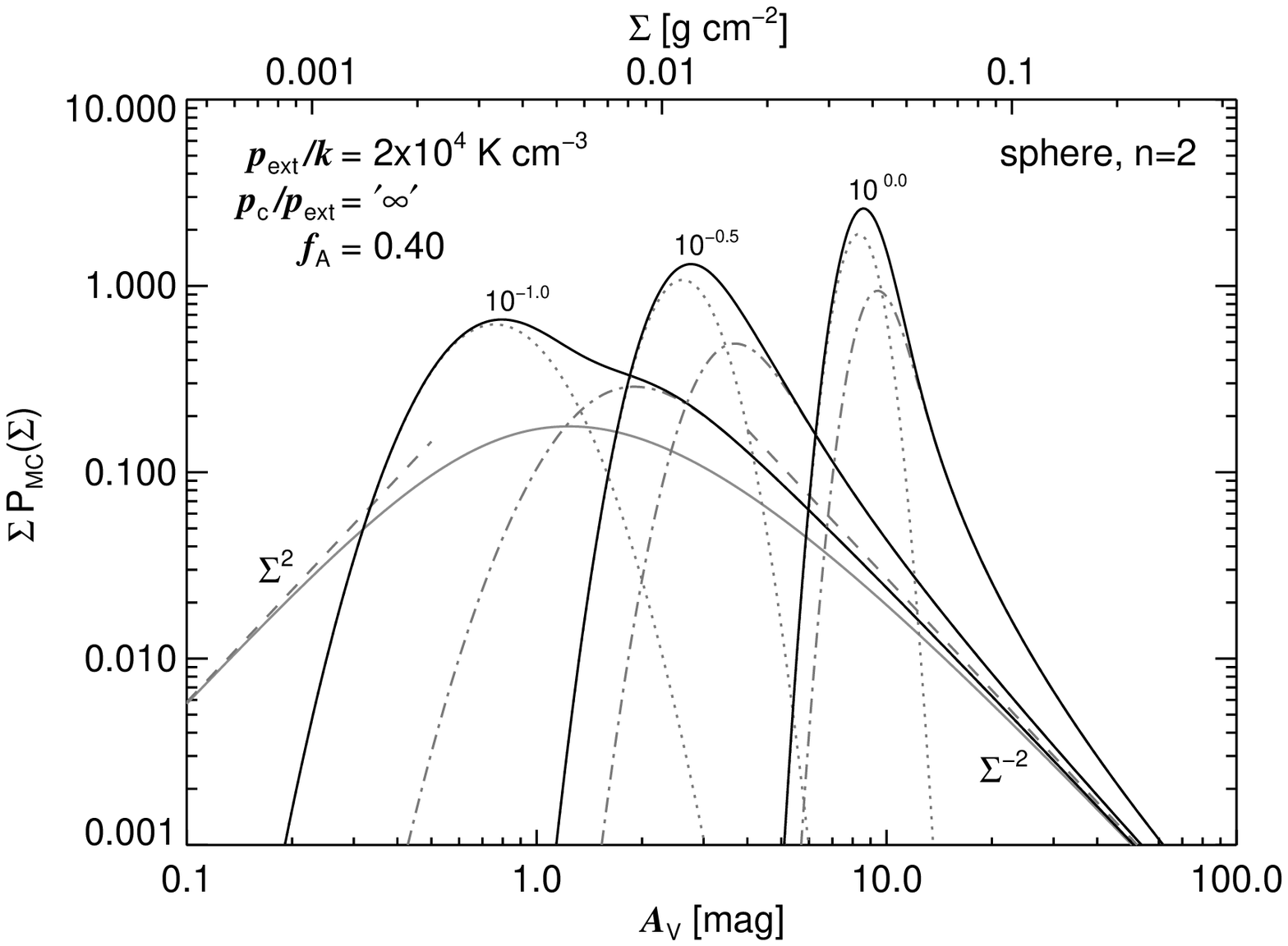}
        \caption{\label{fig_pdfbgrd}Combined PDF of a molecular cloud consisting of 
        a turbulent medium (dotted curves) and uniformly distributed condensed cores (dashed-dotted curves).
        The cores are either critically stable (left) or in a state of collapse (right) where the density 
        profiles are described using the corresponding approximations for Bonnor-Ebert spheres.
        The turbulent medium is modeled through a simple log-normal function 
        and the pressure in the turbulent gas is assumed to be $p_{\rm ext}/k=2\times 10^4~{\rm K\,cm^{-3}}$. 
        The spheres do not overlap and cover $f_{\rm A}=40\%$ of the area of the molecular cloud.
        The distributions are labeled with the assumed ratio $\left<\Sigma_{\rm b}\right>/\Sigma(0)$ of the 
        mean background $\left<\Sigma_{\rm b}\right>$ and
        the central mass surface density $\Sigma(0)$ through a critically stable Bonnor-Ebert sphere.
        The scaled intrinsic non-background-averaged PDFs of Bonnor-Ebert spheres (see Fig.~\ref{fig_pdfisosphere})
        are shown as solid gray lines. Also given are the corresponding power-law asymptotes at low 
        (Eq.~\ref{eq_pdfsphapproxsmall}) 
        and at high (Eq.~\ref{eq_pdfsphapproxlargen3} and Eq.~\ref{eq_pdfsphapproxlargen2})
        mass surface densities.
        }
\end{figure*}

\section{\label{sect_discussion}Discussion}

We have analyzed the PDF of the projected density 
for several different aspects related to isothermal and self-gravitating structures that
are pressurized by the ambient gas.
In the following we qualitatively compare the theoretical predictions
with the basic properties of observed PDFs of star and non-star-forming 
molecular cloud as derived by \citet{Kainulainen2009}.  Paper~IV \citep{Fischera2014d} 
will present a quantitative analysis of
the PDF at low mass surface densities, including the broad maximum referred to as the turbulent part
and the tail at high mass surface densities.

\subsection{High-end tail of the global PDF}

The tail at high mass surface densities often has a power-law form. 
For the PDFs with a maximum at $A_V\sim 1~{\rm mag,}$
the index $\beta$ of the power law approximation $\Sigma^{-\beta}$ 
of the tail of the logarithmic PDF lies for most cases in between 1 and 2. 
An unusually steep slope with $\beta>4$ can be found for the global PDF of the Pipe molecular cloud,
which is also characterized by an unusually high maximum position around $A_V\approx 3~{\rm mag}$.

We showed that a distribution 
of inclination angles of not overlapping cylinders would produce a steep slope at high mass surface densities.
The flattest tail with  $\beta= 1$ is expected for randomly distributed cylinders.
In principle, the slope would be steeper if the filaments had a certain alignment, but 
this would also be accompanied by a strong feature that is not detected in the observed PDFs. In this case,
additional effects need to be considered that might reduce or even remove the feature
as a distribution of gravitational states, as shown in this study, or 
distortions along filaments possibly associated to turbulence, which will be analyzed in Paper~III.
Still, the tail is expected to show features caused by overlapping filaments that are
not accounted for.

We showed that a power-law distribution $P(q^{-1})\propto {1/(q^{-1})}^{k_{2}}$ of the gravitational
states of Bonnor-Ebert spheres or cylinders would lead to a power-law asymptote in the limit of high mass surface
densities with $\beta=2 k_2$ for spheres or $\beta=2 k_2-1$ for
cylinders. A steeper
distribution of $q^{-1}$ would be related to a steeper power-law asymptote. 

The simplest model for the tail at high mass surface densities are single Bonnor-Ebert spheres. 
We showed that spheres with low overpressure $<\sim 100$ have a rather flat asymptote at high
mass surface densities with $\beta= 1$ and that the slope above the PDF maximum steepens 
with overpressure. 
Clouds with regions essentially in gravitational collapse and in the process of forming new stars are expected to have a 
tail with a slope close to $\beta=2,$ as observed for the Ophiuchus cloud or Taurus. A flatter tail
would be expected for clouds without any star formation activity. In this case, the tail is expected to
have a knee or a break at high mass surface densities. These characteristics can be found
in the tail of the PDF derived for the Musca molecular cloud, which is classified as non-star-forming. However, the
agreement may be coincidental, considering its elongated structure.

A break can also be observed in some other PDFs such as Lupus~III, Lupus~I, LDN~1228, LDN~204,
Ori A, the Perseus cloud, or Ori B. It occurs around $A_V=6-15~{\rm mag,}$ which
approximately coincides with the central extinction through critically stable spheres
pressurized by a medium with $p_{\rm ext}/k\sim 2\times 10^4~{\rm K\,cm^{-3}}$. 
However, it cannot be excluded that
the break may also be caused by the resolution of the map where the 
high mass surface densities in the central regions of collapsing clouds 
maybe smoothed over a larger area. These observational problems need to
be addressed and can be solved using maps with higher resolution. 

The shape of the tail suggests that the condensed structures are seen against a certain background
and that the slope of the tail at high mass surface densities steepens with the highest position of the global PDF, as expected for cores that are randomly distributed within a turbulent medium (Sect.~\ref{sect_pdfsphmeanbgrd}).

We showed that the tail would appear in the case of a low background
close to the power law, which is otherwise only expected as an asymptote in the limit of high mass surface densities. 
For the PDF of the Taurus complex the power law continues almost down to the highest position
without a clear separation of the two components.

The rather steep slope of the PDF tail ($\beta>4$) of the Pipe molecular cloud might be caused by 
a rather high background, as suggested by the high maximum position at $A_V\sim 3~{\rm mag}$.
The same may apply to the global PDF of the Rosette molecular cloud.
The logarithmic PDF derived by \citet{Schneider2012, Schneider2013} using Herschel
observations does not only have a relatively high maximum at $A_V=2~{\rm mag,}$ but also
an unusually steep tail with $\beta\sim 3.1$ (Fig.~6,
\citet{Schneider2012}). Relating this slope to a power-law density profile $\rho(r)\propto r^{-n}$
using the relation $n=(2/\beta)+1$ \citep{Kritsuk2011, Federrath2013, Fischera2014a} would
indicate a rather flat density profile with power $n=1.65$.\footnote{
The even flatter power law given in the study of \citet{Schneider2012} is 
inaccurate and
has been corrected \citep{Schneider2013}.}
However, the steeper slope may equally well be a direct consequence of condensed structures seen against a 
relatively high background and not directly related to the density profile. 

The PDF of Ophiuchus has a broad peak or a shoulder at the low end of the tail around $A_V\sim 3~{\rm mag,}$
similar to the broad feature in the theoretical mean PDF of condensed cores located on a low background (Fig.~\ref{fig_pdfbgrd}). 
The feature at the low end of the PDF tail of the Pipe molecular cloud might
have a similar explanation.

\subsection{Turbulent part of the global PDF}

Most of the derived PDFs of molecular clouds studied by \citet{Kainulainen2009} have
a maximum around $A_V = 1~{\rm mag}$,  which is  close to the maximum of the PDFs of 
Bonnor-Ebert spheres pressurized by the assumed ISM pressure. 
Although it may also be coincidental, this would suggest
that the molecular gas as a whole is not only pressurized, but also self-gravitating.

At mass surface densities below the maximum, the observed PDFs show
an obvious deviation to a simple log-normal function.
For a number of cases the PDF at low mass surface densities 
has a functional form close to a power law. The deviation may be related to an unsymmetrical nature of the local density PDF, as seen in more current simulations of forced 
turbulence \citep{Federrath2008b, Konstandin2012, Federrath2013}.
The behavior in the limit of low mass surface density might also be
related to a truncated density profile of the larger cloud if it is pressurized by a low dense medium.
We showed that in the case of a negligible background,
the PDF of pressurized nonturbulent spheres and cylinders 
asymptotically approaches a power law 
$\Sigma P(\Sigma)\propto \Sigma^2$ at low mass surface densities. The actual shape may be flatter because of the turbulent nature
of the gas, even if the PDF of the local density is log-normal, as will be shown in Paper~III. 
It still needs to be considered that the shape in particular at low mass
surface densities will be affected by additional material along the line of sight that is not related to the
molecular cloud, which will shift the probabilities to higher mass surface densities.

A first estimate of the density fluctuation in the turbulent cloud surrounding the condensed
structures can be derived using Eq.~\ref{eq_msurfcontrast}, which
is based on the infinite turbulent slab approximation. The estimate of the density contrast 
would be uncertain by the unknown power spectrum of the density structure and the number
of turbulent length scale. A statistical analysis of molecular clouds should therefore be combined
with studies of the power spectrum of the projected density.
A better estimate of the density contrast might be obtained 
by taking geometric effects and a possible background into account (Papers~III and IV).

It appears to be reasonable to assume that most of the star-forming gas lies in the central region of the cloud
where the gas is shielded from the interstellar radiation field to allow the gas to cool. The gas
also experiences the highest pressure in the center, so that in the case of dynamic equilibrium the gas 
becomes most easily gravitationally unstable because the 
critical mass varies as $M_{\rm crit} \propto T^2/\sqrt{p_{\rm ext}}$. This would suggest a threshold for star
formation as indicated by observational studies \citep{Kirk2006, Foster2009}.
The model described in this paper should allow an independent determination of the central pressure in the 
turbulent gas and might therefore provide a better understanding of the star formation process.

\section{\label{sect_summary}Summary and conclusion}

We have analyzed the properties of the PDF of the mass surface density
of pressurized isothermal self-gravitating spheres, known as Bonnor-Ebert spheres,
and cylinders where we applied the
results obtained in Paper~I for a simple analytical density profile as a generalization of
physical density profiles given by $\rho(r) = \rho_{\rm c}/(1+(r/r_0)^2)^{n/2}$.
We showed that for $\rho(r)/\rho_{\rm c}>0.01$ the density profile of Bonnor-Ebert spheres
is well approximated
by the analytical profile with power $n=3$. At larger cloud radius the radial density profile fluctuates
asymptotically to the analytical profile with $n=2$.
Emphasis was given on the dependence of the PDF of given geometry on the pressure ratio $p_{\rm ext}/p_{\rm c}$
and the external pressure assumed to be $p_{\rm ext}/k=2\times 10^4~{\rm K\,cm^{-3}}$.

The main properties of the PDFs of individual spheres and cylinders are found to be as follows:
\begin{enumerate}
        \item The PDF of critically stable spheres
        is truncated at the central mass surface density or extinction value 
        $A_V\approx 8.3\sqrt{(p_{\rm ext}/k)/2\times 10^4~{\rm K\,cm^{-3}}}~{\rm mag}$.
        \item Below an overpressure of $\sim 100,$ the PDF of a Bonnor-Ebert sphere can be well approximated by the PDF of
        a sphere with an analytical density profile with $n=3$. At fixed mass surface density, the PDF of spheres 
        fluctuates asymptotically with overpressure to the value of a sphere with an analytical sphere with $n=2$ in the limit of 
        infinite overpressure.
        \item   For the reference
        pressure the maximum of the logarithmic PDF of a sphere lies between $A_V\sim 1~{\rm mag}$  and 3~mag.  
        For spheres with an overpressure $<\sim 100$,
        the highest position of the PDF shifts proportionally to $q^{1/6}$ to lower mass surface densities. For highly
        supercritical clouds the highest position of the logarithmic PDF asymptotically approaches
        a constant given by $[A_V]_{\rm max}\sim 1.24\sqrt{(p_{\rm ext}/k)/2\times 10^4~{\rm K\,cm^{-3}}}$ mag. 
        \item The flattening of the radial density profile of spheres with overpressure 
        is related to a steepening of the PDF at mass surface densities above the peak maximum.       
        The slope $\beta$ of the power-law asymptote $\Sigma P_{\rm sph}(\Sigma)\propto \Sigma^{-\beta}$ 
        varies from $\beta=1$ for spheres with overpressures below $\sim 100$
        to $\beta =2$ for highly supercritical spheres.
        \item The highest position of the logarithmic asymptotic PDF (underlying PDF without a pole) 
        of an isothermal self-gravitating pressurized cylinder decreases with $q^{1/4}$. 
        For overpressures below 100, the maxima of cylinders and spheres agree within a factor of two.
        \item For cylinders with high overpressure, the PDF at high mass surface densities
        is approximately given by $P_{\rm cyl}(\Sigma)\propto \Sigma^{-4/3}(1-(\Sigma/\Sigma(0))^{2/3})^{-1/2}$.
        \item At low mass surface densities the PDF of both spheres and cylinders
        approaches a power law $P_{\rm cl}(\Sigma)\propto\Sigma$.
\end{enumerate}

In addition to individual clouds,
an ensemble of spheres or cylinders with a distribution of overpressures $q^{-1}=p_{\rm c}/p_{\rm ext}$ 
given by 
$P(q^{-1})\propto q^{-k_1}/(1+(q_0/q)^\gamma)^{\frac{1}{\gamma}(k_1+k_2)}$
were considered. Studied were distributions with fixed parameters $k_1=2$ and $\gamma=2$ and 
different values for $q_0$ and $k_2$. The corresponding mean PDF has the following properties:
\begin{enumerate}
        \item The distribution does not change the asymptotic  behavior of the PDF at low
        mass surface density, which is, as in case of individual clouds, proportional to a power law 
                $\left<P(\Sigma)\right>\propto \Sigma$.
        \item At high mass surface densities, the PDF asymptotically
approaches a power law that is independent
                of the radial density profile. This asymptote is either $\left<P_{\rm sph}(\Sigma)\right>\propto \Sigma^{-2 k_2-1}$
                for spheres or $\left<P_{\rm cyl}(\Sigma)\right>\propto \Sigma^{-2 k_2}$ for cylinders.
        \item For cylinders, the distribution effectively  decreases
                the high probabilities at the highest mass surface density
                of single cylinders.
        \item For $q_0\gg 1$ the mean PDF maximum of spheres and cylinders
                lies at $A_V=1-2~{\rm mag}$.
\end{enumerate}

The mean PDF of an ensemble of cylinders with the same overpressure, but a distribution of inclination angles, 
decreases the high probabilities at the highest mass surface density of single cylinders.
Randomly oriented and not overlapping cylinders with a narrow distribution of gravitational states will produce a PDF 
                with a power-law tail $\Sigma P(\Sigma)\propto \Sigma^{-1}$ at high mass surface densities.

A simple model of the global PDF of molecular clouds was presented based on a combination
of a turbulent medium and embedded randomly distributed Bonnor-Ebert spheres.
The model apparently reproduces the basic features of many of the observed PDFs 
derived by \citet{Kainulainen2009}.
\begin{enumerate}
        \item The combined model produces a log-normal function around the peak and 
        a tail at high mass surface densities where the relative ratio of the two components
        is related to the covering factor of the condensed cores.
        \item At low covering factor, the combined PDF shows a break between the two components.
        \item The functional form of the tail at high mass surface densities is affected by the background level
        (PDF maximum of the turbulent medium) relative to the PDF maximum of the cores.\\
        a) A low background can produce a tail that across the whole range is approximately
                a power law.\\
        b) A high background will lead to a curved tail, where the probabilities at the low end are higher than a power
        law.
\end{enumerate}
For the assumed reference pressure the background can be considered to be low for $A_V<\sim 1~{\rm mag}$
and high for $A_V> \sim 1~{\rm mag}$. The curvature for a high background might not be easily detectable
because of limiting resolution and noise in the observed data, so that the assumption of a simple power law
would imply a steeper slope than expected for the radial density profile.

\bibliographystyle{aa} 
\bibliography{fischera_23647_reference}

\begin{thebibliography}{49}
\expandafter\ifx\csname natexlab\endcsname\relax\def\natexlab#1{#1}\fi

\bibitem[{{Arzoumanian} {et~al.}(2011){Arzoumanian}, {Andr{\'e}}, {Didelon},
  {K{\"o}nyves}, {Schneider}, {Men'shchikov}, {Sousbie}, {Zavagno}, {Bontemps},
  {di Francesco}, {Griffin}, {Hennemann}, {Hill}, {Kirk}, {Martin}, {Minier},
  {Molinari}, {Motte}, {Peretto}, {Pezzuto}, {Spinoglio}, {Ward-Thompson},
  {White}, \& {Wilson}}]{Arzoumanian2011}
{Arzoumanian}, D., {Andr{\'e}}, P., {Didelon}, P., {et~al.} 2011, \aap, 529, L6

\bibitem[{{Beetz} {et~al.}(2008){Beetz}, {Schwarz}, {Dreher}, \&
  {Grauer}}]{Beetz2008}
{Beetz}, C., {Schwarz}, C., {Dreher}, J., \& {Grauer}, R. 2008, Physics Letters
  A, 372, 3037

\bibitem[{{Bohlin} {et~al.}(1978){Bohlin}, {Savage}, \& {Drake}}]{Bohlin1978}
{Bohlin}, R.~C., {Savage}, B.~D., \& {Drake}, J.~F. 1978, \apj, 224, 132

\bibitem[{{Bonnor}(1956)}]{Bonnor1956}
{Bonnor}, W.~B. 1956, \mnras, 116, 351

\bibitem[{{Boulares} \& {Cox}(1990)}]{Boulares1990}
{Boulares}, A. \& {Cox}, D.~P. 1990, \apj, 365, 544

\bibitem[{{Brunt}(2010)}]{Brunt2010a}
{Brunt}, C.~M. 2010, \aap, 513, A67

\bibitem[{{Brunt} {et~al.}(2010{\natexlab{a}}){Brunt}, {Federrath}, \&
  {Price}}]{Brunt2010c}
{Brunt}, C.~M., {Federrath}, C., \& {Price}, D.~J. 2010{\natexlab{a}}, \mnras,
  405, L56

\bibitem[{{Brunt} {et~al.}(2010{\natexlab{b}}){Brunt}, {Federrath}, \&
  {Price}}]{Brunt2010b}
{Brunt}, C.~M., {Federrath}, C., \& {Price}, D.~J. 2010{\natexlab{b}}, \mnras,
  403, 1507

\bibitem[{{Curry} \& {McKee}(2000)}]{Curry2000}
{Curry}, C.~L. \& {McKee}, C.~F. 2000, \apj, 528, 734

\bibitem[{{Ebert}(1955)}]{Ebert1955}
{Ebert}, R. 1955, Zeitschrift fur Astrophysik, 37, 217

\bibitem[{{Federrath} \& {Klessen}(2013)}]{Federrath2013}
{Federrath}, C. \& {Klessen}, R.~S. 2013, \apj, 763, 51

\bibitem[{{Federrath} {et~al.}(2008){Federrath}, {Klessen}, \&
  {Schmidt}}]{Federrath2008b}
{Federrath}, C., {Klessen}, R.~S., \& {Schmidt}, W. 2008, \apjl, 688, L79

\bibitem[{{Federrath} {et~al.}(2010){Federrath}, {Roman-Duval}, {Klessen},
  {Schmidt}, \& {Mac Low}}]{Federrath2010}
{Federrath}, C., {Roman-Duval}, J., {Klessen}, R.~S., {Schmidt}, W., \& {Mac
  Low}, M.-M. 2010, \aap, 512, A81

\bibitem[{{Federrath} {et~al.}(2011){Federrath}, {Sur}, {Schleicher},
  {Banerjee}, \& {Klessen}}]{Federrath2011}
{Federrath}, C., {Sur}, S., {Schleicher}, D.~R.~G., {Banerjee}, R., \&
  {Klessen}, R.~S. 2011, \apj, 731, 62

\bibitem[{{Fiege} \& {Pudritz}(2000)}]{FiegePudritz2000a}
{Fiege}, J.~D. \& {Pudritz}, R.~E. 2000, \mnras, 311, 85

\bibitem[{{Fischera}(2011)}]{Fischera2011}
{Fischera}, J. 2011, \aap, 526, A33+

\bibitem[{{Fischera}(2014)}]{Fischera2014a}
{Fischera}, J. 2014, \aap, 565, A24

\bibitem[{Fischera(2014{\natexlab{a}})}]{Fischera2014c}
Fischera, J. 2014{\natexlab{a}}, submitted for publication in A\&A

\bibitem[{Fischera(2014{\natexlab{b}})}]{Fischera2014d}
Fischera, J. 2014{\natexlab{b}}, submitted for publication in A\&A

\bibitem[{Fischera \& Dopita(2004)}]{Fischera2004a}
Fischera, J. \& Dopita, M. 2004, ApJ, 611, 911

\bibitem[{Fischera \& Dopita(2008)}]{Fischera2008}
Fischera, J. \& Dopita, M. 2008, ApJS, 176, 164

\bibitem[{{Fischera} \& {Martin}(2012{\natexlab{a}})}]{Fischera2012b}
{Fischera}, J. \& {Martin}, P.~G. 2012{\natexlab{a}}, \aap, 547, A86

\bibitem[{{Fischera} \& {Martin}(2012{\natexlab{b}})}]{Fischera2012a}
{Fischera}, J. \& {Martin}, P.~G. 2012{\natexlab{b}}, \aap, 542, A77

\bibitem[{Fitzpatrick(1999)}]{Fitzpatrick1999}
Fitzpatrick, E.~L. 1999, PASP, 111, 63

\bibitem[{{Foster} {et~al.}(2009){Foster}, {Rosolowsky}, {Kauffmann}, {Pineda},
  {Borkin}, {Caselli}, {Myers}, \& {Goodman}}]{Foster2009}
{Foster}, J.~B., {Rosolowsky}, E.~W., {Kauffmann}, J., {et~al.} 2009, \apj,
  696, 298

\bibitem[{Joos \& Richter(1978)}]{JoosRichter1978}
Joos \& Richter. 1978, H{\"o}here Mathematik f{\"u}r den Praktiker (Verlag
  Harri Deutsch - Thun - Frankfurt)

\bibitem[{{Kainulainen} {et~al.}(2009){Kainulainen}, {Beuther}, {Henning}, \&
  {Plume}}]{Kainulainen2009}
{Kainulainen}, J., {Beuther}, H., {Henning}, T., \& {Plume}, R. 2009, \aap,
  508, L35

\bibitem[{{Kainulainen} \& {Tan}(2013)}]{Kainulainen2013}
{Kainulainen}, J. \& {Tan}, J.~C. 2013, \aap, 549, A53

\bibitem[{{Kandori} {et~al.}(2005){Kandori}, {Nakajima}, {Tamura}, {Tatematsu},
  {Aikawa}, {Naoi}, {Sugitani}, {Nakaya}, {Nagayama}, {Nagata}, {Kurita},
  {Kato}, {Nagashima}, \& {Sato}}]{Kandori2005}
{Kandori}, R., {Nakajima}, Y., {Tamura}, M., {et~al.} 2005, \aj, 130, 2166

\bibitem[{{Keto} \& {Caselli}(2010)}]{Keto2010}
{Keto}, E. \& {Caselli}, P. 2010, \mnras, 402, 1625

\bibitem[{{Keto} {et~al.}(2014){Keto}, {Rawlings}, \& {Caselli}}]{Keto2014}
{Keto}, E., {Rawlings}, J., \& {Caselli}, P. 2014, \mnras, 440, 2616

\bibitem[{{Kirk} {et~al.}(2006){Kirk}, {Johnstone}, \& {Di
  Francesco}}]{Kirk2006}
{Kirk}, H., {Johnstone}, D., \& {Di Francesco}, J. 2006, \apj, 646, 1009

\bibitem[{{Konstandin} {et~al.}(2012){Konstandin}, {Girichidis}, {Federrath},
  \& {Klessen}}]{Konstandin2012}
{Konstandin}, L., {Girichidis}, P., {Federrath}, C., \& {Klessen}, R.~S. 2012,
  \apj, 761, 149

\bibitem[{{Kritsuk} {et~al.}(2007){Kritsuk}, {Norman}, {Padoan}, \&
  {Wagner}}]{Kritsuk2007}
{Kritsuk}, A.~G., {Norman}, M.~L., {Padoan}, P., \& {Wagner}, R. 2007, \apj,
  665, 416

\bibitem[{{Kritsuk} {et~al.}(2011){Kritsuk}, {Norman}, \&
  {Wagner}}]{Kritsuk2011}
{Kritsuk}, A.~G., {Norman}, M.~L., \& {Wagner}, R. 2011, \apjl, 727, L20

\bibitem[{{Lazarian} \& {Pogosyan}(2000)}]{Lazarian2000}
{Lazarian}, A. \& {Pogosyan}, D. 2000, \apj, 537, 720

\bibitem[{{McCrea}(1957)}]{McCrea1957}
{McCrea}, W.~H. 1957, \mnras, 117, 562

\bibitem[{Nordlund \& Padoan(1999)}]{Nordlund1999}
Nordlund, {\AA}.~P. \& Padoan, P. 1999, in Insterstellar Turbulence, ed.
  J.~Franco \& A.~Carrami{\~{n}}ana (Cambridge University Press), 218

\bibitem[{Ostriker {et~al.}(2001)Ostriker, Stone, \& Gammie}]{Ostriker2001}
Ostriker, E.~C., Stone, J.~M., \& Gammie, C.~F. 2001, ApJ, 546, 980

\bibitem[{{Ostriker}(1964)}]{Ostriker1964}
{Ostriker}, J. 1964, \apj, 140, 1056

\bibitem[{{Padoan} {et~al.}(1997{\natexlab{a}}){Padoan}, {Jones}, \&
  {Nordlund}}]{Padoan1997b}
{Padoan}, P., {Jones}, B.~J.~T., \& {Nordlund}, A.~P. 1997{\natexlab{a}}, \apj,
  474, 730

\bibitem[{{Padoan} {et~al.}(1997{\natexlab{b}}){Padoan}, {Nordlund}, \&
  {Jones}}]{Padoan1997a}
{Padoan}, P., {Nordlund}, A., \& {Jones}, B.~J.~T. 1997{\natexlab{b}}, \mnras,
  288, 145

\bibitem[{Passot \& V\'azquez-Semadeni(1998)}]{Passot1998}
Passot, T. \& V\'azquez-Semadeni, E. 1998, Phys. Rev. E, 58, 4501

\bibitem[{{Price} {et~al.}(2011){Price}, {Federrath}, \& {Brunt}}]{Price2011}
{Price}, D.~J., {Federrath}, C., \& {Brunt}, C.~M. 2011, \apjl, 727, L21

\bibitem[{{Schneider} {et~al.}(2012){Schneider}, {Csengeri}, {Hennemann},
  {Motte}, {Didelon}, {Federrath}, {Bontemps}, {Di Francesco}, {Arzoumanian},
  {Minier}, {Andr{\'e}}, {Hill}, {Zavagno}, {Nguyen-Luong}, {Attard},
  {Bernard}, {Elia}, {Fallscheer}, {Griffin}, {Kirk}, {Klessen}, {K{\"o}nyves},
  {Martin}, {Men'shchikov}, {Palmeirim}, {Peretto}, {Pestalozzi}, {Russeil},
  {Sadavoy}, {Sousbie}, {Testi}, {Tremblin}, {Ward-Thompson}, \&
  {White}}]{Schneider2012}
{Schneider}, N., {Csengeri}, T., {Hennemann}, M., {et~al.} 2012, \aap, 540, L11

\bibitem[{{Schneider} {et~al.}(2013){Schneider}, {Csengeri}, {Hennemann},
  {Motte}, {Didelon}, {Federrath}, {Bontemps}, {Di Francesco}, {Arzoumanian},
  {Minier}, {Andr{\'e}}, {Hill}, {Zavagno}, {Nguyen-Luong}, {Attard},
  {Bernard}, {Elia}, {Fallscheer}, {Griffin}, {Kirk}, {Klessen}, {K{\"o}nyves},
  {Martin}, {Men'shchikov}, {Palmeirim}, {Peretto}, {Pestalozzi}, {Russeil},
  {Sadavoy}, {Sousbie}, {Testi}, {Tremblin}, {Ward-Thompson}, \&
  {White}}]{Schneider2013}
{Schneider}, N., {Csengeri}, T., {Hennemann}, M., {et~al.} 2013, \aap, 551, C1

\bibitem[{{Stod{\'o}lkiewicz}(1963)}]{Stodolkiewicz1963}
{Stod{\'o}lkiewicz}, J.~S. 1963, Acta Astronomica, 13, 30

\bibitem[{{Vazquez-Semadeni}(1994)}]{Vazquez1994}
{Vazquez-Semadeni}, E. 1994, \apj, 423, 681

\bibitem[{{V{\'a}zquez-Semadeni} \& {Garc{\'{\i}}a}(2001)}]{Vazquez2001}
{V{\'a}zquez-Semadeni}, E. \& {Garc{\'{\i}}a}, N. 2001, \apj, 557, 727

\end{thebibliography}

\begin{acknowledgement}
        This work was supported by grants from the Natural
        Sciences and Engineering Research Council of Canada and the Canadian
        Space Agency. I would like to thank Richard Tuffs for his support and
        Jouni Kainulainen for helpful discussions. Furthermore, I would like to thank the
        referee for a careful reading of the manuscript and the suggestions
        that helped to improve the manuscript.
        Many thanks go to my parents who enabled me to finish the project.
\end{acknowledgement}

\appendix

\section{\label{app_isocomparison}Approximations of a Bonnor-Ebert sphere}

Here, the mass, radius, and mean mass surface density of spheres with an analytical 
profile given by Eq.~\ref{eq_densityprofile} are compared with the
correct values of Bonnor-Ebert spheres. They are shown in Fig.~\ref{fig_massradius1}.
Characteristic parameters are listed in Table~\ref{table_cloudparameters}.

\begin{figure}[htbp]
	\hfill
        \includegraphics[width=0.92\hsize]{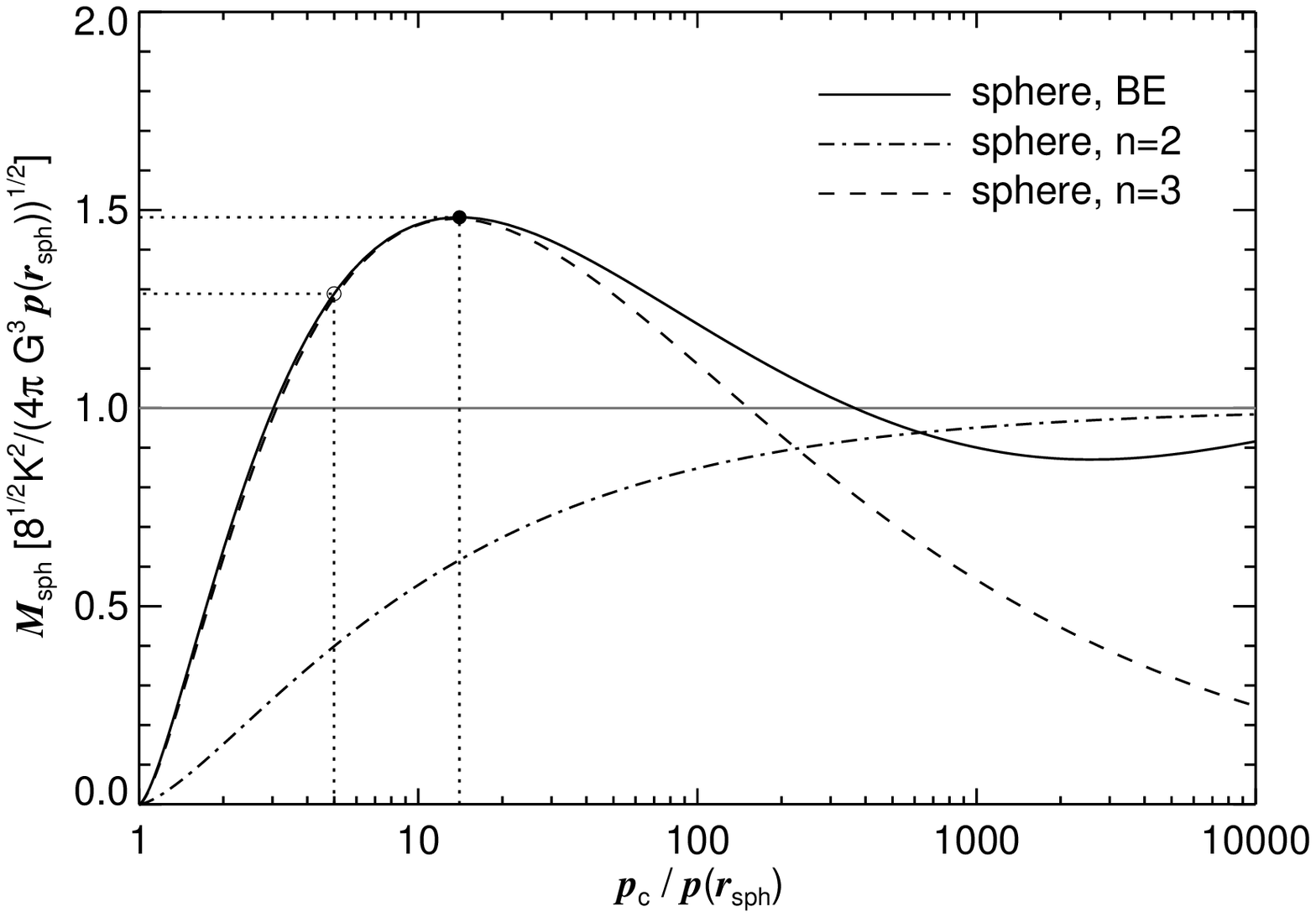}
	\hfill\hfill

	\hfill
        \includegraphics[width=0.92\hsize]{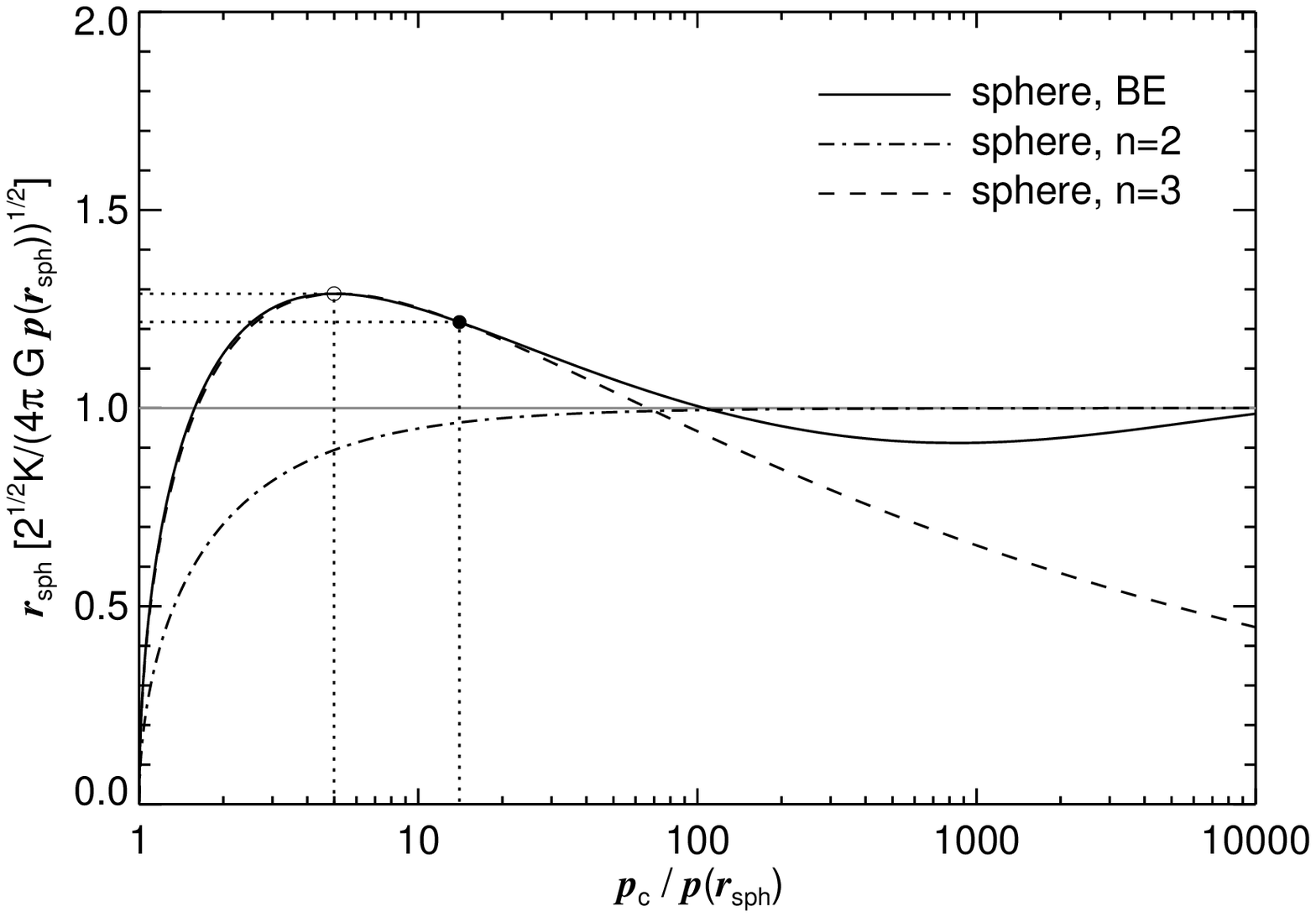}     
	\hfill\hfill

	\hfill
        \includegraphics[width=0.92\hsize]{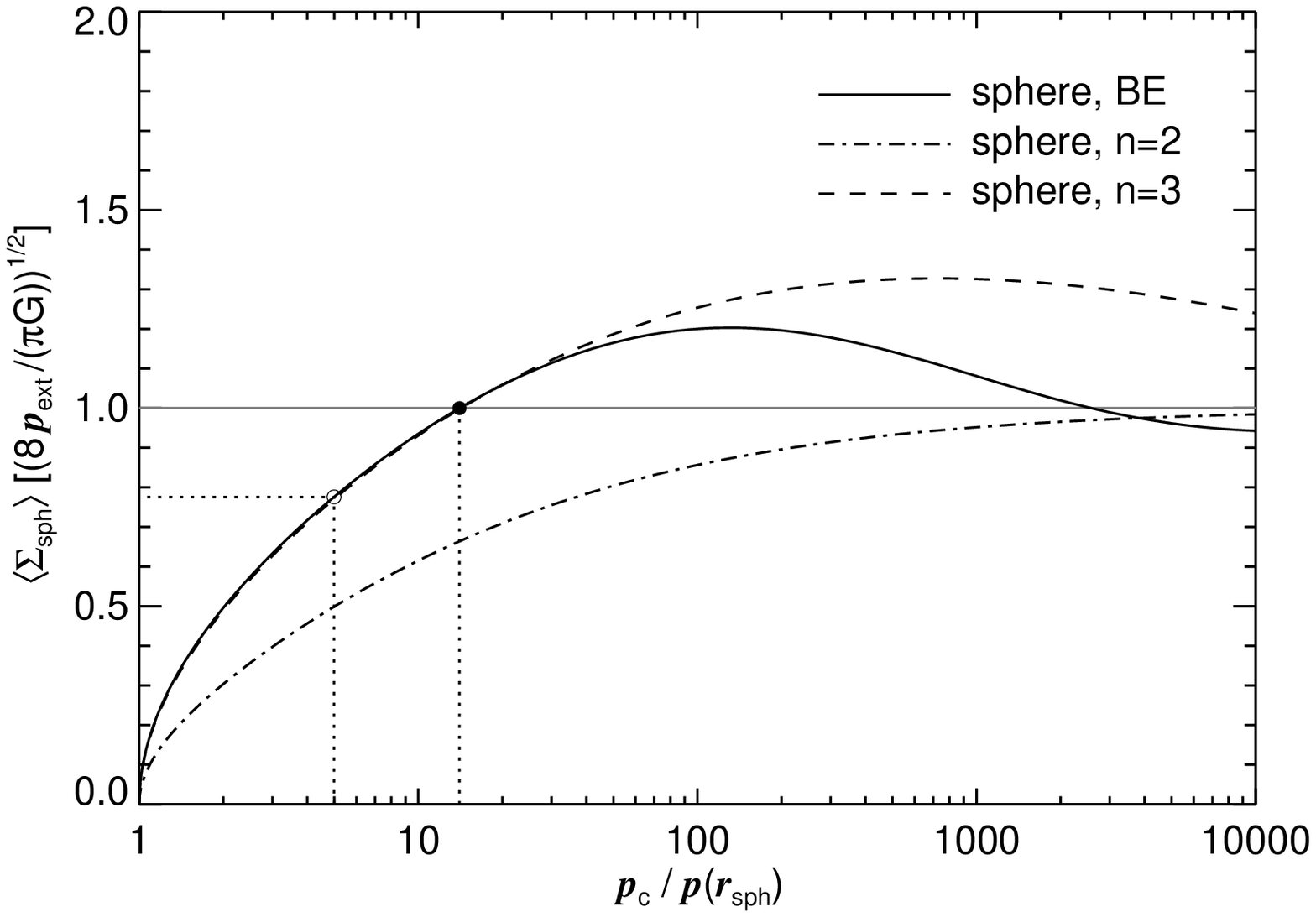}
	\hfill\hfill
        \caption{\label{fig_massradius1}Mass, radius, and mean mass surface density
        of spherical clouds with fixed bounding pressure $p(z_{\rm sph})$ and temperature
        as a function of the pressure ratio $p_{\rm c} / p(z_{\rm sph})$. The values of a Bonnor-Ebert
        sphere are compared with the corresponding values of spherical clouds with smooth density profiles as 
        given in Eq.~\ref{eq_densityprofile} with $n=2$ and $n=3$ (dashed-dotted and dashed line, respectively). 
        The dotted vertical and horizontal lines mark
        the values at maximum radius (open circle) and at critical stability (filled circle) of Bonnor-Ebert spheres.
        The   mass, radius, and mean mass surface density of a Bonnor-Ebert sphere in the limit of infinite
        overpressure are shown as gray horizontal lines.}
\end{figure}

\subsection{\label{app_cloudmass}Mass of Bonnor-Ebert spheres}
The mass of a  Bonnor-Ebert sphere is given by (\citealt{Fischera2008})
\begin{equation}
        M_{\rm BE}(\theta_{\rm cl}) =  
        \frac{K^2}{\sqrt{4\pi G^3 p_{\rm ext}}}
        e^{-\omega(\theta_{\rm cl})/2}\int_0^{\theta_{\rm cl}}{\rm d}\theta\,\theta^2\,e^{-\omega(\theta)},
\end{equation}
where the size $\theta_{\rm cl}$ is determined by pressure equilibrium 
$p_{\rm c}e^{-\omega(\theta_{\rm cl})}=p(\theta_{\rm cl})=p_{\rm ext}$.

In Fig.~\ref{fig_massradius1} the mass of Bonnor-Ebert spheres
is compared with the masses of spheres using the approximation $n=3$ and
$n=2,$ which are given by
\begin{eqnarray}
        \label{eq_massn3}
        M_{\rm sph,3}(q) &=& \frac{K^2\xi^{3/2}_3 \sqrt{q}}{\sqrt{4\pi G^3 p_{\rm ext}}}\left\{\ln\left[q^{-\frac{1}{3}}+q^{-\frac{1}{3}}\sqrt{1-q^{\frac{2}{3}}}\right]\right.\nonumber\\
                                &&\quad\quad \left.-\sqrt{1-q^{\frac{2}{3}}}\right\},
\end{eqnarray}
and
\begin{equation}
        M_{\rm sph,2}(q) = \frac{K^2\xi^{3/2}_2 \sqrt{q}}{\sqrt{4\pi G^3 p_{\rm ext}}}\left\{\sqrt{\frac{1-q}{q}}
                -\tan^{-1}\sqrt{\frac{1-q}{q}}\right\}.
\end{equation}
At given external pressure and temperature, the mass of a Bonnor-Ebert sphere increases with increasing overpressure
up to the critical value where the sphere has the highest mass. 
For higher overpressure the mass is lower
than the critical value and fluctuates to an asymptotic value at infinite overpressure given by
\begin{equation}
        M_{\rm BE} \rightarrow \frac{\sqrt{8}K^2}{\sqrt{4\pi G^3 p_{\rm ext}}}.
\end{equation}

\subsubsection{\label{sect_critmass}Critical stability}

Bonnor-Ebert spheres have a well-known critically stable configuration
related to the response of the gas pressure at cloud radius in relation to compression.
As long as the compression leads to a pressure increase (${\rm d} p(r_{\rm cl})/{\rm d r_{\rm cl}}<0$), a cloud
is considered stable, otherwise as unstable.
Critical stability is given for
\begin{equation}
        \frac{{\rm d} p(r_{\rm cl})}{{\rm d}r_{\rm cl}} =\frac{{\rm d} p(r_{\rm cl})}{{\rm d}q} \left(\frac{{\rm d}r_{\rm cl}}{{\rm d}q}\right)^{-1}= 0.
\end{equation}
As the derivative of the radius {\rm with respect to q} is nonzero at the pressure maximum, 
the critical condition is 
\begin{equation}
        \frac{d p(r_{\rm cl})}{dq} =0.
\end{equation}

For Bonnor-Ebert spheres, a critically stable sphere of the first pressure maximum, which
is also the global maximum, is characterized through an overpressure of 14.04. 
When we apply the same criterion to spheres with  
an analytical profile as given by Eq.~\ref{eq_densityprofile} with $n=3,$ 
we find from Eq.~\ref{eq_massn3} for fixed mass and fixed $K$ 
a critical overpressure $q_{\rm crit}^{-1}=13.46$, close to the critical value
of a Bonnor-Ebert sphere. The physical parameters corresponding to the critical
value of a sphere with $n=3$ are listed in Table~\ref{table_cloudparameters}.

As pointed out in the paper of \citet{Fischera2012a}, the critical mass of a Bonnor-Ebert sphere 
is also the highest possible mass for fixed $K$ and external pressure $p_{\rm ext}$ 
, but varying pressure ratio $q$. The same applies for a sphere with $n=3$. A cylinder, by
comparison, has no critical stability with regard to compression. The highest mass line density
is related to infinite overpressure. The situation for spheres with a $n=2$-profile would be
the same as for cylinders.

\subsection{\label{app_cloudradius} Radius of Bonnor-Ebert spheres}

The radius of the Bonnor-Ebert sphere is given by 
\begin{equation}
        r_{\rm BE}(\theta_{\rm cl}) = \theta_{\rm cl} \frac{K}{\sqrt{4\pi G p_{\rm ext}}}e^{-\omega(z_{\rm cl})/2}.
\end{equation}
For a cloud with a truncated analytical density profile  we find from Eq.~\ref{eq_densityprofile} and the
inner radius $r_0=\sqrt{\xi_n}/A,$ where $A$ can be expressed through $A^2 = 4\pi G p_{\rm ext}/K^2/q,$
\begin{equation}
        \label{eq_cloudradius}
        r_{\rm cl,n} = \frac{\sqrt{\xi_n} K}{\sqrt{4\pi G p_{\rm ext}}} q^{1/2-1/n}\sqrt{1-q^{2/n}}.
\end{equation}
For a given $q$ and $p_{\rm ext}$ the radius increases proportionally
to $K$. Clouds with the same $q$ and $K$ will be smaller in higher pressure regions.
Clouds with $n>2$ have the greatest extension at an overpressure
\begin{equation}
        q^{-1}_{\rm max} = \left(\frac{n}{n-2}\right)^{\frac{n}{2}}.
\end{equation}

Figure~\ref{fig_massradius1} shows that the radius of the Bonnor-Ebert sphere
behaves similarly as a function of overpressure
as does the mass. However, the cloud with the greatest extension is subcritical with an overpressure of 4.990.
The greatest extension of a sphere with n=3
corresponds to an overpressure
\begin{equation}
        q_{\rm max}^{-1} = 3^{3/2}\approx 5.196,
\end{equation}
which is close to the highest value of Bonnor-Ebert spheres. 
For a higher overpressure, the radius of a Bonnor-Ebert sphere is generally smaller 
than the radius at critical stability, but at a given $q$ it
is larger than a sphere with 
an analytical density profile with $n=3$. In the limit of high overpressure the radius of the Bonnor-Ebert
sphere fluctuates asymptotically toward 
\begin{equation}
        r_{\rm BE} \rightarrow \frac{\sqrt{2} K}{\sqrt{4\pi G p_{\rm ext}}},
\end{equation}
which is identical with the radius of a sphere with $n=2$ in the limit of infinite overpressure.

\subsection{\label{app_msurfmean}Mean mass surface density}

It has been shown \citep{Fischera2012a} that for overpressures where the gas pressure at the cloud boundary 
of a sphere with fixed mass has local maxima for varying overpressure
the mean mass surface density of Bonnor-Ebert spheres is simply given by
\begin{equation}
        \left<\Sigma_{\rm BE}\right> = \frac{M_{\rm cl}}{\pi r_{\rm cl}^2}=\sqrt{\frac{8 p_{\rm ext}}{\pi G}}.
\end{equation}
The mean value applies, for example, to a critically stable sphere and for a sphere with infinite 
overpressure (Fig.~\ref{fig_massradius1}). For a sphere with infinite overpressure this can be easily verified
by replacing the density ratio $e^{-\omega(\theta)}$ by the asymptotic density profile at large scales $\theta$,
which is shown in Sect.~\ref{sect_isosphere} to be $2/\theta^2$.

\begin{table}[htbp]
        \caption{\label{table_cloudparameters}Characteristic cloud parameters}
        \begin{tabular}{c|c|c|c|c}
                                &       & BE & $n=3$ & $n=2$ \\
        \hline
                $\xi_n$ &       &       & $8.63$ & $2$ \\
        \hline
        \multicolumn{5}{c}{critical mass}\\
        \hline
        $\theta_{\rm crit}$ &           &       $6.451$ & $6.340$  &  --- \\
        $q_{\rm crit}^{-1}$ &           &14.04 & $13.46$ & --- \\
        $M_{\rm crit}$  & $[K^2/\sqrt{4\pi G^3 p_{\rm ext}}]$ & $4.191$        & $4.180$ & ---\\
        $r_{\rm crit}$  & $[K/\sqrt{4\pi G p_{\rm ext}}]$ & $1.721$ & $1.728$  & --- \\
        $\left<\Sigma\right>$ & $[\sqrt{ 8p_{\rm ext}/(\pi G)}]$ & 1 & 0.990 & --- \\
        $[\Sigma(0)]_{\rm crit}$ & $[\sqrt{p_{\rm ext}/(\pi G)}]$ & 10.07 & 9.778 & --- \\
        \hline
        \multicolumn{5}{c}{largest radius}\\
        \hline
        $\theta_{\rm max}$      &               & 4.071 & 4.155& $'\infty'$\\
        $q_{\rm max}^{-1}$ &    & $4.990$ & $3^{3/2}$ & $'\infty'$ \\
        $M_{\rm sph}$ & $[K^2/\sqrt{4\pi G^3 p_{\rm ext}}]$ & 3.645 & 3.667 & $\sqrt{8}$\\
        $r_{\rm max}$   & $[K/\sqrt{4\pi G p_{\rm ext}}]$ & $1.8226$ & 1.8226 & $\sqrt{2}$\\
        $\left<\Sigma\right>$ & $[\sqrt{ 8p_{\rm ext}/(\pi G)}]$ & 0.776 & 0.781 & 1 \\
        \hline
        \multicolumn{5}{c}{parameters in the limit $\theta \rightarrow \infty$, $q\rightarrow 0$}\\
        \hline
        $M_{\rm lim}$ & $[K^2/\sqrt{4\pi G^3 p_{\rm ext}}]$ & $\sqrt{8}$ & 0 & $\sqrt{8}$ \\
        $r_{\rm lim}$   & $[K/\sqrt{4\pi G p_{\rm ext}}]$ & $\sqrt{2}$ & 0 & $\sqrt{2}$\\
        $\left<\Sigma\right>$ & $[\sqrt{8 p_{\rm ext}/(\pi G)}]$ & 1 & 0 & 1 \\
        \end{tabular}
\end{table}

\section{\label{app_pdfdens}Probability distribution function of the local density}

In this section we provide for the sake of completeness the PDF of the local density of spheres
and cylinders with a truncated density profile as given by Eq.~\ref{eq_densityprofile}. 

The logarithmic PDF is given by
\begin{equation}
        \label{eq_pdfdensity}
        \rho P(\rho) = \frac{{\rm d}V(r)}{{\rm d}r\,V_{\rm cl}}\left(-\frac{1}{\rho}\frac{{\rm d}\rho}{{\rm d}r}\right)^{-1},
\end{equation}
where $V_{\rm cl}$ is the cloud volume and where $V(r)$ is the cloud volume within radius $r$. 

For a truncated density profile with a pressure ratio $p_{\rm ext}/p_{\rm c}=q$ 
we obtain for spheres
\begin{equation}
        \rho\,P_{\rm sph}(\rho) = \frac{3}{n} \left(\frac{\rho}{\rho(r_{\rm sph})}\right)^{-\frac{3}{n}}\frac{\sqrt{1-\left(\frac{\rho}{\rho_{\rm c}}\right)^{2/n}}}{(1-q^{2/n})^{3/2}}.
\end{equation}
In the limit of high overpressure this becomes for $(\rho/\rho_{\rm c})^{2/n}\ll 1$ a power law
\begin{equation}
        \rho\,P_{\rm sph}(\rho) \approx \frac{3}{n}\left(\frac{\rho}{\rho(r_{\rm sph})}\right)^{-\frac{3}{n}},
\end{equation}
or $\rho P_{\rm sph}(\rho)\propto \rho^{-3/n}$ , as expected for simple radial density 
profiles $\rho\propto r^{-n}$ \citep{Kritsuk2011,Federrath2011}.

For cylinders the PDF of the local density is given by
\begin{equation}
        \rho\,P_{\rm cyl}(\rho) = \frac{2}{n} \frac{1}{1-q^{2/n}} \left(\frac{\rho}{\rho(r_{\rm cyl})}\right)^{-\frac{2}{n}}.
\end{equation}
The PDF for a given $q$ over the full range is a power law. For isothermal self-gravitating cylinders
we obtain $\rho P_{\rm cyl}(\rho)\propto \rho^{-\frac{1}{2}}$.

\subsection{\label{app_pdfdensbonnor}Local density PDF of Bonnor-Ebert spheres}

\begin{figure}[htbp]
        \includegraphics[width=\hsize]{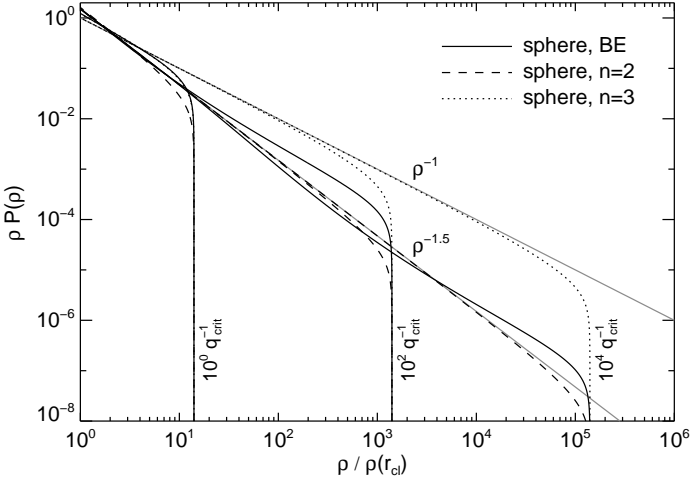}
        \caption{\label{fig_pdfdensity} Local density PDF of critical and supercritical 
        Bonnor-Ebert spheres for three different pressure ratios, given in units of the overpressure of a critically stable
        Bonnor-Ebert sphere. They are directly compared with the PDFs of spheres with density profiles
        as given in Eq.~\ref{eq_densityprofile} with n=3 and n=2. Their corresponding power-law asymptotes 
        in the limit of infinite overpressure are shown as gray lines.
        }
\end{figure}

Replacing in Eq.~\ref{eq_pdfdensity}
the density by the solution of the Lane-Emden equation $\rho = \rho_{\rm c}e^{-\omega(\theta)}$ and the radius
by the unit-free size $\theta=r A$ provides for the local density PDF of a Bonnor-Ebert sphere
\begin{equation}
        \rho\,P_{\rm BE}(\rho) = \frac{3 \theta^2}{\theta_{\rm cl}^3} \left(\frac{{\rm d}\omega}{{\rm d}\theta}\right)^{-1}.
\end{equation}
The PDFs for critically stable and supercritical Bonnor-Ebert spheres (here defined as a self-gravitating sphere
with an overpressure $p_{\rm c}/p_{\rm ext}>14.04$) are shown in Fig.~\ref{fig_pdfdensity}. 
We showed in Sect.~\ref{sect_densityprofile} that the inner part up to $p_{\rm c}/p(r)<\sim 100$ 
of the density profile of self-gravitating
spheres can be approximated by an analytical profile as given by Eq.~\ref{eq_densityprofile} with n=3. The 
shape of the PDF of the Bonnor-Ebert spheres at high density values is consequently determined by this 
profile. For example,
the PDF of Bonnor-Ebert spheres at densities close to the central density value is approximately
a power law with $\rho^{-1}$. 
For supercritical Bonnor-Ebert spheres, the PDF of density values $\rho\ll\rho_{\rm c}$ 
asypmtotically approaches the PDF of a sphere with an analytical density profile with n=2. 
The PDF becomes close to a power law $\rho^{-1.5}$.

The values are close to the values obtained with collapse models. \citet{Kritsuk2011} studied the theoretical
density distribution in star-forming
interestellar clouds. The collapse model provided a PDF that can be approximated over a wide range by a power
law $\rho^{-1.695}$. At highest densities the PDF showed a flatter shape approximated by a power law $\rho^{-1}$
and therefore the same as for Bonnor-Ebert spheres.
However, the physical explanation given are rotationally supported cores.

\section{\label{app_pdfmaximum}Estimating the maxima of $P_{\rm BE}(\Sigma)$ and $\Sigma P_{\rm BE}(\Sigma)$}

The maxima positions for linear and logarithmic PDFs of Bonnor-Ebert spheres 
were derived by estimating the zero points of the first derivatives. 
The highest positions of the linear and logarithmic PDF fulfill the conditions
\begin{eqnarray}
        \label{eq_pdfmaximumlin}
        \frac{{\rm d}}{{\rm d}\theta_\bot}[P_{\rm BE}(\Sigma)] &=& 0,\\
        \label{eq_pdfmaximumlog}
         \frac{{\rm d}}{{\rm d}\theta_\bot}[\Sigma\,P_{\rm BE}(\Sigma)] &=& 0,
\end{eqnarray}
where the linear PDF of the Bonnor-Ebert sphere is given by 
\begin{equation}
        \label{eq_pdfbonnor}
        P_{\rm BE}(\Sigma) = \frac{2r_{\bot}}{r_{\rm cl}^2}\left(-\frac{{\rm d}\Sigma}{{\rm d}r_\bot}\right)^{-1}
                =\frac{2\theta_\bot}{\theta_{\rm cl}^2}\left(-\frac{{\rm d}\Sigma}{{\rm d}\theta_\bot}\right)^{-1}.
\end{equation}

\subsection{Linear PDF}

Inserting the expression \ref{eq_pdfbonnor} into the condition Eq.~\ref{eq_pdfmaximumlin} for the maxima, we obtain
\begin{equation}
        \frac{{\rm d}}{{\rm d}\theta_\bot}[P_{\rm BE}(\Sigma)] 
        = \frac{2}{\theta_{\rm cl}^2} \left(\frac{1}{\theta_\bot}\frac{{\rm d}\Sigma}{{\rm d}\theta_\bot}\right)^{-2}
        \frac{{\rm d}}{{\rm d}\theta_\bot} \left[\frac{1}{\theta_\bot}\frac{{\rm d}\Sigma}{{\rm d}\theta_\bot}\right] = 0.
\end{equation}
The condition becomes
\begin{equation}
        \frac{{\rm d}}{{\rm d}\theta_\bot} \left[\frac{1}{\theta_\bot}\frac{{\rm d}\Sigma}{{\rm d}\theta_\bot}\right] = 0,
\end{equation}
where the expression in brackets is referred to as $g(\theta_\bot)$.
From the derivative of the mass surface density 
Eq.~\ref{eq_dmsurfiso}, it follows that
\begin{eqnarray}
        g(\theta_\bot) & = & -2\frac{\rho_{\rm c}}{A}\Bigg\{
                        \frac{e^{-\omega(\theta_{\rm cl})}}{\sqrt{\theta_{\rm cl}^2-\theta_\bot^2}}+\nonumber\\
                        &&\int\limits_0^{\sqrt{\theta_{\rm cl}^2-\theta_\bot^2}}
                        {\rm d}\theta_{\|}\,\frac{e^{-\omega(\sqrt{\theta_{\|}^2+\theta_\bot^2})}}{\sqrt{\theta_{\|}^2+\theta_\bot^2}}
                        \left.\frac{{\rm d}\omega}{{\rm d}\theta}\right|_{\sqrt{\theta_{\|}^2+\theta_\bot^2}}\Bigg\},
\end{eqnarray}
where the integration along the radius $\theta$ has been changed to the integration along the
depths $\theta_{\|}=\sqrt{\theta^2-\theta_\bot^2}$ at impact radius $\theta_\bot$. The change of $g(\theta_\bot)$
by an infinitely small increase ${\rm d}\theta_\bot$ is obtained by expanding the upper limit of the integral and the
functions depending on $\theta_\bot$ into 
Taylor series. To first order, we find
\begin{eqnarray}
        &&g(\theta_\bot+{\rm d}\theta_\bot) = g(\theta_\bot)-2\frac{\rho_{\rm c}}{A}{\rm d}\theta_\bot\Bigg\{\nonumber\\
                &&      \int\limits_0^{\sqrt{\theta^2_{\rm cl}-\theta_{\bot}^2}}{\rm d}\theta_{\|}\,
                        \frac{\theta_\bot e^{-\omega(\sqrt{\theta_{\|}^2+\theta_\bot^2})}}{{\theta_{\|}^2+\theta_\bot^2}}
                                h(\sqrt{\theta_{\|}^2+\theta_\bot^2})\nonumber\\
                && - \frac{e^{-\omega(\theta_{\rm cl})}}{\sqrt{\theta_{\rm cl}^2-\theta_\bot^2}}
                        \left[\frac{\theta_\bot}{\theta_{\rm cl}}\left.\frac{{\rm d}\omega}{{\rm d}\theta}\right|_{\theta_{\rm cl}}
                                -\frac{\theta_\bot}{\theta_{\rm cl}^2-\theta_\bot^2}\right]\Bigg\},
\end{eqnarray}
where
\begin{equation}
        \label{eq_htheta}
        h(\theta)=\frac{{\rm d}^2\omega}{{\rm d}\theta^2}\bigg|_{\theta} - \frac{1}{\theta} \,\frac{{\rm d}\omega}{{\rm d}\theta}\bigg|_{\theta}
                -\left(\frac{{\rm d}\omega}{{\rm d}\theta}\bigg|_{\theta}\right)^2.
\end{equation}
For the derivative we obtain
\begin{eqnarray}
        \frac{{\rm d}g(\theta_\bot)}{{\rm d}\theta_\bot} &=& \frac{g(\theta_\bot+{\rm d}\theta_\bot)-g(\theta_\bot)}{{\rm d}\theta_\bot}\nonumber\\
                &=& -\frac{2\rho_{\rm c}}{A}\Bigg\{\int\limits_{\theta_\bot}^{\theta_{\rm cl}}{\rm d}\theta\,
                        \frac{\theta_\bot\,e^{-\omega(\theta)}}{\theta\sqrt{\theta^2-\theta_\bot^2}}
                        h(\theta) -\nonumber\\          
                && \frac{e^{-\omega(\theta_{\rm cl})}}{\sqrt{\theta_{\rm cl}^2-\theta_\bot^2}}
                        \left[
                        \frac{\theta_\bot}{\theta_{\rm cl}}\frac{{\rm d}\omega}{{\rm d}\theta}\bigg|_{\theta_{\rm cl}}
                                -\frac{\theta_\bot}{\theta_{\rm cl}^2-\theta_\bot^2}\right]\Bigg\}.
\end{eqnarray}

\subsection{Logarithmic PDF}

The condition for the highest positions of the logarithmic PDF is derived in a similar manner. 
Eq.~\ref{eq_pdfmaximumlog} provides 
\begin{equation}
        -\left(\frac{1}{\theta_\bot}\frac{{\rm d}\Sigma}{{\rm d}\theta_\bot}\right)^2
                +\frac{\Sigma}{\theta_\bot}\frac{{\rm d}}{{\rm d}\theta_\bot}\left(\frac{1}{\theta_\bot}\frac{{\rm d}\Sigma}{{\rm d}\theta_\bot}\right) = 0.
\end{equation}

\section{\label{app_pdfisoasymptote}Approximation of the $P_{\rm BE}(\Sigma)$}

In this appendix we show for the sake of completeness that for the approximations of Bonnor-Ebert spheres 
 expression \ref{eq_pdfisosph} of the PDF is identical to  Eqs.~\ref{eq_pdfsphere} 
and~\ref{eq_pdfsphere2}.
Replacing in Eq.~\ref{eq_pdfisosph} the density profile $e^{-\omega(\theta)}=\rho(\theta)/\rho_{\rm c}$ by the analytical density
profile as given in Eq.~\ref{eq_densityprofile} and the potential by $\omega = -\ln [\rho(\theta)/\rho_{\rm c}],$ we obtain 
\begin{eqnarray}
        \label{eq_pdfbealternative}
        &&P_{\rm BE}(\Sigma(\theta_\bot)) \sim \sqrt{\frac{4\pi G}{p_{\rm ext}}}\frac{\sqrt{q}}{\theta_{\rm cl}^2}\Bigg\{
                \frac{q}{\sqrt{\theta_{\rm cl}^2-\theta_\bot^2}}                \nonumber\\
                        &&
                        +\int\limits_{0}^{\sqrt{\theta_{\rm cl}^2-\theta_\bot^2}}{\rm d}\theta_{\|}\,
                                \frac{n/\xi_n}{(1+(\theta_{\|}^2+\theta_{\bot}^2)/\xi_n)^{\frac{n}{2}+1}}\Bigg\}^{-1}.
\end{eqnarray}
where we have changed the integration along the radius $\theta$ to the integration along the
depth $\theta_{\|}=\sqrt{\theta^2-\theta_\bot^2}$. By introducing the integration variable
\begin{equation}
        u = \frac{\theta_{\|}/\sqrt{\xi_n}}{\sqrt{1+\theta_\bot^2/\xi_n}}
\end{equation}
and by replacing the projected radius and the cloud radius by $\theta_\bot = x\theta_{\rm cl}$ and
$\theta_{\rm cl}=\sqrt{\xi_n} q^{-1/n}\sqrt{1-q^{2/n}}$ the expression \ref{eq_pdfbealternative} can be transformed into
\begin{eqnarray}
        P_{\rm BE}(\Sigma(y_n)) &\sim& \sqrt{\frac{4\pi G}{\xi_n p_{\rm ext}}}\frac{q^{\frac{2-n}{2n}}}{1-q^{2/n}} \Bigg\{
                        \frac{1}{\sqrt{y_n}}+\nonumber\\
                        &&\frac{n}{(1-y_n)^{\frac{1+n}{2}}}\int\limits_0^{u_{\rm max}}\frac{{\rm d}u}{(1+u^2)^{\frac{n}{2}+1}}\Bigg\}^{-1},
\end{eqnarray}
where $y_n = (1-q^{2/n})(1-x^2)$ and where the upper limit is given by $u_{\rm max} = \sqrt{y_n/(1-y_n)}$.
The exponent of the integrand can be reduced by using (simplified version of Eq.~6-57 of \citet{JoosRichter1978})
\begin{eqnarray}
        \int{\rm d}u\,\frac{1}{(1+u^2)^m}&=&\frac{u}{2(m-1)(u^2+1)^{m-1}}\nonumber\\
        &&+\frac{2m-3}{2(m-1)}\int{\rm d}u\frac{1}{(1+u^2)^{m-1}},
\end{eqnarray}
which provides 
\begin{eqnarray}
        P_{\rm BE}(\Sigma(y_n)) & \sim & \sqrt{\frac{4\pi G}{\xi_n p_{\rm ext}}}\frac{q^{\frac{2-n}{2n}}}{1-q^{2/n}} \Bigg\{
                \frac{1}{\sqrt{y_n}(1-y_n)}+\nonumber\\
                &&\frac{n-1}{(1-y_n)^{\frac{1+n}{2}}} \int\limits_0^{u_{\rm max}}\,\frac{{\rm d}u}{(1+u^2)^{n/2}}\Bigg\}^{-1}.
\end{eqnarray}
After applying Eq.~\ref{eq_defxn2} for the unit-free mass surface density, we obtain
\begin{eqnarray}
        P_{\rm BE}(\Sigma(y_n))  &\sim&  \sqrt{\frac{4\pi G}{\xi_n p_{\rm ext}}}\frac{q^{\frac{2-n}{2n}}}{1-q^{2/n}}\nonumber\\
                &&\times \frac{\sqrt{y_n}(1-y_n)}{[1+(n-1)\sqrt{y_n}X_n(y_n)]},
\end{eqnarray}
which is identical with Eqs.~\ref{eq_pdfsphere} and \ref{eq_pdfsphere2}.

\section{\label{app_meanpdfasymptotes}Asymptotes at high and low mass surface densities for $q-averaged$ PDFs}

 In this appendix we provide the asymptotes at high and low mass surface 
densities of the mean PDF for a considered distribution 
of the pressure ratios $q$ of an ensemble of isothermal, self-gravitating, and pressurized 
spheres or cylinders as shown in Figs.~\ref{fig_pdfisosphmean} and \ref{fig_pdfcylmean}. 
For the sake of simplicity, for Bonnor-Ebert spheres the analytical 
approximations for low and high overpressure are considered. 
The external pressure $p_{\rm ext}$ is considered the same for all clouds in the sample.

For convenience, we introduce the unit-free radius defined by
\begin{equation}
        \hat r_{{\rm cl},n} = \left(\frac{K\sqrt{\xi_n}}{\sqrt{4\pi G p_{\rm ext}}}\right)^{-1} r_{{\rm cl},n}
                = q^{-\frac{1}{n}+\frac{1}{2}}\sqrt{1 - q^{2/n}}.
\end{equation}
The mean PDF as defined by Eq.~\ref{eq_pdfmean} at a given mass surface density $\Sigma_n$ is then given by the integral
\begin{equation}
        \label{eq_pdfmeanq}
                \left<P_{\rm cl}(\Sigma_n)\right> = \frac{1}{\tilde C}\int_{q_{\rm min}^{-1}}^{\infty} {\rm d}q^{-1}\frac{1}{C}
                P(q^{-1})\,\hat r_{{\rm cl},n}^\kappa P_{\rm cl}(\Sigma_n,q),
\end{equation}
where $P(q^{-1})$ is the distribution function of the overpressures $q^{-1}$ given by Eq.~\ref{eq_probstates},
$P_{\rm cl}(\Sigma_n,q)$  the conditional PDF under the pressure ratio $q$, and 
where $\tilde C$ is a normalization constant given by
\begin{equation}
        \label{eq_normconstant}
        \tilde C =  \int_{1}^{\infty}{\rm d}q^{-1}\,\frac{1}{C} P(q^{-1})\,\hat r^\kappa_{{\rm cl},n},
\end{equation}
where $\kappa=2$ for spheres and $\kappa=1$ for cylinders.
The lowest overpressure $q_{\rm min}^{-1}$ is related to the central mass surface density 
with $\Sigma_n(0) = \Sigma_n$.

\subsection{\label{app_meanpdfsphasymptotes}Sphere}

\subsubsection{\label{app_msurfapproxlargesphere}Asymptote at high mass surface densities}

As shown in Paper~I and in Sect.~\ref{sect_pdfsphapproxlarge}, 
the PDF of spheres with high overpressure 
asymptotically approaches within the limit of high mass surface densities 
 the power laws given by Eq.~\ref{eq_pdfsphasymptotelarge2}.
For the product we derive\begin{equation}
        \hat r_{{\rm cl},n}^2 P_{\rm sph}(\Sigma_n) \sim
        \frac{2^{\frac{n+1}{n-1}}}{n-1} \Sigma_n^{-\frac{n+1}{n-1}}\zeta_n^{\frac{2}{n-1}} q^{\frac{n-2}{n-1}}
                                                                \left(\frac{\xi_n p_{\rm ext}}{4\pi G}\right)^{\frac{1}{n-1}}.
\end{equation}
Because the probabilities at high mass surface densities are related to
spheres with high overpressure (Sect.~\ref{sect_pdfisosphere}), 
we can simplify the probability distribution Eq.~\ref{eq_probstates}
of the overpressure $q^{-1}$ through the 
corresponding power-law asymptote in the limit $q^{-1} \gg q^{-1}_0${, so that
$P(q^{-1}) \sim C q_0^{-k_1-k_2} q^{k_2}$}. For the integral \ref{eq_pdfmeanq} we obtain
\begin{eqnarray}
        \left<P_{\rm sph}(\Sigma_n)\right>  
                &\approx & \frac{1}{\tilde C}\Sigma_n^{-2 k_2-1} \frac{2 q_0^{-k_1-k_2}}{(n-1)k_2-1} \nonumber\\
                        &&\times \left(\frac{\xi_n p_{\rm ext}}{4\pi G}\right)^{k_2}
                        {\rm B}\left(\frac{n-1}{2},\frac{1}{2}\right)^{2 k_2},
\end{eqnarray}
where we have replaced the lowest overpressure $q_{\rm min}^{-1}$ by the central mass surface density
using Eq. \ref{eq_msurfapproxhigh}.

The estimate of the normalization constant $\tilde C$ (Eq.~\ref{eq_normconstant}) 
is straightforward and is\begin{eqnarray}
        \tilde C  &=& \frac{q_0^{-k1}}{\gamma}\Big\{q_0^{-\frac{2}{n}}
                                {\rm B}\left(a ,b \right){\rm I}_\psi\left(a,b \right)\nonumber\\
                &&\quad\quad- {\rm B}\left(a',b'\right){\rm I}_\psi\left(a',b'\right)\Big\},
\end{eqnarray}
where 
\begin{equation}
        {\rm I}_\psi(a,b) = \frac{1}{{\rm B}(a,b)}\int_0^\psi{\rm d}t\,t^{a-1}(1-t)^{b-1}
\end{equation}
is the normalized incomplete beta-function with
$\psi = 1/(1+q_0^\gamma)$ and where
\begin{equation}
        a = \frac{k_2-\frac{2}{n}}{\gamma},\, b = \frac{k_1+\frac{2}{n}}{\gamma},
                \, a' = \frac{k_2}{\gamma},\, b' = \frac{k_1}{\gamma}.
\end{equation}

\subsubsection{\label{app_msurfapproxsmallsphere}Asymptote at low mass surface densities}

In a similar manner as in the previous section, we can derive an asymptote of the mean PDF for
low mass surface densities.
As discussed in Sect.~\ref{sect_pdfsphapproxsmall}, the
PDF of individual clouds approaches in the limit of  low mass surface densities
a simple power law given by Eq.~\ref{eq_pdfsphapproxsmall}. We find for 
the product of the square of the unit-free radius and the PDF the approximation
\begin{equation}
        \hat r^2_{{\rm cl},n}P_{\rm sph}(\Sigma_n)\sim \Sigma_n\frac{1}{2} \frac{4\pi G}{\xi_n p_{\rm ext}}.
\end{equation}
The value does not depend on the overpressure $q^{-1}$. 

In the limit of low mass surface densities we can approximate $q^{-1}_{\rm min}\approx 1$.
The mean asymptotic value of the PDF is then given by
\begin{eqnarray}
        \left<P_{\rm sph}(\Sigma_n)\right> &\sim & \frac{1}{\tilde C} \Sigma_n \frac{q_0^{-k_1-1}}{2\gamma} 
                                        \frac{4\pi G}{\xi_n p_{\rm ext}} {\rm B}(a,b){\rm I}_\psi(a,b)
\end{eqnarray}
with $\psi \approx 1/(1+q_0^\gamma)$ and 
\begin{equation}
        a = \frac{1}{\gamma}(k_2-1),\quad b = \frac{1}{\gamma}(k_1+1).
\end{equation}
The slope of the asymptotic power law for low 
mass surface densities is the same as for individual clouds.

\subsection{\label{app_meanpdfcylasymptotes}Cylinder}

\subsubsection{Asymptote at high mass surface densities}
To estimate the mean value of the PDF for high mass surface densities we can  apply the approximation
Eq.~\ref{eq_msurfcylapproxhigh}.
For the product of radius and PDF we obtain
\begin{eqnarray}
        \hat r_{{\rm cl},4}P(\Sigma_4)& \approx & \frac{1}{3}
                        \left(\frac{\pi}{2}\frac{1}{\cos i}\right)^{\frac{1}{3}}
                                \left(\frac{\xi_4 p_{\rm ext}}{4\pi G}\right)^{\frac{1}{6}}\nonumber\\
                                &&\times\Sigma_4^{-\frac{4}{3}} q^{\frac{1}{3}}\left[1-(q^{-1}_{\rm min}/ q^{-1})^{\frac{1}{3}}\right]^{-\frac{1}{2}},
\end{eqnarray}
where we have replaced the central mass surface density by the asymptotic value 
given by Eq.~\ref{eq_msurfapproxhigh_n4} valid in the limit of high overpressure. 
The same replacement has been made for the fixed mass surface density $\Sigma_4$ in the denominator,
which is related to the lowest overpressure $q_{\rm min}^{-1}$.
To estimate the integral \ref{eq_pdfmeanq} we can make the same simplification
as   in App.~\ref{app_msurfapproxlargesphere}
by replacing the probability distribution Eq.~\ref{eq_probstates}
by the corresponding power law $P(q^{-1})\sim C q_0^{-k_1-k_2}q^{k_2}$. 
For the mean PDF of self-gravitating cylinders
we obtain as the asymptote at high mass surface densities
\begin{eqnarray}
        \left<P_{\rm cyl}(\Sigma_4)\right> 
        &\approx & \frac{1}{\tilde C}\Sigma_4^{-2 k_2}
                 \,{\rm B}\left(3 k_2-2,\frac{1}{2}\right) q_0^{-k_1-k_2}
                \nonumber\\
                        &&\times \left(\frac{\xi_4 p_{\rm ext}}{4\pi G}\right)^{k_2-\frac{1}{2}}
                        \left(\frac{\pi}{2}\frac{1}{\cos i}\right)^{2 k_2-1}.
\end{eqnarray}

To provide a simple analytical estimate for the normalization constant $\tilde C$
we can consider the two extreme cases we discussed here. 
For a distribution in $q^{-1}$ with $q^{-1}_0\gg 1$
the radius can be simplified to $\hat r_4\approx q^{\frac{1}{4}}(1-0.5q^{1/2})$. 
The estimate of the integral~\ref{eq_normconstant} is again straightforward
and the normalization constant becomes
\begin{eqnarray}
        \tilde C  & \approx & 
                  \frac{q_0^{-k_1-\frac{3}{4}}}{\gamma}\Big\{
                                {\rm B}\left(a ,b \right){\rm I}_\psi\left(a,b \right)\nonumber\\
                                &&\quad\quad\quad
                - \frac{q_0^{\frac{1}{2}}}{2}{\rm B}\left(a',b'\right){\rm I}_\psi\left(a',b'\right)\Big\},
\end{eqnarray}
where 
\begin{equation}
        a = \frac{k_2-\frac{3}{4}}{\gamma},\, b = \frac{k_1+\frac{3}{4}}{\gamma},
                \, a' = \frac{k_2-\frac{1}{4}}{\gamma},\, b' = \frac{k_1+\frac{1}{4}}{\gamma}.
\end{equation}
For $q^{-1}_0\ll 1$, the distribution of overpressure is a simple power law
$P(q^{-1})\sim C q_0^{-k_1-k_2} q^{k_2}$ and the integral can be solved without further
simplification. The normalization constant becomes
\begin{equation}
        \tilde C = 2 q_0^{-k_1-k_2} {\rm B}\left(2k_2-\frac{3}{2},\frac{3}{2}\right).
\end{equation}

\subsubsection{Asymptote at low mass surface densities}

In the limit of low mass surface densities it follows from Eqs.~\ref{eq_pdfsphapproxsmall} 
and~\ref{eq_pdfcylinder} with \ref{eq_pdfcylinder2}
that the asymptote of the PDF in the limit of low mass surface densities is given by
\begin{equation}
        P_{\rm cyl}(\Sigma_4) \approx \frac{1}{1-q^{1/2}}\frac{\pi G}{\xi_4 p_{\rm ext}}\frac{1}{x} q^{-\frac{1}{2}}\Sigma_4.
\end{equation}
For low mass surface densities the mean PDF is dominated by $x\approx 1$.
For the product of radius and PDF we obtain
\begin{equation}
        \hat r_{{\rm cl},4} P_{\rm cyl}(\Sigma_4) \approx \frac{\pi G}{\xi_4 p_{\rm ext}}\Sigma_4q^{-\frac{1}{4}}\frac{1}{\sqrt{1-q^{1/2}}}.
\end{equation}

For large $q^{-1}_0 \gg 1$ the contribution to the mean PDF is related to high overpressures so that
we can simplify the expression by removing the square root using $1/\sqrt{1-q^{1/2}}\approx 1+0.5\,q^{1/2}$.
The asymptote of the mean PDF at low mass surface densities becomes
\begin{eqnarray}
        \left<P_{\rm cyl}(\Sigma_4)\right> &\approx & \frac{1}{\tilde C}\Sigma_4\frac{\pi G}{\xi_4 p_{\rm ext}}
                \frac{q^{-k_1-\frac{5}{4}}}{\gamma}
                \Big\{{\rm B}(a,b){\rm I}_\psi(a,b)\nonumber\\
                        &&\quad+\frac{q^{\frac{1}{2}}}{2}{\rm B}(a',b'){\rm I}_\psi(a',b')\Big\},
\end{eqnarray}
where
\begin{equation}
        a = \frac{k_2-\frac{5}{4}}{\gamma},\, b = \frac{k_1+\frac{5}{4}}{\gamma},
                \, a' = \frac{k_2-\frac{3}{4}}{\gamma},\, b' = \frac{k_1+\frac{3}{4}}{\gamma}.
\end{equation}

For the limit $q^{-1}_0\ll 1$, where the probability distribution $P(q^{-1})$ becomes a simple
power law, we find
\begin{equation}
        \left<P_{\rm cyl}(\Sigma_4)\right> \approx \frac{1}{\tilde C} \Sigma_4\frac{\pi G}{\xi_4 p_{\rm ext}}
                2 q_{\rm 0}^{-k_1-k_2}{\rm B}\left(2 k_2-\frac{5}{2},\frac{1}{2}\right).
\end{equation}

\end{document}